\documentclass[reprint,prx,aps,amsmath,amssymb,floatfix,superscriptaddress,longbibliography]{revtex4-2}

\usepackage{mathtools,amsmath}
\usepackage{graphicx} 
\usepackage[breaklinks=true,colorlinks,citecolor=teal,linkcolor=teal,urlcolor=teal]{hyperref}
\usepackage{siunitx}
\usepackage{accents}
\usepackage{bbold}
\usepackage{soul}
\usepackage{bm}
\usepackage{multirow}
\usepackage[normalem]{ulem}
\usepackage{times}
\usepackage{enumerate}  
\usepackage{float}
\usepackage{amssymb}
\usepackage{xcolor} 
\usepackage{booktabs}

\begin{document}


\title{Tunable Superconducting Quantum Interference Device Coupler for Fluxonium Qubits}
\author{Abhishek Chakraborty}
\affiliation{Department of Physics and Astronomy, University of Rochester, Rochester, New York 14627, USA}
\affiliation{Institute for Quantum Studies, Chapman University, Orange, California 92866, USA}
\author{Bibek Bhandari}
\affiliation{Institute for Quantum Studies, Chapman University, Orange, California 92866, USA}

\author{D. Dominic Brise\~no-Colunga}
\affiliation{Institute for Quantum Studies, Chapman University, Orange, California 92866, USA}
\author{Noah Stevenson}
\affiliation{Quantum Nanoelectronics Laboratory, Department of Physics, University of California, Berkeley, California 94720, USA}

\author{Zahra Pedramrazi}
\affiliation{Quantum Nanoelectronics Laboratory, Department of Physics, University of California, Berkeley, California 94720, USA}
\affiliation{Applied Mathematics and Computational Research Division, Lawrence Berkeley National Laboratory, Berkeley, California 94720, USA}
\author{Chuan-Hong Liu}
\affiliation{Quantum Nanoelectronics Laboratory, Department of Physics, University of California, Berkeley, California 94720, USA}
\affiliation{Applied Mathematics and Computational Research Division, Lawrence Berkeley National Laboratory, Berkeley, California 94720, USA}

\author{David I. Santiago}
\affiliation{Quantum Nanoelectronics Laboratory, Department of Physics, University of California, Berkeley, California 94720, USA}
\affiliation{Applied Mathematics and Computational Research Division, Lawrence Berkeley National Laboratory, Berkeley, California 94720, USA}
\author{Irfan Siddiqi}
\affiliation{Quantum Nanoelectronics Laboratory, Department of Physics, University of California, Berkeley, California 94720, USA}
\affiliation{Applied Mathematics and Computational Research Division, Lawrence Berkeley National Laboratory, Berkeley, California 94720, USA}

\author{Justin Dressel}
\affiliation{Institute for Quantum Studies, Chapman University, Orange, California 92866, USA}
\affiliation{Schmid College of Science and Technology, Chapman University, Orange, California 92866, USA}

\author{Andrew N. Jordan}
\affiliation{Department of Physics and Astronomy, University of Rochester, Rochester, New York 14627, USA}
\affiliation{Institute for Quantum Studies, Chapman University, Orange, California 92866, USA}
\affiliation{Schmid College of Science and Technology, Chapman University, Orange, California 92866, USA}
\affiliation{The Kennedy Chair in Physics, Chapman University, Orange, California 92866, USA}

\date{\today}

\begin{abstract}
Tunable couplers enable high-fidelity two-qubit gates leveraging high on/off coupling ratios and reduced crosstalk within a single design. We investigate a galvanically connected direct-current superconducting quantum interference device (dc SQUID) as a minimal tunable coupling element for fluxonium qubits. Comparing grounded and floating fluxonium designs, we find that the latter contains an extra sloshing mode that strongly hybridizes with the qubit modes, leading to significant static ZZ crosstalk. In contrast, the grounded design avoids this issue and allows suppression of static ZZ crosstalk using a shunting capacitor. Leveraging fast flux control, we present two schemes to implement two-qubit gates, predicting a $\sqrt{i\text{SWAP}}$-like gate with $99.9\%$ open-system fidelity in less than 6 nanoseconds assuming modest relaxation and dephasing rates.
\end{abstract}

\maketitle
\section{Introduction}\label{sec:introduction}

Superconducting circuits \cite{vool2017introduction,Blais2020circuit,kjaergaard2020superconducting,krantz2019quantum,rasmussen2021superconducting} are a leading platform for building scalable quantum processors, enabled by the realization of high-fidelity single and two-qubit gates across different qubit encodings~\cite{grimm2020stabilization,reglade2024quantum,hajr2024high,bhandari2024symmetrically, ding2023high}, critical for quantum error correction \cite{google2023suppressing, livingston2022experimental} beyond break-even. Their engineering flexibility in terms of energy spectra and interactions \cite{nguyen2024programmable, kounalakis2018tuneable} enables quantum simulation and the study of complex many-body quantum phenomena \cite{frey2022realization,bernien2017probing} at an unprecedented scale. 

Fluxonium qubits \cite{manucharyan2009fluxonium, zhu2013circuit}, especially in the low-frequency regime \cite{earnest2018realization} have achieved hundreds of microseconds \cite{nguyen2019high,thibodeau2024the} to millisecond \cite{somoroff2023milli, ding2021free, ding2023high} coherence times. Their inherent protection against flux-induced dephasing at their sweet spot makes them a promising candidate for the development of scalable quantum processors \cite{long2022blueprint}. Futhermore, fluxonium displays strikingly different properties compared to the transmon \cite{koch2007charge}. The phase matrix elements for fluxonium are significantly larger than the charge matrix elements in the computational subspace, making this qubit an ideal testbed for flux-controlled \cite{zhang2021universal} inductive coupling and readout \cite{smith2016quantization}. 

For two-qubit gates, direct coupling can be realized using simple circuit designs but introduces crosstalk which limits gate performance and, consequently, scalability. Nevertheless, there are many recent proposals for two-qubit gates between capacitively coupled fluxonium qubits \cite{chen2022fast, nesterov2018microwave, nesterov2022cnot, nesterov2021proposal}, as well as a demonstration of a cross-resonance CNOT gate with $99.49\%$ gate fidelity~\cite{dogan2023twofluxonium}. Direct (galvanic) inductive coupling has been less explored but a recent design with a shared junction between the two fluxonium circuits \cite{lin2025verifying} demonstrated a CNOT gate with over $99.9\%$ fidelity \cite{lin202524days}. 

Tunable couplers enable qubit-qubit interactions during gate operations, while leaving the qubits decoupled otherwise, but require more complex designs and control. Encouraged by the success of tunable couplers in transmon-based architectures \cite{sete2021floating, liang2023tunable, campbell2023modular}, a variety of tunable coupler designs have been proposed for the fluxonium qubits, inspired by the scheme presented in Ref.~\cite{yan2018tunable}. In this scheme, the coupler mode interacts dispersively with the qubits and parametric modulation of the coupler frequency enables high-fidelity gate operations. Refs.~\cite{moskalenko2021tunable, moskalenko2022high} demonstrated two-qubit gates with fidelity greater than $99\%$ using a capacitively-coupled tunable fluxonium coupler. With an additional harmonic mode enabling multi-path cancellation of the static ZZ interaction,  Ref.~\cite{ding2023high} demonstrated a \textsc{CZ} gate between two fluxonium qubits with a capacitively coupled tunable transmon coupler with fidelity greater than $99.9\%$. Inspired by the fluxonium molecule circuit \cite{kou2017fluxonium} and the gmon coupler for transmons \cite{chen2014qubit}, Refs.~\cite{weiss2022fast,  zhang2024tunable} realized a $\sqrt{i\text{SWAP}}$ gate with fidelity exceeding $99.7\%$ between two grounded fluxonium qubits galvanically connected to a shared Josephson junction using strong flux modulation. 
 
In this paper, we continue exploring designs for tunable couplers by considering a galvanically connected direct-current superconducting quantum interference device (dc SQUID) as a minimal tunable coupling element for fluxonium qubits. The dc SQUID was previously analyzed to control the hopping and non-linear cross-Kerr interaction between two floating transmon qubits~\cite{kounalakis2018tuneable}. A shared and grounded dc SQUID was also used to demonstrate microwave-activated parametric two-qubit gates \cite{jin2023fast} and pulsed parametric readout \cite{noh2023strong} in grounded transmon devices. However, its utility for flux-dominant qubits like fluxonium has not been fully explored.

In our design, one circuit node from each qubit is galvanically connected to a shared SQUID loop on opposite sides. We consider both the grounded and floating fluxonium circuit designs and compare both their energy spectra and static ZZ coupling when the coupler is off. We quantify the impact of SQUID junction asymmetry on the interaction at the off-point and propose a simple mitigation strategy for the design with grounded qubits. In the floating design, extra circuit modes hybridize strongly with the qubit modes, making the design more susceptible to environmental noise and limiting its feasibility. Finally, we propose two schemes to realize fast, high-fidelity two-qubit gates in the grounded design using flux drives on the qubit and the coupler. 

The paper is structured as follows: In Sec.~\ref {sec:squid-coupler}, we describe the two different circuit designs, the energy spectra and static ZZ interaction strength. In Sec.~\ref{sec:two-qubit-gates}, we describe an effective two-level model and present numerical simulations of a two-qubit $\sqrt{i\text{SWAP}}$-like gates for the grounded design. Finally, we summarize our findings and outline future directions for improving this tunable coupler design in Sec.~\ref{sec:conclusions}.

\section{SQUID-Coupled Fluxonium}\label{sec:squid-coupler}
We consider two different fluxonium designs: grounded and floating. In each design, the two qubit modes are labeled ${\mathrm {A}}$ and ${\mathrm {B}}$, with external fluxes $\Phi_{\mathrm{A}}$ and $\Phi_{\mathrm{B}}$. The external magnetic flux threading the dc SQUID loop is labeled $\Phi_{\mathrm{S}}$ which we will call ``coupler flux" for brevity.

\subsection{Grounded Fluxonium Qubits}
\label{sec:grounded}

\begin{figure}[tbp]
    \centering
    \includegraphics[width=\linewidth]{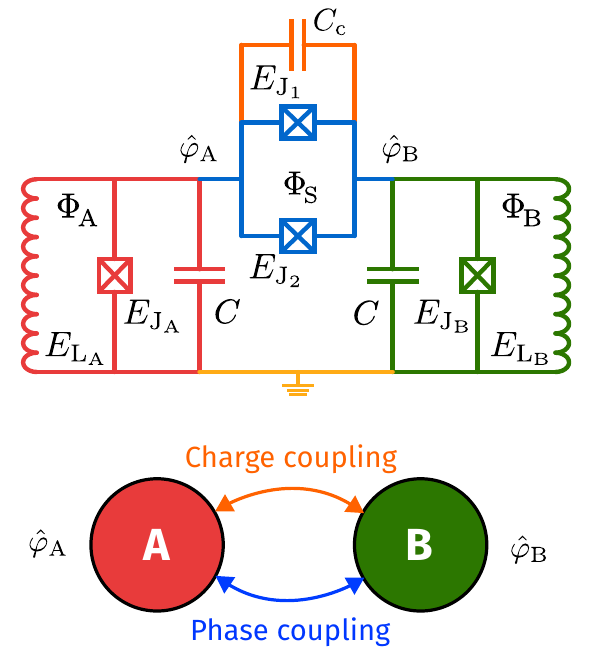}
    \caption{Circuit diagram for two grounded fluxonium qubits galvanically connected to a floating dc SQUID. The capacitor $C_{\mathrm c}$ is the effective capacitance shunting the dc SQUID that includes contributions from the junctions. For the grounded design, $\hat \varphi_{\mathrm {A/B}}$ are the relevant modes for the fluxonium A/B.}
    \label{fig:grounded-design}
\end{figure}

The circuit Hamiltonian for the grounded design, shown in Fig.~\ref{fig:grounded-design}, is
\begin{equation}
    \hat{H}_{\mathrm{gr}} = \hat{H}_{\mathrm {A}} + \hat{H}_{\mathrm {B}} + \hat{H}_{\mathrm C} +  \hat{H}_{\mathrm S},
    \label{eq:full-ham}
\end{equation}
where $\hat{H}_{\mu = {\{\mathrm {A, B}}\}}$ are Hamiltonians for the two fluxonium modes, given by \cite{manucharyan2009fluxonium}   
\begin{equation}
    \hat{H}_\mu = 4E_{{\mathrm C}_{\mu}} \hat{n}^2_\mu + \frac{1}{2}E_{{\mathrm L}_\mu} \left(\hat{\varphi}_\mu + \frac{2\pi\Phi_\mu}{\Phi_0}\right)^2 -E_{\mathrm{J_\mu}}\cos\hat{\varphi}_{\mu}, 
    \label{eq:indiv_flux_ham}
\end{equation}
We present a full derivation of the circuit Hamiltonian in Appendix~\ref{app:grounded-cqed}. The parameters $E_{{\mathrm C}_\mu}, E_{{\mathrm L}_\mu},\text{ and } E_{{\mathrm J}_\mu}$ denote the charging, inductive and Josephson energies of the fluxonium qubit mode labeled with subscript $\mu$. $\Phi_{\mu}$ is the magnetic flux threading the inductive loop, where $\Phi_0 = h/2e$ is the superconducting flux quantum. $\Phi_{\mu}$ is written in the inductor term, as appropriate for time-dependent flux drives \cite{riwar2022circuit, you2019circuit}. $\hat{\varphi}_\mu$ and $\hat{n}_\mu$ are reduced flux and canonically-conjugate reduced charge operators, respectively, for the qubit modes that satisfy the commutation relations $ \left[\hat{\varphi}_\mu, \hat{n}_\nu\right] = i\delta_{\mu\nu}$ \cite{krantz2019quantum}. The term $\hat{H}_{\mathrm{C}}$ in Eq.~\eqref{eq:full-ham} represents charge coupling between the two qubits effected by the capacitance $C_{\mathrm c}$ shunting the SQUID and is given by
\begin{equation}
\hat{H}_{\mathrm C} = J_{\mathrm c}\hat{n}_{\mathrm {A}}\hat{n}_{\mathrm {B}},
\label{eq:charge-coupling-ham}
\end{equation}
with interaction strength $J_{\mathrm c}={4e^2 C_{\mathrm c}}/{C (C + 2C_{\mathrm c})} $. 

In this design, we ideally have symmetric SQUID junctions, but realistic fabrication variance leads us to treat the general case allowing for asymmetry in the junctions. We define the total Josephson energy of the SQUID $E_{\mathrm J_\Sigma} =(E_{\mathrm J_1} + E_{\mathrm J_2}) $ and asymmetry parameter $d = (E_{\mathrm J_1} - E_{\mathrm J_2})/E_{\mathrm J_\Sigma}$, where $E_{\mathrm J_1}$ and $E_{\mathrm J_2}$ are the Josephson energies of the two SQUID junctions. We further assume that the inductive loop formed via ground can be flux-biased at a static value while still allowing independent flux control of the qubit and SQUID loops. The effects of the SQUID are then described using the Hamiltonian
\begin{equation}
    \hat H_{\mathrm S}= -E_{\mathrm J_\Sigma} \left[\cos\left( \frac{\pi\Phi_{\mathrm{S}}}{\Phi_0}\right)\cos\hat\varphi_{-} +d \sin\left( \frac{\pi\Phi_{\mathrm{S}}}{\Phi_0}\right) \sin\hat\varphi_{-}\right], 
    \label{eq:Squid_ham_gr}
\end{equation}
where $\Phi_{\mathrm S}$ is the coupler flux and $ \hat\varphi_{-} \equiv \hat \varphi_{\mathrm {A}} - \hat \varphi_{\mathrm {B}}$. The first term is the symmetric contribution, and the second term which is proportional to $d$ results from the asymmetry in the SQUID junctions. The effects of the SQUID can be separated into two parts 
\begin{equation}   
\hat{H}_{\mathrm S} =\hat H_{\mathrm {S,c}}+\sum\nolimits_{\mu = \{\mathrm {A, B}\}}\hat{H}_{{\mathrm S},\mu},
\label{eq:squid_terms_split}
\end{equation}
where each $\hat{H}_{{\mathrm S},\mu}$ is a correction to the corresponding  qubit Hamiltonian and $\hat H_{\mathrm {S,c}}$ denotes purely two-qubit coupling. We will now examine each term in detail. 

Fluxonium is protected from flux-induced dephasing at its sweet spot $(\Phi_{\mathrm S} = \Phi_0/2)$ where it has a symmetric double-well potential \cite{manucharyan2009fluxonium, lin2018demonstration}. 
The single-qubit correction terms only contain reduced flux operators for a single fluxonium qubit
\begin{equation}
\hat{H}_{{\mathrm S},\mu } = - E_{\mathrm{J_\Sigma}}\left[\cos\left( \frac{\pi\Phi_{\mathrm{S}}}{\Phi_0}\right)\cos{\hat{\varphi}_{\mu}}
\pm d \sin\left( \frac{\pi\Phi_{\mathrm{S}}}{\Phi_0}\right)\sin{\hat{\varphi}_{\mu}}\right],
\label{eq:fluxonium-ham-with-correction}
\end{equation}
where the second term in the square brackets is positive for qubit A and negative for qubit B. We refer to the first term on the right-hand side of Eq.~\eqref{eq:fluxonium-ham-with-correction} as the symmetric correction which is an even function in $\hat{\varphi}_\mu$. We call the second term, which is an odd function in $\hat{\varphi}_\mu$, the asymmetric correction. Thus, the effective Hamiltonian for each qubit becomes $\hat{H}_{\mu,\,\mathrm{ eff}} = \hat{H}_{\mu} + \hat{H}_{{\mathrm S},\mu} $ after incorporating the single-qubit corrections.

For $d = 0$, only the symmetric correction remains. The coupler flux $\Phi_{\mathrm S}$ changes the depth of the two degenerate wells but, keeps both qubits at their sweet spots, enabling high-fidelity two qubit gates protected from dephasing. 
To leading order in $\hat{\varphi}$'s, the two-qubit coupling contribution from the SQUID is
\begin{equation}
\hat H_{\mathrm {S,c}} = -E_{\mathrm J_\Sigma} \cos \left(
\frac{\pi\Phi_{\mathrm S}}{\Phi_0}\right)\hat \varphi_{\mathrm {A}}\hat \varphi_{\mathrm {B}} + \mathcal{O}(\hat \varphi^3) \quad (d=0).
\label{eq:gr_flux_coupling}
\end{equation}
The coupling between the qubits can be turned off by setting $\Phi_{\mathrm S} = \Phi_0/2$, which we call the ``off-point'' for the symmetric case. 

For $d \neq 0$, the asymmetric correction lifts the degeneracy between the two potential wells upon moving $\Phi_{\mathrm S}$ away from the symmetric off-point, rendering the qubits sensitive to flux noise. 
Importantly, coupling terms due to junction asymmetry appear only at third order in the flux operators and higher [see Appendix~\ref{app:grounded-cqed} for a higher-order expansion of Eq.~\eqref{eq:gr_flux_coupling}]. 


\subsection{Floating Fluxonium Qubits}
\label{sec:floating}

\begin{figure}[tbp]
    \centering
    \includegraphics[width=\linewidth]{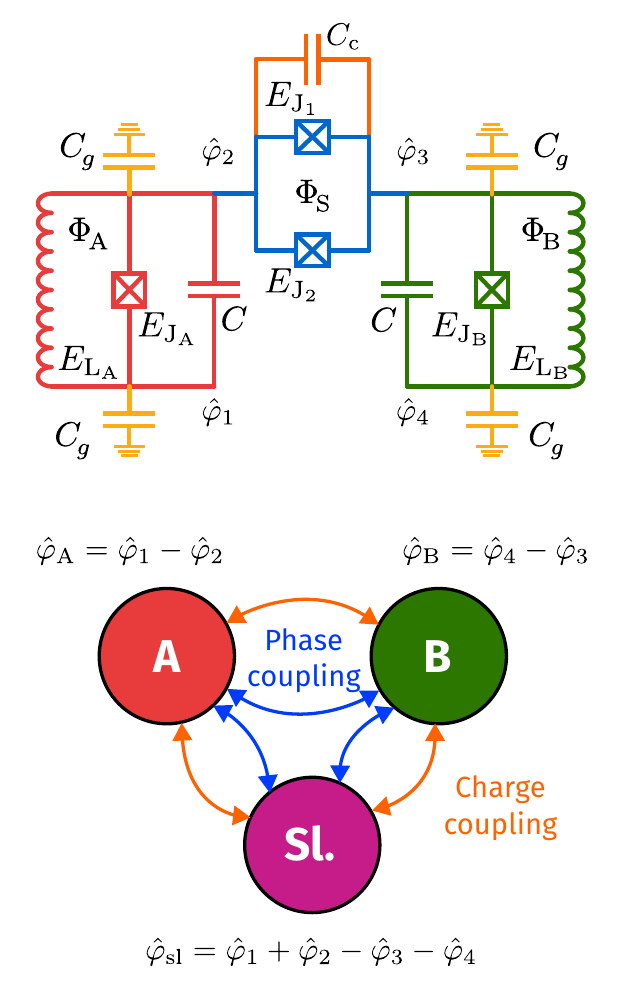}
    \caption{Circuit diagram for two floating fluxonium qubits galvanically connected to a floating dc SQUID. The capacitor $C_{\mathrm c}$ is the effective capacitance shunting the dc SQUID including contributions from the junctions. The relevant modes for fluxonium A and B are given by $\hat \varphi_{\mathrm {A}} = \hat \varphi_{\mathrm 1} - \hat \varphi_{\mathrm 2}$ and $\hat \varphi_{\mathrm {B}} = \hat \varphi_{\mathrm 3} - \hat \varphi_{\mathrm 4}$, respectively.}
    \label{fig:floating-design}
\end{figure}

We show the circuit for the floating fluxonium design in Fig.~\ref{fig:floating-design}. This circuit has two differential fluxonium modes denoted by the reduced flux operators $\hat \varphi_{\mathrm {A}} = \hat \varphi_1-\hat \varphi_2$ and  $\hat \varphi_{\mathrm {B}} = \hat \varphi_4-\hat \varphi_3$. Additionally, we expect a ``sloshing" mode $\hat \varphi_{\mathrm {sl}} = \hat \varphi_1 + \hat \varphi_2 - \hat \varphi_3 - \hat \varphi_4$ corresponding to Cooper-pairs tunneling across the SQUID loop \cite{kounalakis2018tuneable} with conjugate charge operator $\hat{n}_{\text{sl}}$. Finally, the overall circuit common mode $\hat \varphi_{\mathrm \Sigma} = \hat \varphi_1 + \hat \varphi_2 + \hat \varphi_3+ \hat \varphi_4$ decouples from all other modes if all grounding capacitors $C_g$ are approximately the same \cite{kounalakis2018tuneable} and can safely be ignored. The circuit Hamiltonian for this configuration is (see Appendix~\ref{app:floating-cqed} for a full derivation)
\begin{equation}
    \hat{H}_{\mathrm{fl}} = \hat H_{\mathrm {A}} + \hat H_{\mathrm {B}} + \hat H_{\mathrm C} +\hat H_{\text{sl}} +\hat H_{\text{C,\,sl}} +  \hat{H}^{\mathrm{fl}}_{\mathrm S} ,
    \label{eq:full-ham_fl}
\end{equation}
The first three terms in Eq.~\eqref{eq:full-ham_fl} are the same as in Eq.~\eqref{eq:full-ham}. The fourth term is the sloshing mode Hamiltonian
\begin{multline}
\hat{H}_{\mathrm {sl}} = 4E_{\mathrm {C_{sl}}}  (\hat{n}_{\mathrm {sl}}-n_{\mathrm g})^2 \\
- E_{\mathrm J_\Sigma} \left[ \cos \left(
\frac{\pi\Phi_{\mathrm S}}{\Phi_0}\right) \cos \hat{\varphi}_{\mathrm {sl}} - d\sin \left(
\frac{\pi\Phi_{\mathrm S}}{\Phi_0}\right) \sin \hat{\varphi}_{\mathrm {sl}}\right ],
\label{eq:ham_breath}
\end{multline}
which has the same form as a Cooper-pair box \cite{nakamura1999coherent, bouchiat1998quantum} but with an effective Josephson energy that can be tuned with the coupler flux $\Phi_S$.  We include a gate charge $n_g$ explicitly since the sloshing mode is far from the transmon regime \cite{koch2007charge} and is thus expected to show charge dispersion for typical experimental parameters. For simplicity, we assume that this mode can be biased at the charge-degeneracy point $n_g=1/2$, although implementing such control might be challenging in practice. When biased at other points, the frequency of the sloshing mode is sensitive to charge noise and leads to additional dephasing of the qubits. The third term in Eq.~\eqref{eq:full-ham_fl} represents charge coupling between the qubits and the sloshing mode
\begin{equation}
\hat H_{\mathrm { C, sl}}  = J_{\mathrm {sl}} (\hat{n}_{\mathrm {A}} -\hat{n}_{\mathrm {B}})\hat{n}_{\mathrm {sl}}.
\label{eq:charge_br}
\end{equation}
The interaction strength $J_{\mathrm {sl}}$ and the charging energy $E_{\mathrm {C_{sl}}}$ are defined in the Appendix~\ref{app:floating-cqed}. We set $J_{\mathrm {sl}} = 0$ throughout this paper, assuming that the flux coupling is the dominant interaction mechanism. The interplay of charge and phase coupling in the floating design is outside the scope of this work.

Once again, we can distinguish two kinds of terms coming from the last term in Eq.~\eqref{eq:full-ham_fl}, $\hat H^{\mathrm{fl}}_{\mathrm S} = H^{\mathrm{fl}}_{\mathrm {S,c}} + \sum_\mu H^{\mathrm{fl}}_{\mathrm S,\mu} $, with $H^{\mathrm{fl}}_{\mathrm S,\mu}$ denoting corrections to each qubit potential and $H^{\mathrm{fl}}_{\mathrm {S,c}}$ representing pairwise phase coupling terms between all three modes. To leading order in flux operators, this inductive coupling has the form 
\begin{multline}
    H^{\mathrm{fl}}_{\mathrm {S,c} } = -E_{\mathrm J_\Sigma} \cos \left(\frac{\pi\Phi_{\mathrm S}}{\Phi_0}\right) \left[\frac{\hat \varphi_{\mathrm {A}} \hat \varphi_{\mathrm {B}}}{4} + \frac {(\hat{\varphi}_{\mathrm {A}} - \hat{\varphi}_{\mathrm {B}})}{2}\hat{\varphi}_{\mathrm {sl}}\right]\\
    + \mathcal{O}(\hat{\varphi}^3).
    \label{eq:coupl_float}
\end{multline}
Like the grounded design, the effects of junction asymmetry on the coupling strength appear only at higher order. The symmetric $(d=0)$ off-point for this design is $\Phi_{\mathrm S} = \Phi_0/2$, where all modes are decoupled from each other. For finite junction asymmetry $(d\neq0)$, parasitic couplings exist at this point and can be challenging to mitigate, as shown in \ref{app:floating-cqed}.

\subsection{Coupled Energy Spectra}
\label{sec:energy_spectra}

\begin{ruledtabular}
\begin{table}[tbp]
\centering
\begin{tabular}{>{\centering\arraybackslash}m{2cm}ccccc}
\toprule
        & $E_\mathrm{J}/h$ & $E_\mathrm{C}/h$ & $E_\mathrm{L}/h$ & $\omega/2\pi$ \\
        & \unit{\GHz}& \unit{\GHz}& \unit{\GHz}& \unit{\MHz} \\
\midrule
Qubit A & 3.8   & 1.0     & 1.0   & 0.646\\
Qubit B & 3.2   & 1.0     & 1.0   & 0.876\\
\midrule
        & $E_{\mathrm{J_\Sigma}}/h$ & $E_{\mathrm{C_{sl}}}/h$ & &\\
\midrule
& 7   & 3.4 &  &\\
\bottomrule
\end{tabular}
\caption{Default circuit parameters used for numerical simulations in this work. All quantities are in units of \unit{\GHz}.}
\label{tab:params}
\end{table}
\end{ruledtabular}

\begin{figure}
    \centering
    \includegraphics[width=\linewidth]{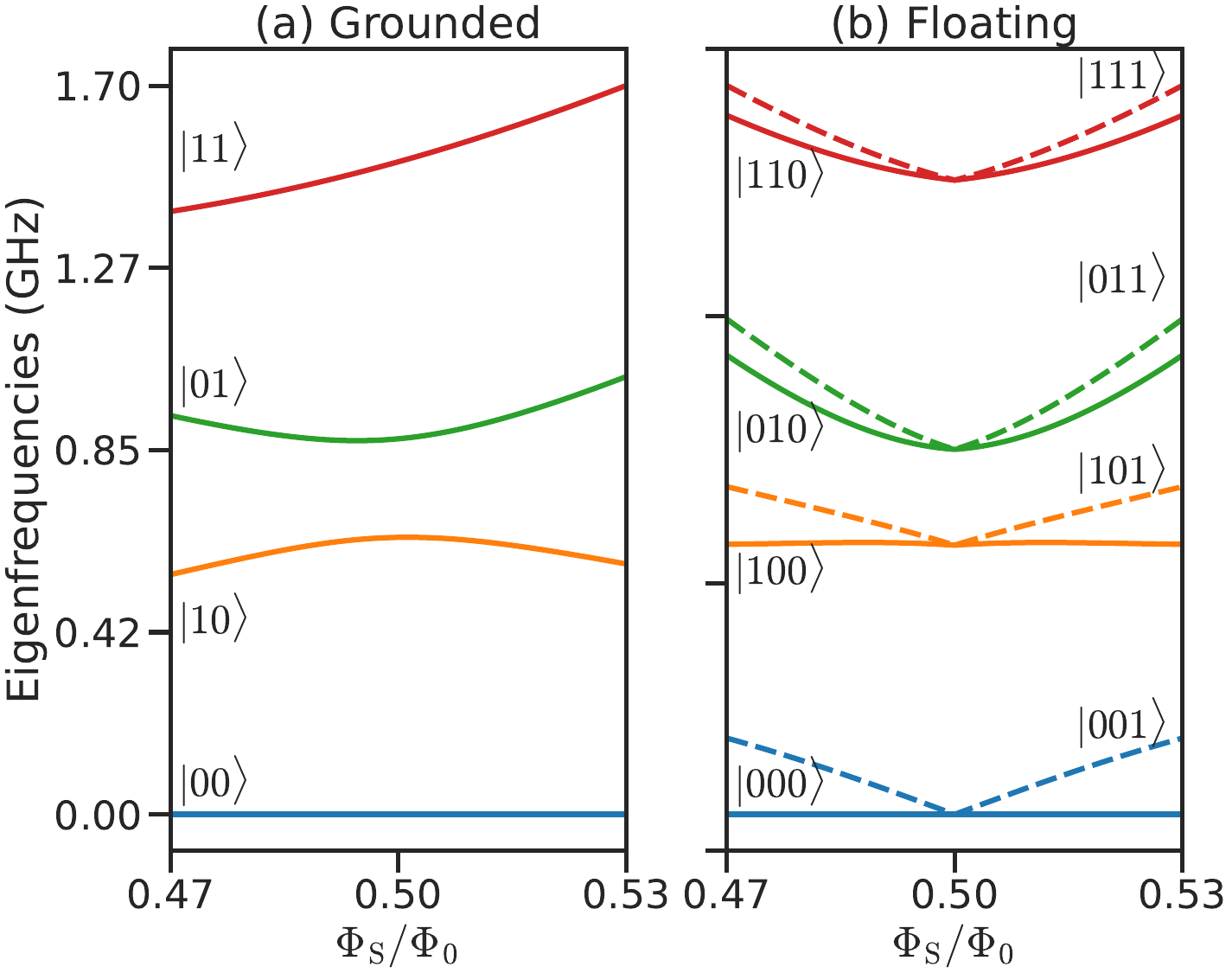}
    \caption{Energy spectrum as a function of external SQUID flux for two fluxonium qubits for (a) grounded and  (b) floating designs of the SQUID coupler for parameters in Table~\ref{tab:params}. In the floating design, levels where the sloshing mode is excited are shown as dashed lines. For both designs, we consider $E_{\mathrm J_{\Sigma}}/h = $\qty{7}{\GHz} with symmetric junctions. In the floating design, we considered $n_{\mathrm g} = 1/2$ for the sloshing mode.}
    \label{fig:spectrum}
\end{figure}

Next, we will investigate the energy spectra of the grounded and floating designs and highlight the effect of the sloshing mode on the energy spectra of individual fluxonium and the coupling between them. Figure~\ref{fig:spectrum} shows the energy spectrum for both grounded and floating designs with varying coupler flux $\Phi_{\mathrm S}$ for the parameters shown in Table.~\ref{tab:params}. Our circuit parameters were picked for the high-coherence fluxonium \cite{long2022blueprint} architecture. $E_\mathrm{J_\Sigma}$ and $E_\mathrm{C_{sl}}$ were chosen based on typical Joshephson energies and capacitance values seen in experiment. For both designs, we bias the fluxonium qubits at their respective sweet spots $\Phi_{\mathrm {A/B}} = \Phi_0/2$. We use $E_{lm(n)}$ to label eigenenergies corresponding to the eigenstates $|l_{\mathrm {A}} m_{\mathrm {B}} (n_{\mathrm {sl}})\rangle$ of Eq.~\eqref{eq:full-ham} for the grounded design and those of Eq.~\eqref{eq:full-ham_fl} for the floating design. The variables $l$, $m$ and $n$ indicate the number of excitations in the mode denoted by the corresponding subscript. The parentheses indicate that the sloshing mode index is only considered for the floating design. In both cases, we label the dressed states according to their overlap with the bare states, constructed by taking tensor products of individual eigenstates $|l_{\mathrm{A}}\rangle$, $|m_{\mathrm{B}}\rangle$ and $|n_{\mathrm{sl}}\rangle$ of $\hat{H}_{\mathrm {A}}$, $\hat{H}_{\mathrm {B}}$ and $\hat{H}_{\mathrm {sl}}$ respectively, where $|n_{\mathrm{sl}}\rangle$ eigenstates are only considered for the floating design. The overlap is calculated using the absolute value of the inner product $\left[\langle l_\text{A}|\otimes \langle m_\text{B} |\otimes (\langle n_\text{sl}|)\right]|l_{\mathrm {A}} m_{\mathrm {B}} (n_{\mathrm {sl}})\rangle$. The labeling away from the symmetric off-point indicates that the eigenstates adiabatically continue from the off-point eigenstates according to the labels, but strong hybridization away form the off-point means that the label indices are only useful in distinguishing the various dressed energy levels. 
Due to the large anharmonicity of fluxonium at its sweet spot, the computational states are well-separated from higher-energy states by a large gap, thereby suppressing leakage. At the symmetric off-point $\Phi_{\mathrm S} = \Phi_0/2$, all modes are effectively decoupled and the qubit eigenstates reduce to those of uncoupled fluxonium. 

For the grounded design spectrum in Fig.~\ref{fig:spectrum}(a), the splitting between $|01\rangle$ and $|10\rangle$ increases as we move away from the off-point. In addition, the $|11\rangle$ level clearly shows that the SQUID interaction is not symmetric about the off-point, leading to different dynamical ZZ interaction strengths on either side. This feature affects how we choose the flux offset for two-qubit gates, which will be described later in this article.

For the floating design spectrum in Fig.~\ref{fig:spectrum}(b), the dashed lines indicate a single excitation in the sloshing mode, which interacts with the qubit states when the coupler is activated $(\Phi_{\mathrm S} \neq \Phi_0/2)$. At the symmetric off-point, the potential energy for this mode vanishes (see Appendix~\ref{app:floating-cqed}), rendering all sloshing mode levels degenerate. In this configuration, the lowest energy scale in the system is set by the sloshing mode. A similar effect was observed with floating transmon qubits galvanically connected to a dc SQUID~\cite{kounalakis2018tuneable}. However, the effects are not as pronounced in that case since transmon qubit frequencies are far detuned from the extra mode levels. Since fluxonium qubits, especially heavy fluxonium, operate at significantly lower frequencies compared to transmons, we find that excited states of the sloshing mode hybridize more strongly with the fluxonium modes away from the off-point in our case. As a result, this mode poses a greater challenge for low-frequency qubits such as fluxonium. Away from the symmetric off-point, the strong coupling hybridizes the fluxonium modes into symmetric and antisymmetric superpositions of the bare eigenstates~\cite{kounalakis2018tuneable}. The $|100\rangle$ level now corresponds to the antisymmetric superposition, corresponding to no charge oscillations across the SQUID, and hence does not tune with the coupler flux $\Phi_{\mathrm S}$. The $|010\rangle$ mode corresponds the symmetric superposition and remains flux-tunable. 

\subsection{Static ZZ Interaction}
\label{sec:static_ZZ}
Tunable couplers leverage a high on-off ratio to enable high-fidelity gates. Consequently, it is important to ensure that the off-point configuration is free of parasitic interactions \cite{yan2018tunable, sung2021realization}. One such parasitic interaction is static ZZ coupling between qubits, which appear as energy shifts in one qubit conditioned on the state of the other, leading to correlated phase errors that complicate idling and limit the fidelity of gate operations \cite{fors2024comprehensive}. Below, we examine the static ZZ at the symmetric off-point of both designs. 

In the grounded design, the static ZZ interaction originates from coupling between the $|11\rangle$ state and the $|02\rangle$ and $|20\rangle$ states. Fluxonium qubits typically have a large positive anharmonicity, so the higher excited stated are far-detuned and weakly coupled to the computational states. The galvanic coupling of the dc SQUID yields stronger coupling between these levels despite high detuning. We quantify the static ZZ interaction strength in the grounded circuit with the energy difference
\begin{equation}
\label{eq:static_zz_grounded}
    \zeta^{\mathrm {gr}}_{\mathrm {ZZ}} = (E_{11} - E_{01} - E_{10} + E_{00})/h. 
\end{equation} 

For the floating circuit, the sloshing mode also interacts with the two fluxonium modes, leading to multiple ZZ-type interactions. For now, we only consider the static ZZ interaction between the two fluxonium modes (assuming the sloshing mode remains in its ground state) given which we quantify with the energy difference
\begin{equation}
\label{eq:static_zz_floating}
    \zeta^{\mathrm{fl}}_{\mathrm {ZZ}} = (E_{110} - E_{010} - E_{100} + E_{000})/h.
\end{equation} 
In both designs, static ZZ interaction is absent at the symmetric off-point in the case of perfectly symmetric SQUID junctions $(d=0)$.
\begin{figure}
    \centering
    \includegraphics[width=\linewidth]{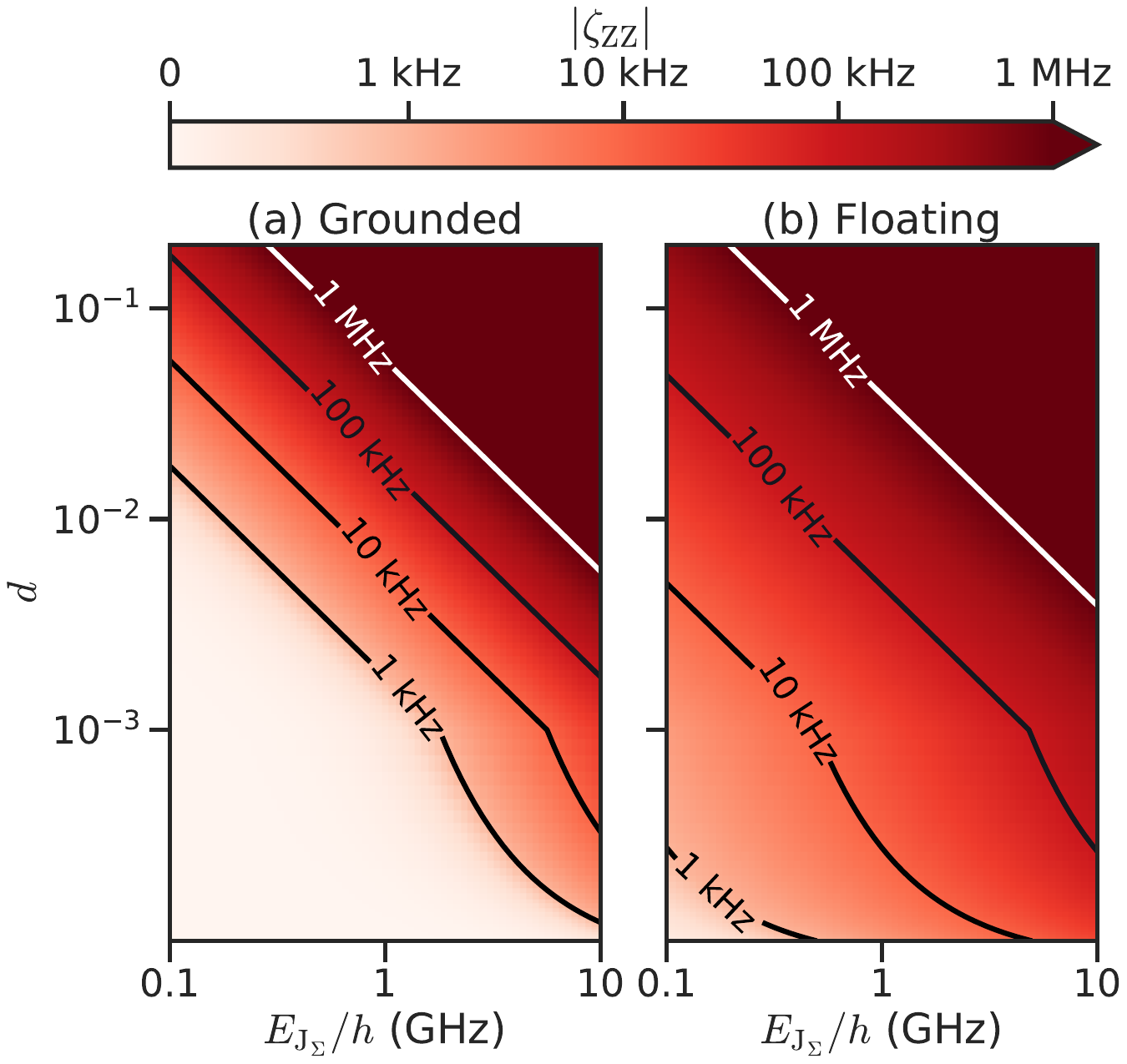}\\
    \caption{Static ZZ interaction strength at the symmetric off-point $\Phi_{\mathrm{S}}=\Phi_0/2$ for (a) grounded and (b) floating design plotted as a function of the total Josephson energy of the SQUID junction $E_{\mathrm J_\Sigma}$ and the asymmetry in the SQUID junctions $d$. We assume that the charge coupling is negligible $(J_{\mathrm{c}}=0)$ compared to the phase coupling.} 
    \label{fig:static-zz-asym}
\end{figure}

When junction asymmetry is present $(d\neq0)$ and the charge coupling is also negligible $(J_{\mathrm{c}}=0)$ to the phase coupling, a finite residual coupling exists at the symmetric off-point due to the asymmetry terms in the coupling Hamiltonian (see Appendix~\ref{app:circuit-qed}). This residual coupling shifts the energy of the eigenstates and leads to a static ZZ interaction. We numerically calculate the strength of this interaction with varying $E_{\mathrm J_\Sigma}$ and $d$ for both designs and compare them in Fig.~\ref{fig:static-zz-asym} using parameters given in Table~\ref{tab:params}.  Note that the static ZZ interaction strength increases along either axis since the coefficient of the asymmetry term is a product of both $d$ and $E_{\mathrm J_\Sigma}$.

The static ZZ interaction is suppressed at lower $E_{\mathrm J_\Sigma}$. Following the contours of constant $|\zeta_{\mathrm{ZZ}}|$ in Fig.~\ref{fig:static-zz-asym}, we see that maintaining a certain $\zeta_{\mathrm {ZZ}}$ budget permits greater asymmetry when $E_{\mathrm J_\Sigma}$ is smaller. Notably, the grounded circuit yields nearly an order of magnitude smaller static ZZ coupling with asymmetry compared to the floating design, Moreover, the area under the \qty{1}{\kHz} contour in panel (a) is substantially larger than in panel (b), indicating that the grounded architecture maintains low $\zeta_{\mathrm {ZZ}}$ over a broader range of fabrication imperfections. The worse performance of the floating design stems from the sloshing mode, which introduces additional coupling channels between the fluxonium modes. Specifically, the coupling mediated by the sloshing mode is twice as strong as the direct qubit-qubit interaction (see App.~\ref{app:floating-cqed}). Consequently, the floating architecture imposes stricter constraints on junction symmetry for $\zeta_{\mathrm {ZZ}}$ suppression, particularly at large $E_{\mathrm J_\Sigma}$ where the deleterious effects of SQUID junction asymmetry become magnified.

\begin{figure}
    \centering
    \includegraphics[width=\linewidth]{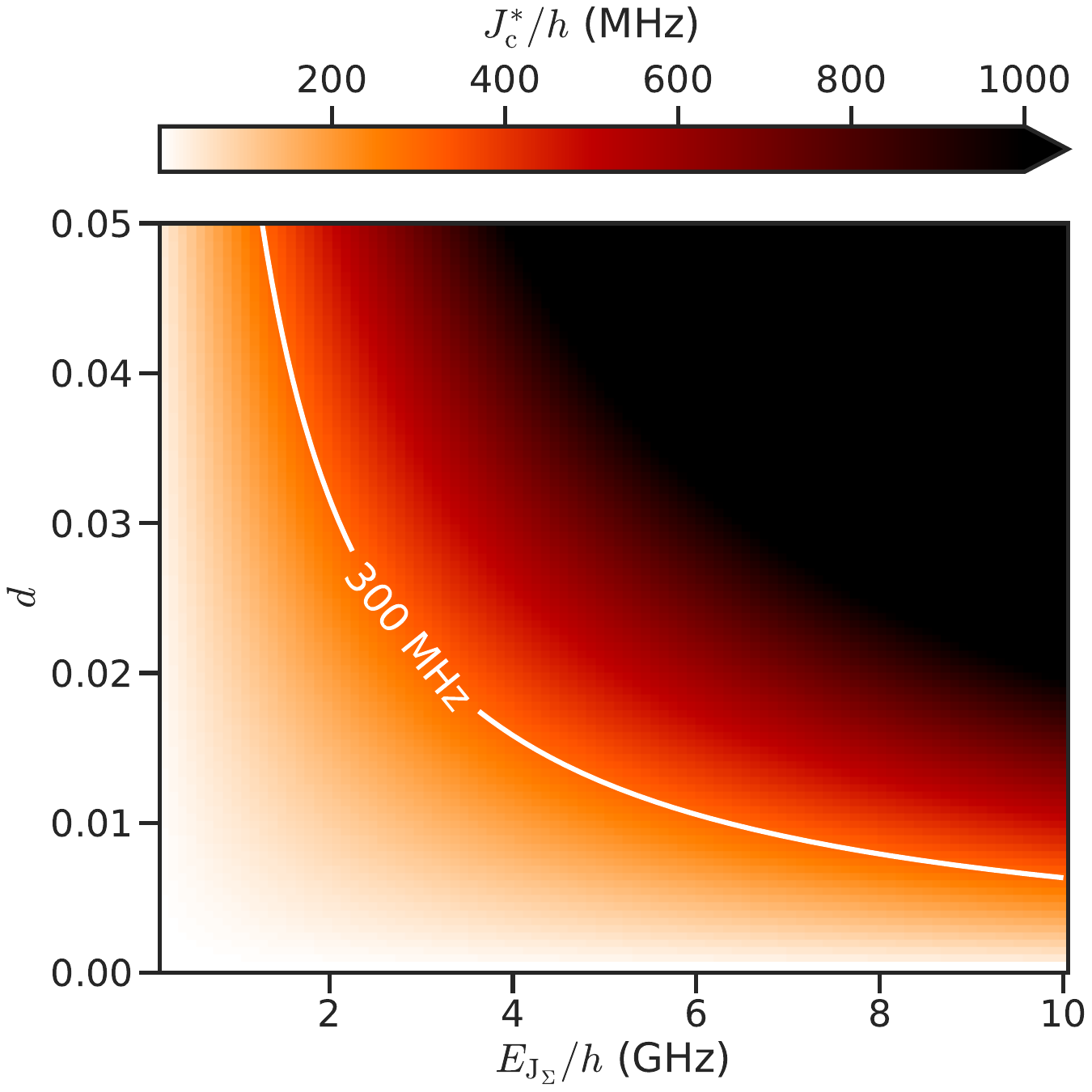}\\
    \caption{Effective charge coupling strength $J_{\mathrm c}^*$ required to mitigate static ZZ at the symmetric off-point to less than $1{\text{ kHz}}$, plotted as a function of the asymmetry $d$ and the total Josephson energy $E_{\mathrm J_\Sigma}$ of the SQUID, for the grounded design. The white curve represents $J_{\mathrm c}^*/h =$ \qty{300}{\MHz}, a typical value realized in experiment. \cite{dogan2023twofluxonium}. }
    \label{fig:static-zz-cap}
\end{figure}

In the grounded design, we can compensate for this residual ZZ coupling at the symmetric off-point by including a charge coupling by varying the capacitance shunting the SQUID. Figure~\ref{fig:static-zz-cap} explores the role of the shunting capacitor, which mediates charge coupling at a rate $J_{\mathrm c}$ as described by Eq.~\eqref{eq:charge-coupling-ham}. An appropriately chosen value of charge coupling, which we denote using the coefficient $J_{\mathrm c}^*$, can reduce $\zeta_{\mathrm {ZZ}}$ to sub-kilohertz levels for the parameters in Table~\ref{tab:params}. However, this mitigation strategy has practical limitations. The shunting capacitor limits the capacitance budget available for the rest of the circuit, a known concern in grounded fluxonium designs \cite{ding2023high}. The white contour in Fig.~\ref{fig:static-zz-cap} shows $J_{\mathrm c}^*/h =$ \qty{300}{\MHz}, which has been considered in previous theoretical~\cite{chen2022fast} and experimental~\cite{dogan2023twofluxonium} work. These findings underline a critical trade-off in our coupler design: while larger $E_{\mathrm J_\Sigma}$ (i.e., stronger inductive coupling) enhances the qubit-qubit interaction, thus enabling faster gate operations, it also increases parasitic interactions due to junction asymmetry. These imperfections intensify static ZZ interactions, and may require impractically large values of $J_{\mathrm c}^*$ to mitigate. Nevertheless, for the parameters considered here, adding a charge coupling using $J_{\mathrm c}^*$ does not qualitatively affect the gate performance. During any gate operations, the coupler flux is tuned away from the symmetric off-point, thereby enhancing the effect of the asymmetry terms in the Hamiltonian. 

This ZZ mitigation mechanism is inherently absent in the floating design. Due to additional interactions with the sloshing mode, a more complicated mitigation scheme might be required. The parameters $J_{\mathrm c}$ and $J_{\mathrm {sl}}$ both depend strongly on the various capacitances in the circuit and cannot be independently varied to suppress the static ZZ interaction(see Appendix~\ref{app:circuit-qed}). This highlights a fundamental limitation of the floating design. Due to the strong all-to-all coupling present in the floating design, we expect pairwise static ZZ coupling between all three modes (A, B, and sl) as well as three-body ZZZ interactions. We leave these aspects of the the floating design to be explored in more depth in future work. \hfill

In considering the optimal architecture for SQUID-coupled fluxonium qubits, we are motivated to explore the grounded fluxonium design over the floating design. In floating transmon architectures, such as demonstrated in Ref.~\cite{kounalakis2018tuneable}, the (transmon) qubit frequencies were far-detuned from the sloshing mode by several gigahertz at the operating point, suppressing unwanted hybridization. However, in our case, the extra mode lies too close to the qubit frequencies, typically $\sim$\qty{1}{\GHz}  for the parameters considered in the floating fluxonium setup. One potential workaround is to consider fluxonium parameters that lead to higher qubit frequencies, for instance, in the light fluxonium regime, to recover the large detuning between the qubit and sloshing mode as in Ref.~\cite{kounalakis2018tuneable}. Other mitigation strategies might also be considered, such as introducing an inductor in series with the SQUID, diluting the effective impedance of the coupler. The smaller effective coupling would reduce the sensitivity to fabrication asymmetry, but would also slow down gates. Such an inductance could be realized using a Josephson junction array, but higher Josephson harmonics \cite{kim2025emergent} could arise from the introduction of a series inductance. 

On the other hand, the grounded design has no extra modes and offers improved static ZZ mitigation capability compared to the floating design. However, the large ground loop in this design can introduce additional control challenges. For simplicity, we have assumed that the flux through this loop can be maintained at a constant value through careful flux-crosstalk calibration, but doing so in practice adds additional overhead to the complexity of the tuneup procedure.

\section{Two-Qubit gates in the grounded design}
In this section, we study $\sqrt{i\text{SWAP}}$-like gates, focusing exclusively on the grounded design for reasons discussed in the previous section. We first illustrate the the mechanism of the gate operations using an effective two-level description for SQUID-coupled fluxonium qubits. We then numerical simulate gates using the full Hamiltonian Eq.~\eqref{eq:full-ham}, keeping the 4 lowest fluxonium eigenstates for each qubit in our simulations. 

\subsection{Effective two-level description}\label{sec:two-level-description}
In this subsection, we focus only on the lowest-lying eigenstates of the Hamiltonian for individual fluxonium qubits ($H_{\mu}$ in Eq.~\eqref{eq:indiv_flux_ham}) in the grounded design. We park the individual fluxonium at the sweet spot where the large anharmonicity allows for a simple and effective two-level description. We label the qubit states in the eigenbasis of individual fluxonium (Eq.~\eqref{eq:indiv_flux_ham}) as ${|00\rangle}_{\pi},~{|01\rangle}_\pi,~{|10\rangle}_\pi$ and ${|11\rangle}_\pi$, where the $\pi$ subscript indicates that the fluxonia are parked at the sweet-spot $\Phi_{\mathrm {A/B}} = \Phi_0/2$. We consider a symmetric SQUID coupler with $d=0$ such that the fluxonium Hamiltonian has no contribution from the SQUID at the off-point.  
 
In the mentioned qubit eigenbasis, the phase operators can be expressed in terms of Pauli operators as~\cite{chen2022fast}
\begin{equation}
\hat{{\varphi}}_\mu = a^\mu_{\mathrm {I}} \hat{\mathrm{I}} + a^\mu_{\mathrm x} \hat \sigma_{\mathrm x}^{\mu},\label{eq:flux_operator_two_level_approx}
\end{equation}
where the diagonal term, $a^\mu_{I}=-\pi$, which results from the choice of gauge ($\Phi_{\mathrm {A/B}} = \Phi/2$ allocated to the inductor), and $a^\mu_\mathrm{x}$ is the off-diagonal coefficient of the phase operator which depends on the circuit parameters for the corresponding fluxonium. Hence, for symmetric SQUID coupling, the total Hamiltonian in the two-dimensional subspace for the grounded circuit becomes
\begin{equation}
\hat{H}_{\mathrm{qbt}}^{\mathrm{sym}} = \sum_{\mu=A,B} \omega_{\mu}\sigma_{\mathrm z}^{\mu} + g_{\mathrm{SQ}} \sigma_{\mathrm x}^{\mathrm{A}}\sigma_{\mathrm x}^{\mathrm{B}} + g_{\mathrm C} \sigma_{\mathrm y}^{\mathrm{A}}\sigma_{\mathrm y}^{\mathrm{B}},  
\label{eq:simple2x2}
\end{equation} written in the eigenbasis of the individual qubits (see \ref{eq:two_level_SQUID_hamiltonian}) where $\omega_{\mu}$ is the qubit gap of the fluxonium, $g_{\mathrm {SQ}} = -(E_{\mathrm J_\Sigma}/4) a^{\mathrm {A}}_\mathrm{x} a^{\mathrm {B}}_\mathrm{x}\cos\left(\pi\Phi_{\mathrm S}/\Phi_0\right)$ is the flux coupling strength [see Eq.~\eqref{eq:gr_flux_coupling}], and $g_{c} = \vert J_{c}{{\langle 0_{\mathrm{A}}}|}_{\pi} \hat{n}_{\mathrm {A}}{|1_{\mathrm{A}}\rangle}_\pi{\langle 0_{\mathrm{B}}}|_\pi\hat{n}_{\mathrm {B}}{ |1_{\mathrm{B}}}\rangle_\pi\vert $ is the charge coupling strength.

Under these approximations, the two fluxonium qubits are coupled exclusively through exchange-type interactions, which are well-suited for the implementation of a two-qubit $\sqrt{i\text{SWAP}}$ gate. While the target coupling between the two fluxonia is an ideal XX interaction, in practice, the Hamiltonian contains additional non-negligible terms that introduce additional types of coupling. These arise both from the higher energy levels of the individual fluxonia and from asymmetries in the SQUID junctions. As a result, the effective two-qubit interaction deviates from the idealized XX form, complicating the implementation of high-fidelity $\sqrt{i\text{SWAP}}$-type gates. In the following, we will discuss the impact of junction asymmetry and higher energy levels on gate performance.

We now analyze the lowest-order effect of SQUID junction asymmetry. In this case Eq.~\eqref{eq:simple2x2} takes the following form
\begin{multline}
\hat H_{\mathrm {qbt}} =  \sum_\mu \left(\omega_{\mu}\sigma_{\mathrm z}^{\mu}+ \Delta^{\mathrm {SQ}}_{\mu}\hat \sigma_{\mathrm x}^{\mu}\right)\\
+ g_{\mathrm {SQ}} \sigma_{\mathrm x}^{\mathrm A}\sigma_{\mathrm x}^{\mathrm B}+ g_{c} \sigma_{\mathrm y}^{\mathrm A}\sigma_{\mathrm y}^{\mathrm B},
\label{eq:ham_qbt_full}
\end{multline}
where $\Delta^{\mathrm {SQ}}_{\mu} = (-1)^{\mu}a^\mu_{\mathrm x}d(E_{\mathrm J_\Sigma}/2)\sin\left(\pi\Phi_{\mathrm S}/\Phi_0\right)$ results from the asymmetric part of the SQUID Hamiltonian. At the coupler off-point, $|\Delta_{\rm{SQ}}^{\mu}|$ takes its maximal value. Thus, in the two-level approximation, junction asymmetry introduces undesired X rotations to each qubit with opposite directions of precession between the qubits. For small asymmetries (${d\times E_{\rm{J}_\Sigma}}\ll \omega_\mu$), these rotations can be corrected with single qubit X gates applied to individual qubits.

\subsection{$\sqrt{i{\text{SWAP}}}$-type gates}\label{sec:two-qubit-gates}

The tunable inductive coupling in our design enables the implementation of two-qubit gates from the fermionic simulation (fSim) family--a class of operations critical for quantum chemistry simulations \cite{krantz2019quantum, foxen2020demonstrating}. fSim gates are formally defined as a two-parameter family of unitary operators 
\begin{equation}
\hat{U}_{\mathrm {ideal}}(\theta, \xi) = \begin{pmatrix}
e^{-i\xi/2} & 0 & 0 & 0 \\
0 & \cos\theta/2 & -i\sin\theta/2 & 0 \\
0 & -i\sin\theta/2 & \cos\theta/2 & 0 \\
0 & 0 & 0 & e^{-i\xi/2}
\end{pmatrix}.\label{eq:uideal}
\end{equation}

The swap angle $\theta$ quantifies the $|01\rangle \leftrightarrow|10\rangle$ population exchange. The angle $\xi$ encodes the conditional phase accumulation specific to the $|11\rangle$ state due to a dynamical ZZ interaction introduced when the coupler is on in addition to static ZZ, which also limits idling fidelity.  This unitary is locally equivalent to any two-qubit gate that combines state exchange and phase modulation \cite{foxen2020demonstrating}. fSim gates can be realized in our design using a combination of flux drives on the qubits $(\Phi_{\mathrm {A/B}})$ and coupler $(\Phi_{\mathrm S})$. Specifically, we explore $\sqrt{i\text{SWAP}}$-like gates, corresponding to $\theta=\pi/2$. All such gates, represented by $\hat{U}_{\mathrm {ideal}}(\pi/2, \xi)$ for $\xi \in (0, 2\pi)$ have the same entangling power \cite{nesterov2021proposal, zanardi2000entangling,ma2007matrix}.

We perform numerical gate simulations for the grounded design using device parameters specified in Table~\ref{tab:params}. We solve the Schr\"odinger equation 
\begin{equation}
    i\frac{d}{dt}|\Psi\rangle = \hat{H} (t)|\Psi\rangle
    \label{eq:schrodinger-equation}
\end{equation}
for our closed system simulations with the Hamiltonian given in Eq.~\eqref{eq:full-ham} including time-dependent fast-flux drives for qubit and coupler fluxes. We use these simulations to estimate the coherent error during our gate schemes. 

To quantify the deleterious impact of the environmental noise on the gate fidelity, we also perform open systems simulations by numerically solving the Lindblad master equation for the time-evolution of the system density matrix $\hat{\rho}$ coupled to a thermal environment considering both relaxation and dephasing processes
\begin{equation}
\frac{d \hat{\rho}}{dt} = -i\left[\hat{H}(t), \hat{\rho} \right]
+\sum_{\substack{j=\{\mathrm {1, \phi}\} \\ \alpha=\{\mathrm {A, B}\}}} \left[\hat{L}_{j}^{\alpha} \hat{\rho} {\hat{L}^{\alpha}_{j}}^{\dagger} -\frac{1}{2}\left\{ {\hat{L}_j^{\alpha}}^{\dagger} \hat{L}^{\alpha}_j, \hat{\rho}\right\}\right],
\end{equation}
where $\{\hat{\mathrm P}, \hat{\mathrm Q}\} = \hat{\mathrm P}\hat{\mathrm Q} + \hat{\mathrm Q}\hat{\mathrm P}$ denotes the anti-commutator of two operators $\hat{\mathrm P}$ and $\hat{\mathrm Q}$. The summation index $\alpha$ iterates over the two qubits and $j$ iterates over relaxation and dephasing processes modeled using the jump operators
\begin{equation}
\begin{split}
    \hat{L}^{\alpha}_1 &= \frac{1}{\sqrt{T_1}}|0_\alpha\rangle_\pi\langle 1_\alpha|_\pi, \\
    \hat{L}^{\alpha}_\phi &= \sqrt{\frac{2}{T_\phi}}\left(|0_\alpha\rangle_\pi\langle 0_\alpha|_\pi-|1_\alpha\rangle_\pi\langle 1_\alpha|_\pi\right),
    \label{eq:kraus}
\end{split}
\end{equation} 
respectively, where $|0_\alpha\rangle_\pi$ and $|1_\alpha\rangle_\pi$ denote the ground and first excited states of each qubit at their sweet spot. $T_1$ and $T_\phi$ are called the bit-flip and dephasing lifetimes respectively. We assume that the two qubits have the same values for the respective quantities, and the dephasing lifetime is twice the bit-flip time $T_\phi = 2T_1$, which means decoherence and bit-flips occur over the same effective timescale $T_2=T_1$ for our simulations. We simulate two different sets of lifetime values $T_1 \in \{10, 100 \}$ \unit{\micro\second} and the corresponding $T_\phi$ values. Recent experiments have shown fluxonium devices with similar lifetimes~\cite{nguyen2019high}. We neglect relaxation processes from higher energy levels as the high anharmonicity of fluxonium qubits prevents any significant population leakage outside the qubit subspace, which we have verified in our gate simulations. 
We perform numerical quantum process tomography (QPT) ~\cite{nielsen2002simple, horodecki1999general, chow2009randomized} for our gate simulations and estimate gate fidelity using 
\begin{equation}
F_g = \frac{4{\mathrm{Tr}}(\chi_{\mathrm {ideal}}\chi_{\mathrm {sim}})+{\mathrm {Tr}}(\chi_{\mathrm {sim}})}{5}.
\label{eq:iSWAP_fidelity}
\end{equation}
$\chi_{\mathrm {sim}}$ is numerically calculated from the final-time density matrix (state vector) in our open (closed) system simulation. Eq.~\eqref{eq:iSWAP_fidelity} captures both coherent and incoherent errors, providing a measure of the gate error $1-F_g$. Note that Eq.~\eqref{eq:iSWAP_fidelity} only quantifies coherent errors for the closed system simulations and agrees with the equivalent metric that uses the unitary propagator \cite{pedersen2007fidelity} instead of the process matrix. . We construct $\chi_{\mathrm {ideal}}$ from $\hat{U}_{\mathrm {ideal}}(\pi/2, \xi_{\mathrm {sim}})$, with the $\xi_{\mathrm {sim}}$ measured as discussed in Ref.~\cite{chen2022fast}.

\begin{figure}
    \centering
    \includegraphics[width=\linewidth]{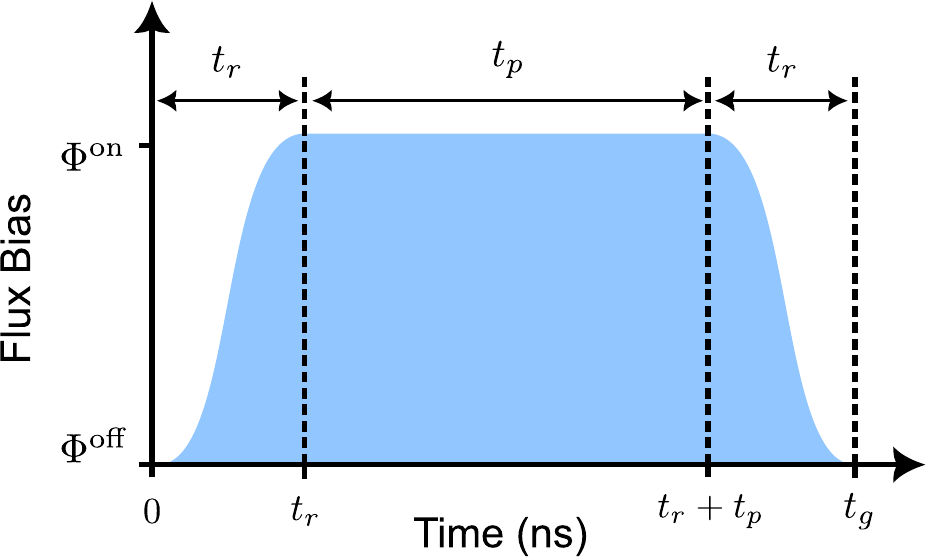}
    \caption{Fast-flux pulse shape used for gate operations. At the beginning of the pulse ($t=0$), the flux bias is $\Phi^{\mathrm {off}} = \Phi_0/2$. The flux bias is tuned, using a cosine ramp of duration $t_r$ to the maximum value $\Phi^{\mathrm {on}}$ which is maintained for a duration $t_p$, then ramped down in duration $t_r$. The total gate duration is $t_g = t_p+2t_r$. We consider different ranges of $\Phi^{\mathrm {on}}$ for gate operations.}
    \label{fig:pulse-shape}
\end{figure}

We use a fast-flux pulse waveform as illustrated in Fig.~\ref{fig:pulse-shape} to realize a $\sqrt{i\text{SWAP}}$-like gate. The pulse starts at value $\Phi^{\mathrm {off}}$, ramps up to a value $\Phi^{\mathrm {on}}$, remains constant for a duration $t_p$, then ramps down again to $\Phi^{\mathrm {off}}$. The on/off notation is used for consistency across all flux drives. We use cosine function ramps with duration $t_r$ for both ramps. A longer ramp time increases the adiabaticity of the beam-splitter interaction realized by the ramps \cite{campbell2020universal}. The total length of the pulse is the gate time $t_g= t_p+2t_r$. 

In the following sections, we describe two different proposals to realize a $\sqrt{i\text{SWAP}}$-like gate using fast-flux control: one where only the coupler flux is tuned during the gate and another where the both coupler and qubit fluxes are tuned simultaneously with the intention of bringing the two qubits into resonance during the gate. 

\subsubsection{Coupler-only fast flux gate}
\begin{figure}
    \centering
    \includegraphics[width=\linewidth]{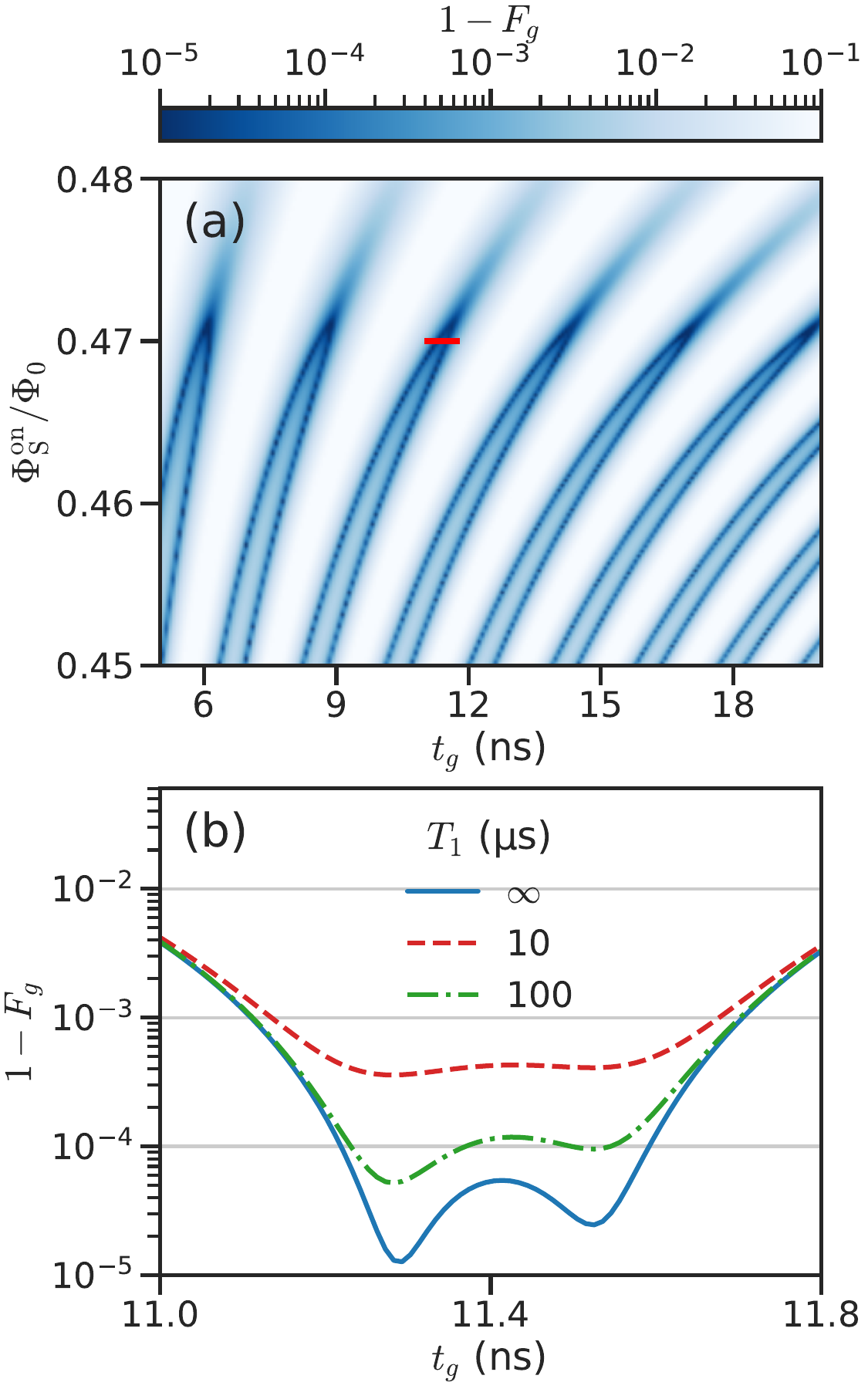}
    \caption{(a) Unitary gate error landscape for $\sqrt{i{\mathrm {SWAP}}}$-like gate with varying gate time ($t_{\mathrm g}$) and coupler flux at the plateau $(\Phi_{\mathrm S}^{\mathrm {on}})/\Phi_0$ for the coupler-only gate. (b) Gate error from open-systems simulations for the solid red line in panel (a) for three different values of $T_1$ and considering $T_\phi = 2T_1$. We choose $t_r =$ \qty{2}{\ns} for these simulations. Other parameters are given in Table~\ref{tab:params}.}
    \label{fig:gate_resonant}
\end{figure}

First, we keep both qubits at their respective sweet spots $(\Phi_{\mathrm {A/B}} = \Phi_0/2)$ throughout the gate operation. Even though the qubits are strongly detuned by approximately \qty{200}{\MHz} at this point, the strong galvanic coupling in our design helps us realize fast, high-fidelity gates in this configuration. Choosing $t_r =$ \qty{2}{\ns}, we quantify the coherent gate error using closed-system simulations for different values of $\Phi_{\mathrm S}^{\mathrm {on}}$ and $t_g$, as shown in Fig.~\ref{fig:gate_resonant}(a). The regions with darker blue color correspond to higher unitary gate fidelity. The red line in Fig.~\ref{fig:gate_resonant}(a) marks one such region with $\Phi_{\mathrm S}^{\mathrm {on}}=0.47\Phi_0$ and $t_g$ ranging from \qtyrange{11}{11.8}{\ns}. We use open-system simulations to characterize fidelity degradation due to relaxation and dephasing, with results shown in Fig.~\ref{fig:gate_resonant}(b) for different values of $T_1$ (dashed and dash-dot lines) with the assumptions discussed in Sec.~\ref{sec:two-qubit-gates}, compared to the unitary simulation (solid line). Gates with fidelity exceeding $99.9\%$ are possible even with a pessimistic estimate for $T_1 = \text{\qty{10}{\us}}$. 

\begin{figure}
    \centering
    \includegraphics[width=\linewidth]{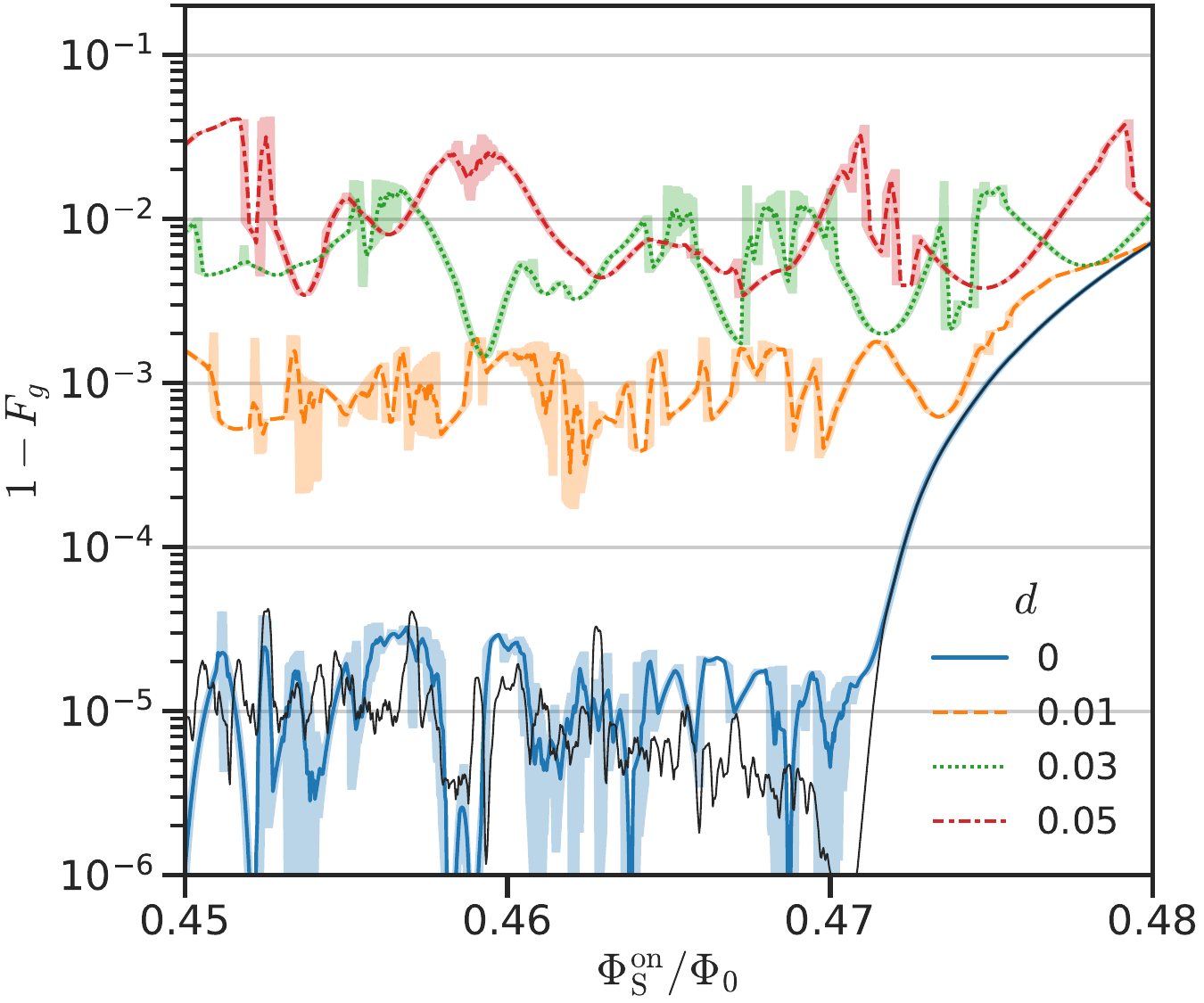}
    \caption{Minimum unitary gate error with varying coupler flux $\Phi_{\mathrm S}^{\mathrm {on}}$ for the coupler-only gate, optimized over gate times $t_{\mathrm g}$ ranging from \qtyrange{5}{20}{\ns}. Different line colors and styles represent different values of the SQUID junction asymmetry $d$ from \numrange{0}{0.05}. For each value of $d$, the shaded region denotes the range of gate error obtained in multiple simulations and the line indicates the mean value. The thin black line denotes the minimum error obtained for each value of coupler flux in Fig.~\ref{fig:gate_resonant}(a).}
    \label{fig:resonant-optim-gate}
\end{figure}
Next, we investigate how junction asymmetry in the SQUID contributes to the coherent gate error for this scheme. The minimum achievable unitary gate error $1-F_g$ with varying $\Phi_{\mathrm S}^{\mathrm {on}}$ is shown in Fig.~\ref{fig:resonant-optim-gate} for different values of $d$ indicated by different line colors and styles. We calculate this by optimizing $t_g$ over the interval spanning the horizontal axis in Fig.~\ref{fig:gate_resonant}(a), with all other parameters held constant. To avoid optimization fine-tuning issues, we sort the optimized gate error values into 100 equal-width bins of $\Phi_{\mathrm S}^{\mathrm {on}}$ and report the mean value (lines) and range (shaded region). The thin black line in Fig.~\ref{fig:resonant-optim-gate} denotes minimum gate error in each $\Phi_{\mathrm S}^{\mathrm {on}}$ bin using the data from Fig.~\ref{fig:gate_resonant}(a), confirming that the optimized gates follow the same trend as the more computationally intensive 2D parameter scans. The gate error rapidly approaches $10^{-5}$ around $\Phi_{\mathrm S}^{\mathrm {on}}=0.47\Phi_0$, then slowly increases with larger flux pulse amplitude (smaller $\Phi_{\mathrm S}^{\mathrm {on}}$). We observe that the minimum gate error is significantly impacted by junction asymmetry in the SQUID. For a modest value of asymmetry $d = 0.01$, the minimum achievable coherent error increases to $10^{-3}$, a hundredfold larger than the symmetric case. Further increasing the asymmetry up to $d = 0.05$ increases the minimum coherent gate error to $10^{-2}$. This reduction in fidelity can be directly attributed to the magnitude of the asymmetry term in Eq.~\eqref{eq:Squid_ham_gr}, which linearly increases with $d$. \hfill

\subsubsection{Detuned-qubit fast flux gate}
In this gate scheme, we bias the coupler flux while also simultaneously flux-tuning the lower frequency qubit (in our case, qubit A) closer into resonance with the other qubit. We choose $\Phi_{\mathrm S}^{\mathrm {on}}=0.49\Phi_0$ to show that this scheme enables two-qubit gates with a lower value of coupler flux $\Phi_{\mathrm S}^{\mathrm {on}}$ compared to the coupler-only gate scheme described in the previous section. For simplicity, we choose the same ramp time $t_r =$ \qty{6}{\ns} for both qubit and coupler flux pulses in this gate scheme. 

\begin{figure}
    \centering
    \includegraphics[width=\linewidth]{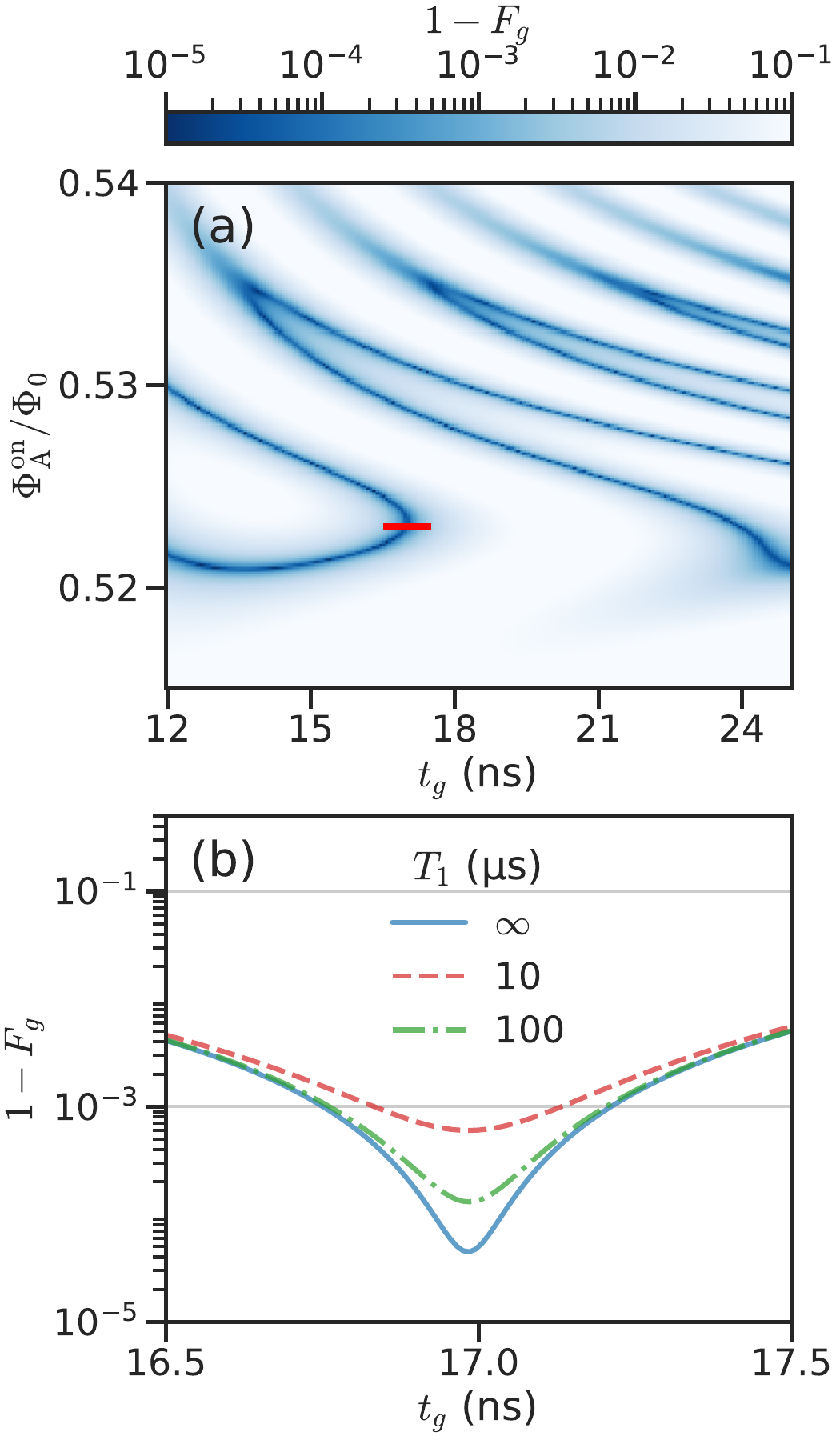}
    \caption{(a) Unitary gate error landscape for $\sqrt{i{\mathrm {SWAP}}}$-like gate with varying gate time $t_{\mathrm g}$ and $2\pi\Phi_\mathrm {A}/\Phi_0^{\mathrm {on}}$, with $\Phi_{\mathrm S}^{\mathrm {on}}=0.49\Phi_0$ and $t_r =$ \qty{6}{\ns}. (b) Gate error from open-systems simulations for the solid red line in panel (a) for three different values of $T_1$ and considering $T_\phi = 2T_1$. Other parameters are given in Table~\ref{tab:params}.}
    \label{fig:gate_detuned}
\end{figure}

We calculate the unitary gate fidelity with varying $2\pi\Phi_\mathrm {A}/\Phi_0^{\mathrm {on}}$ and $t_{\mathrm g}$. The error landscape is shown in Fig.~\ref{fig:gate_detuned}(a). The dark blue regions indicate regions corresponding to high gate fidelity. We perform open-systems simulations for the high-fidelity region near the middle of the chevron pattern around $\Phi_{\mathrm{A}} = 0.523\Phi_0$ and $t_g$ between \qtylist{16.5; 17.5}{\ns}, indicated by the small red line in Fig.~\ref{fig:gate_detuned}(a). Fig.~\ref{fig:gate_detuned}(b) shows the gate error for different $T_1$ in this region. As before, decoherence reduces the gate fidelity. Here, we see an order of magnitude reduction in gate fidelity going from  $T_1 = \infty \rightarrow$ \qty{10}{\us}. Consequently, gates exceeding $99.9\%$ are possible for the devices with $T_1>$ \qty{1}{\ms}, such as those in Ref.~\cite{ding2023high}. 

\begin{figure}
    \centering
    \includegraphics[width=\linewidth]{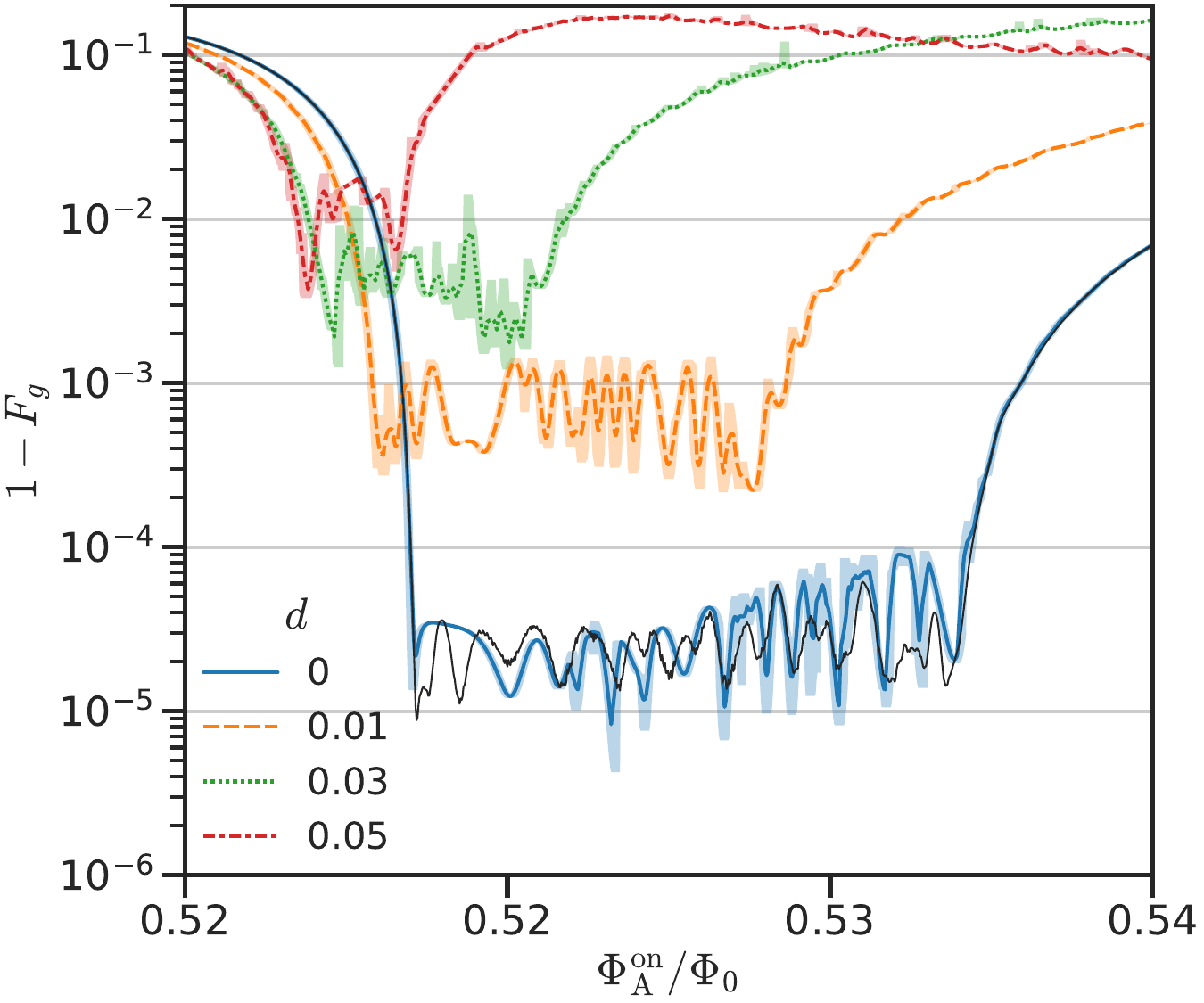}
    \caption{Minimum unitary gate error with varying coupler flux $2\pi\Phi_\mathrm {A}/\Phi_0^{\mathrm {on}}$ for the detuned gate, optimized over gate times $t_{\mathrm g}$ ranging between $12$ and $25$ \unit{\nano\second}. We also choose $\Phi_{\mathrm S}^{\mathrm {on}} = 0.49 \Phi_0$ and $t_r=$ \qty{6}{\ns}. Line colors and styles represent different values of the SQUID junction asymmetry $d$. For each value of $d$, the shaded region denotes the range of gate error obtained in multiple simulations and the line indicates the mean value. The thin black line denotes the minimum error obtained for each value of coupler flux in Fig.~\ref{fig:gate_detuned}(a). }
    \label{fig:detuned-optim-gate}
\end{figure}

Next, we characterize the robustness of the gate fidelity to fabrication uncertainty for this gate scheme. As with the coupler-only gate, we optimize $t_g$ for the lowest gate error over the interval spanning the horizontal axis in Fig.~\ref{fig:gate_detuned}(a), for a range of $2\pi\Phi_\mathrm {A}/\Phi_0^{\mathrm {on}}$ with results plotted in Fig.~\ref{fig:detuned-optim-gate}. The line colors and styles indicate different values of $d$ between \numlist{0;0.05}, reported as mean values (lines) and range (shaded region) for 100 equal-width bins of $2\pi\Phi_\mathrm {A}/\Phi_0^{\mathrm {on}}$. As in the previous section, junction asymmetry increases gate error. In addition to a reduction in fidelity, the size of the high-fidelity region also shrinks with increasing $d$. Even for $d = 0.05$, we predict unitary gate fidelity exceeding $99\%$.

The dual flux control strategy introduces a trade-off: while driving qubit A allows us to perform the gate with the qubits tuned nearly into resonance, its departure from the sweet spot increases susceptibility to flux-induced dephasing. However, the strong coupling in our design allows us to perform this gate in short times, limiting this effect and preserving high fidelity even under realistic noise and asymmetry. The susceptibility to dephasing may be reduced using a parametric gate, like the scheme demonstrated in Refs.~\cite{sete2021parametric, moskalenko2022high}. However, investigations of parametric control are left for future work. 

\section{Conclusions }\label{sec:conclusions}
We have analyzed a galvanically-connected dc SQUID as a minimal tunable coupling element for two designs of fluxonium: grounded and floating. Although the floating design is free from ground-loops, we find that it suffers from strong parasitic couplings to a charge-noise sensitive sloshing mode. For the grounded design, the presence of a ground loop complicates flux crosstalk calibration. 

Provided that the ground loop flux can be maintained at a constant value while other fluxes are driven, fast high-fidelity gates can be realized in this design using flux drives on coupler and qubits. The grounded design is also more robust to junction asymmetry due to fabrication uncertainty. This asymmetry introduces a static ZZ interaction, which can be mitigated using a shunting capacitor in the grounded design, while the floating design requires a more complex mitigation strategy. 

We have demonstrated high-fidelity two-qubit $\sqrt{i\text{SWAP}}$-like gates under \qty{20}{\ns} using two different all-flux control schemes in the grounded design. Our gates have an open-system fidelity exceeding $99.99\%$ for both control schemes even in the presence of modest relaxation and dephasing. While symmetric junctions enable gates with fidelity greater than $99.999\%$ in less than \qty{20}{\ns}, even a modest asymmetry with $d$ from \numrange{0.01}{0.05} can reduce fidelity to the $99- 99.9\%$ range. This sensitivity to junction asymmetry imposes strict constraints on fabrication uncertainty. Nevertheless, the system retains reasonable robustness, maintaining useful gate performance even under moderate imperfections. 

Topics for future work include strategies to decouple the extra mode in the floating circuit and enable high-fidelity gates for floating fluxonium qubits, which have a more generous capacitance budget compared to grounded fluxonium. Pulse shaping and echo sequences may be used to counteract the effects of junction asymmetry in the SQUID and obtain even higher fidelity gates in this design. Near-ultrastrong quartic coupling can be implemented using a quarton \cite{ye2021engineering, ye2025near}, enabling ultra-fast readout \cite{ye2024ultrafast} for our design. 

\begin{acknowledgments}
We acknowledge helpful discussions with Michel Devoret, Jens Koch, David Schuster, Eli Levenson-Falk, Kevin O'Brien and Raymond Simmonds. We also appreciate the continued support of James Kelly from the Chapman University IS\&T Research Computing Support team. This work was supported by the Army Research Office through Grant No. W911NF-22-1-0258. 
\end{acknowledgments}

\appendix

\section{Full Circuit Hamiltonian}\label{app:circuit-qed}
In this section, we derive the circuit Hamiltonian for the two variants of the circuit design considered in the main text. We examine the behavior of the galvanic SQUID coupler for grounded as well as floating fluxonium qubits. 

\subsection{Coupling with Grounded Fluxonium Qubits}\label{app:grounded-cqed}
The circuit design with grounded fluxonium qubits is shown in Fig.~\ref{fig:grounded-design}, where we have ignored stray capacitances that are small and do not qualitatively affect the analysis that follows. 

In this design, the circuit contains only two active nodes. Setting the ground flux to 0, we can directly identify mode labels with the corresponding node, $\hat \varphi_{\mathrm {A}}$ and $\hat \varphi_{\mathrm {B}} $. The circuit Lagrangian using standard circuit QED is then given by

\begin{equation}
    \begin{split}
        \hat{\mathcal{L}}_{\mathrm {gr}} &= \frac{1}{2}\frac{\Phi_0^2}{(2\pi)^2} \dot{\hat{\boldsymbol{{\varphi}}}}^T \mathbf{C} \dot{\hat{\boldsymbol{{\varphi}}}} + E_{\mathrm {J_A}}\cos \hat \varphi_{ \mathrm {A}} + E_{\mathrm {J_B}}\cos \hat \varphi_{\mathrm {B}}\\
& - \frac{1}{2}E_{\mathrm {L_A}} (\hat \varphi_{\mathrm {A}} + 2\pi\Phi_\mathrm {A}/\Phi_0)^2  - \frac{1}{2}E_{\mathrm {L_B}} (\hat \varphi_{\mathrm {B}} + 2\pi\Phi_\mathrm {B}/\Phi_0)^2\\
& + E_{\mathrm J_1}\cos (\hat{\varphi }_{-} + \pi\Phi_\mathrm{S}/\Phi_0) + E_{\mathrm J_2}\cos (\hat{\varphi }_{-} - \pi\Phi_\mathrm{S}/\Phi_0),
    \end{split}
    \label{eq:grounded-design-lagrangian}
\end{equation}
where  $\Phi_\mu$ indicates the external flux piercing the qubit inductive loops $(\mu  = {\mathrm {A, B}})$ and the SQUID loop $(\mu = {\mathrm S})$, respectively, $\hat{\varphi }_-= \hat \varphi_{\mathrm {A}}-\hat \varphi_{\mathrm {B}}$, the flux operator vector $ \hat{\boldsymbol{{\varphi}}} = (\hat \varphi_{\mathrm {A}} , \hat \varphi_{\mathrm {B}})^T$ and the capacitance matrix is
\begin{equation}
    \mathbf{C} = \left[\begin{matrix}C + C_{\mathrm c} & - C_{\mathrm c}\\- C_{\mathrm c} & C + C_{\mathrm c}\end{matrix}\right],
    \label{eq:grounded-capacitance-matrix}
\end{equation}
where we considered identical capacitance for each fluxonium. Next, we can perform a Legendre transformation to obtain the circuit Hamiltonian with mode variables satisfying the commutation relations $\left[\hat \varphi_\mu, \hat n_\mu\right] = i$, where $\hat n_\mu$ is the momentum operator canonically conjugate to $\hat \varphi_\mu$. The Hamiltonian thus separates into distinct terms
\begin{equation}
\hat{H}_{\mathrm {gr}} = \hat H_{\mathrm  A} +\hat H_{\mathrm {B}} + \hat H_{\mathrm C} +  \hat H_{\mathrm S},
    \label{eq:grounded-hamiltonian-terms}
\end{equation}
where
\begin{equation}
    \begin{split}
        \hat H_{\mathrm {A}} =&\; 4E_{\mathrm {C_A}} \hat n_{\mathrm {A}}^2 + \frac{1}{2} E_{\mathrm L_{A}} (\hat \varphi_{\mathrm {A}} + 2\pi\Phi_\mathrm {A}/\Phi_0)^2 - E_{\mathrm {J_A}} \cos \hat \varphi_{\mathrm {A}}\\
    \hat{H}_{\mathrm {B}} =&\; 4E_{\mathrm {C_B}} \hat n_{\mathrm {B}}^2 + \frac{1}{2} E_{\mathrm {L_B}} (\hat \varphi_{\mathrm {B}}+ 2\pi\Phi_\mathrm {B}/\Phi_0)^2 - E_{\mathrm {J_B}} \cos\hat \varphi_{\mathrm {B}}\\
\hat{H}_{\mathrm C} =&\; J_{\mathrm c}  \hat n_{\mathrm {A}} \hat n_{\mathrm {B}}\\
\hat{H}_{\mathrm {S}} =& - E_{\mathrm J_\Sigma} \cos\left(\frac{\pi\Phi_{\mathrm S}}{\Phi_0}\right)\cos \hat{\varphi }_- + d E_{\mathrm J_\Sigma} \sin\left(\frac{\pi\Phi_{\mathrm S}}{\Phi_0}\right)\sin \hat{\varphi }_-.
\end{split}
\label{eq:grounded-hamiltonian-terms-expressions}
\end{equation}
The two terms in $\hat{H}_\mathrm{S}$ correspond to the last line of Eq.~\ref{eq:grounded-design-lagrangian}, rewritten by defining the total Josephson energy of the SQUID, $E_{\mathrm J_\Sigma} = E_{\mathrm J_1} + E_{\mathrm J_2}$, and junction asymmetry fraction, $d = (E_{\mathrm J_1} - E_{\mathrm J_2} )/ E_{\mathrm J_\Sigma}$. 

In Eq.~\eqref{eq:grounded-hamiltonian-terms-expressions}, $E_{\mathrm C_\mu}$ represents the charging energy for the mode $\mu$ and $J_{\mathrm{c}}$ is the direct charge-charge coupling between the two fluxonium modes. We express these quantities in terms of the circuit parameters
\begin{equation}
    \begin{split}
        E_{\mathrm {C}} & = \frac{e^2(C +  C_{\mathrm c})}{2C (C+ 2C_{\mathrm c}) },\\ 
        J_{\mathrm c} &= \frac{4e^2 C_{\mathrm c}}{C (C + 2C_{\mathrm c})},\\
    \end{split}
    \label{eq:grounded-design-charging-energies-couplings}
\end{equation}
where we defined, $E_{\mathrm {C}}=E_{\mathrm {C_A}}  = E_{\mathrm {C_B}}$. We now rewrite this Hamiltonian by collecting terms corresponding to each mode and the dominant couplings between them. To do so, we expand the trigonometric terms from the SQUID and separate terms that contain products of only one kind of flux operator $\hat \varphi_{\mathrm {A/B}}$ from those that contain mixed products:
\begin{equation}
\hat{H}_{\mathrm S} = \sum_\mu \hat{H}_{ \mathrm S,\mu} + \hat{H}_{ \mathrm {S,c}}^{\mathrm{sym}} + \hat{H}_{\mathrm {S,c}}^{\mathrm{asym}},
\label{eq:grounded_squid_terms}
\end{equation}
The $\hat{H}_{\mathrm{S}, \mu}$ terms represent corrections to the fluxonium qubit potentials and are given by
\begin{equation}
\begin{split}
\hat H_{ \mathrm {S,A}} &= - E_{\mathrm J_\Sigma} \left[ \cos \left(\frac{\pi\Phi_{\mathrm S}}{\Phi_0}\right) \cos \hat \varphi_{\mathrm {A}} + d\sin \left(\frac{\pi\Phi_{\mathrm S}}{\Phi_0}\right) \sin \hat \varphi_{\mathrm {A}}\right ],\\
\hat H_{ \mathrm {S,B}} &= - E_{\mathrm J_\Sigma} \left[ \cos \left(\frac{\pi\Phi_{\mathrm S}}{\Phi_0}\right) \cos \hat \varphi_{\mathrm {B}} - d\sin \left(\frac{\pi\Phi_{\mathrm S}}{\Phi_0}\right) \sin \hat \varphi_{\mathrm {B}}\right ].
\end{split}
\label{eq:grounded-coupler-correction-qubits}
\end{equation}
The charge coupling mediated by the capacitance $C_{\mathrm c}$ is given by ${\hat H_{\mathrm C}}$. The inductive coupling terms ${\hat H_{\mathrm {sym}}^{\mathrm {SQ}}}$ and ${\hat H_{\mathrm {asym}}^{\mathrm {SQ}}}$ contain only mixed product terms of flux operators $\hat \varphi_{\mathrm {A}}, \hat \varphi_{\mathrm {B}}$, labeled to distinguish terms arising due to a symmetric SQUID interaction from those due to junction asymmetry. These terms will be considered in detail below. 

\subsubsection{Even-order in flux operator inductive coupling terms}
The even order inductive coupling terms arise from the expansion of the second cosine in $- E_{\mathrm J_\Sigma} \cos\left(\pi\Phi_\mathrm{S}/\Phi_0\right)\cos \hat{\varphi }_-$ (see Eq.~\ref{eq:grounded-hamiltonian-terms-expressions}). For simplicity, we define the coupler-flux dependent coefficient as $g_{\mathrm {SQ}} = -E_{J_\Sigma} \cos\left(\pi\Phi_\mathrm{S}/\Phi_0\right)$, and rewrite ${\hat H^{\mathrm {sym}}_{\mathrm{S, c}}}$ as
\begin{multline}
        {\hat H^{\mathrm{sym}}_{\mathrm{S, c}}}  =g_{\mathrm {SQ}} \left[\cos \hat{\varphi }_{-} - \cos \hat \varphi_{\mathrm {A}} - \cos \hat \varphi_{\mathrm {B}} \right]\\
         =g_{\mathrm {SQ}} \left[\hat \varphi_{\mathrm {A}} \hat \varphi_{\mathrm {B}}  + \frac{\hat \varphi_{A}^{2} \hat \varphi_{\mathrm {B}}^{2}}{4}  - \frac{\hat \varphi_{\mathrm {A}}^{3} \hat \varphi_{\mathrm {B}}}{6} - \frac{\hat \varphi_{\mathrm {A}} \hat \varphi_{\mathrm {B}}^{3}}{6} + \mathcal {O}(\hat \varphi^5)\right].\\
    \label{eq:gr_sym-term-expansion}
\end{multline}
In the second line, we expanded the cosine to include terms up to fourth order in flux operators, which correspond to direct inductive coupling between the two fluxonium modes. The first term is the linear inductive coupling, and the last three terms are fourth-order couplings responsible for the ZZ interaction. Note that all the individual fluxonium terms get cancelled out at each order and only the coupling terms survive.

\subsubsection{Odd-order in flux operator inductive coupling terms}
The odd-order inductive coupling terms arise from the expansion of the sine term, $- dE_{\mathrm J_\Sigma} \sin\left(\pi\Phi_\mathrm{S}/\Phi_0\right)\sin \hat{\varphi }_-$ (see Eq.~\eqref{eq:grounded-hamiltonian-terms-expressions}). As  before, we define $g^{\mathrm {asym}}_{\mathrm {SQ}} = -dE_{\mathrm J_\Sigma} \sin\left(\pi\Phi_\mathrm{S}/\Phi_0\right)$ and rewrite the asymmetric term as \hfill
\begin{equation}
    \begin{split}
        {\hat H^{\mathrm {asym}}_{\mathrm S, c}} & = g^{\mathrm {asym}}_{\mathrm {SQ}} \left[\sin \hat{\varphi }_- - \sin \hat \varphi_{\mathrm {A}} + \sin \hat \varphi_{\mathrm {B}}\right]\\
        & =g^{\mathrm {asym}}_{\mathrm {SQ}} \left[ \frac{\hat \varphi_{\mathrm {A}} \hat \varphi_{\mathrm {B}}}{2} \hat{\varphi }_-+ \mathcal {O}(\hat \varphi^5)\right].\\
    \end{split}
    \label{eq:gr_asym-term-expansion}
\end{equation}

These coupling terms do not qualitatively change the nature of the coupling, but do contribute to coherent errors. While terms beyond fourth-order are not analytically considered here, our numerical simulations use the full expression for coupling. We used Eq.~\eqref{eq:gr_sym-term-expansion} and Eq.~\eqref{eq:gr_asym-term-expansion} to obtain the effective two-level Hamiltonian for the coupled fluxonium and the dominant effect of higher levels on the two-level approximation in Sec.~\ref{sec:two-level-description}.

\subsubsection{Inductive coupling terms in the two-level approximation}\label{app:two_level_inductive_coupling}
If the qubit subspace of the two subsystems is well-isolated, then the phase operator can be approximated by Eq.~\eqref{eq:flux_operator_two_level_approx}. Noting that $\hat \sigma_{x}^A$, $\hat \sigma_x^B$ commute, we can write
\begin{multline}
    e^{i(\hat\sigma_x^A - \hat \sigma_x^B)} = \cos^2(1)\\
    + \sin^2(1) \hat \sigma_x^A\hat \sigma_x^B
    + \frac i 2 \sin(2)\left( \hat \sigma_x^A - \hat \sigma_x^B\right).\notag
\end{multline}
Then, neglecting constant terms, we can write $\cos\hat{\varphi}_{-}$ and $\sin\hat{\varphi}_{-}$ as the real and imaginary parts of the RHS above. Then, $\hat H_{\rm S}$ becomes:
\begin{align}
    \hat H_{\rm S} &= g_{\rm{SQ}}\sin^2(1)\hat \sigma_x^A\hat \sigma_x^B 
    + g^{\rm{asym}}_{\rm{SQ}} \frac{\sin(2)}{2}\left(\hat \sigma_x^A - \hat \sigma_x^B\right)\label{eq:two_level_SQUID_hamiltonian}
\end{align}
The constant coefficients $(\sin^2(1), \sin(2)/2)$ do not scale with any circuit parameters or external fluxes, and do not qualitatively affect the qubit dynamics; thus, they may be absorbed into the constant energy coefficients $(E_{\rm{J}_\Sigma}, d)$, and are omitted from the main text. Note that in the two-level approximation, the asymmetric error term introduces only single-qubit X rotations, which can in principle be corrected by flux pulses applied to the individual fluxonia.

\subsection{Coupling with Floating (Differential) Fluxonium Qubits}\label{app:floating-cqed}
The circuit design with floating (differential) fluxonium qubits is shown in Fig.~\ref{fig:floating-design}. As before, we have ignored insignificant stray capacitances. 

The floating configuration has four active nodes. Using standard circuit quantization techniques \cite{vool2017introduction, Blais2020circuit, krantz2019quantum, rasmussen2021superconducting, you2019circuit}, we can write down the circuit Lagrangian in the node flux basis $\hat{\boldsymbol{\varphi}} = (\hat \varphi_1, \hat \varphi_2, \hat \varphi_3, \hat \varphi_4)^T$ as
\begin{equation}
\begin{split}
\hat{\mathcal{L}}_{\mathrm{fl}} &= \frac{1}{2}\frac{\Phi_0^2}{(2\pi)^2} \dot{\hat{\boldsymbol{{\varphi}}}}^T \mathbf{C} \dot{\hat{\boldsymbol{{\varphi}}}}\\ 
& + E_{\mathrm {J_A}}\cos (\hat \varphi_1 - \hat \varphi_2) + E_{\mathrm {J_B}}\cos (\hat \varphi_3 - \hat \varphi_4)\\
& \quad - \frac{1}{2}E_{\mathrm {L_A}} (\hat \varphi_1 - \hat \varphi_2 + 2\pi\Phi_\mathrm {A}/\Phi_0)^2\\ 
& \quad - \frac{1}{2}E_{\mathrm {L_B}} (\hat \varphi_3 - \hat \varphi_4 + 2\pi\Phi_\mathrm {B}/\Phi_0)^2\\
& \quad \quad + E_{\mathrm J_1}\cos (\hat \varphi_2 - \hat \varphi_3 + \pi\Phi_\mathrm{S}/\Phi_0) \\ &\quad \quad + E_{\mathrm J_2}\cos (\hat \varphi_2 - \hat \varphi_3 - \pi\Phi_\mathrm{S}/\Phi_0),\\
\end{split}
    \label{eq:circuit-lagrangian-node-basis-floating}
\end{equation}
where  $\Phi_\mu$ indicates the external flux piercing the qubit inductive loops $(\mu  = {\mathrm {A, B}})$ and the SQUID loop $(\mu = {\mathrm S})$, respectively. The capacitance matrix for this design is given by 
\begin{equation}
    \mathbf{C} = \left[\begin{matrix}C + C_{\mathrm g} & - C & 0 & 0\\- C & C + C_{\mathrm c} + C_{\mathrm g} & - C_{\mathrm c} & 0\\0 & - C_{\mathrm c} & C + C_{\mathrm c} + C_{\mathrm g} & - C\\0 & 0 & - C & C + C_{\mathrm g}\end{matrix}\right].
\end{equation}
There are four normal modes. First, we can pick the two differential fluxonium modes, $\hat \varphi_{\mathrm {A}} = \hat \varphi_1- \hat \varphi_2$ and  $\hat \varphi_{\mathrm {B}} = \hat \varphi_4-\hat \varphi_3$ as these are the modes which determine the inductive coupling between the two fluxonia. There are two natural choices for the two remaining modes. We may choose the two fluxonium common modes, $\hat \varphi_{\mathrm {A}}^+ = \hat \varphi_1+\hat \varphi_2$ and  $\hat \varphi_{\mathrm {B}}^+ = \hat \varphi_4+\hat \varphi_3$. Alternatively, we can pick the overall circuit common mode $\hat \varphi_\Sigma = \hat \varphi_1 + \hat \varphi_2+ \hat \varphi_3+ \hat \varphi_4$ corresponding to all four grounding capacitors charging up together, and a ``sloshing" mode $\hat \varphi_{\mathrm {sl}} = (\hat \varphi_1 + \hat \varphi_2) - (\hat \varphi_3+ \hat \varphi_4)$ corresponding to Cooper pairs tunneling from the left fluxonium to the right. We pick the latter case because the mode $\hat \varphi_\Sigma$ decouples from the other modes when all grounding capacitances are equal. The case where grounding capacitances are unequal but similar can be shown to be in qualitative agreement with the equal capacitance case~\cite{kounalakis2018tuneable}. 

The basis transformation from node flux basis $\hat{\boldsymbol{\varphi}}$ to the chosen normal mode basis $\tilde{\hat{\boldsymbol{{\varphi}}}} = (\hat \varphi_{\mathrm {A}} , \hat \varphi_{\mathrm {B}}, \hat \varphi_{\mathrm {sl}}, \hat \varphi_\Sigma)^T$ is given by
\begin{equation}
 \tilde{\hat{\boldsymbol{{\varphi}}}} = \mathbf{M} \cdot \hat{\boldsymbol{\varphi}},
    \label{eq:basis-transformation-floating}
\end{equation}
where the transformation matrix $\mathbf{M}$ is given by
\begin{equation}
\mathbf{M} = \left[\begin{matrix}
1 & -1 & 0 & 0\\
0 & 0 & -1 & 1\\
1/2 & 1/2 & - 1/2 & - 1/2\\
1 & 1 & 1 & 1
\end{matrix}\right].
    \label{eq:transformation-matrix-floating}
\end{equation}
The transformed capacitance matrix $\mathbf{K} = (\mathbf{M}^{-1})^T \mathbf{C} \mathbf{M}^{-1}$ is
\begin{equation}
\small{
\mathbf{K} = \left[\begin{matrix}C + C_{\mathrm c} / 4 + C_{\mathrm g} / 2 & - C_{\mathrm c} / 4 & - C_{\mathrm c} / 2 & 0\\- C_{\mathrm c} / 4 & C + C_{\mathrm c} / 4 + C_{\mathrm g} / 2 & C_{\mathrm c} / 2 & 0\\- C_{\mathrm c} / 2 & C_{\mathrm c} / 2 & C_{\mathrm c} + C_{\mathrm g} & 0\\0 & 0 & 0 & C_{\mathrm g} / 4\end{matrix}\right]}.
    \label{eq:transformed-capacitance-matrix-floating}
\end{equation}
The Lagrangian in Eq.~(\ref{eq:circuit-lagrangian-node-basis-floating}) expressed in these modes $\tilde{\hat{\boldsymbol{\varphi}}}$ becomes,
\begin{multline}
        \hat{\mathcal{L}}_{\mathrm{fl}} = \frac{1}{2}\frac{\Phi_0^2}{(2\pi)^2} \dot{\tilde{\hat{\boldsymbol{\varphi}}}}^T \mathbf{K} \dot{\tilde{\hat{\boldsymbol{\varphi}}}}\\
+ \sum\limits_ {\mu \in \{\mathrm {A, B}\}}\Big[ \frac{1}{2}E_{\mathrm {L_\mu}} \left(\hat \varphi_{\mu} + \frac{2\pi\Phi_{\mu}}{\Phi_0}\right)^2
 - E_{\mathrm {J_\mu}}\cos \hat \varphi_{\mu}\Big]\\+ E_{\mathrm J_\Sigma} \cos\left(\frac{\pi\Phi_{\mathrm S}}{\Phi_0}\right)\cos \left(\frac{\hat{\varphi }_-}{2} - \hat \varphi_{\mathrm {sl}} \right)\\
+ d E_{\mathrm J_\Sigma} \sin\left(\frac{\pi\Phi_{\mathrm S}}{\Phi_0}\right)\sin \left(\frac{\hat{\varphi }_-}{2} - \hat \varphi_{\mathrm {sl}} \right),
    \label{eq:fl_circuit-lagrangian-normal-mode-basis-floating}
\end{multline}
where $\hat{\varphi}_{-} \equiv \hat{\varphi}_{\text{A}} -\hat{\varphi}_{\text{B}}$. The last two terms in the last line of  Eq.~\eqref{eq:fl_circuit-lagrangian-normal-mode-basis-floating} can be rewritten in terms of the total Josephson energy of the SQUID, $E_{\mathrm J_\Sigma} = E_{\mathrm J_1} + E_{\mathrm J_2}$, and junction asymmetry fraction, $d = (E_{\mathrm J_1} - E_{\mathrm J_2} )/ E_{\mathrm J_\Sigma}$. 
Next, we perform a Legendre transformation to obtain the circuit Hamiltonian given by 
\begin{equation}
 \hat{H}_{\mathrm{fl}} = {\hat H_{\mathrm {A}} + \hat H_{\mathrm {B}} + \hat H_{\mathrm C}  +  \hat H_{\mathrm {sl}}  + \hat H_{\mathrm {C, fl}}  + \hat H^{\mathrm{fl}}_{\mathrm S}},
     \label{eq:circuit-hamiltonian-terms}
 \end{equation}
where $ \hat H_{\mathrm {A}} + \hat H_{\mathrm {B}} + \hat H_{\mathrm C}$ are the same as in Eq.~(\ref{eq:grounded-hamiltonian-terms-expressions}) and the other contributions are given by
\begin{equation}
\begin{split}
& \hat  H_{\mathrm {sl}} = 4E_{\mathrm {C_{sl}}} (\hat{n}_{\mathrm {sl}} - n_g)^2\\ 
&- E_{\mathrm J_\Sigma} \left[ \cos \frac{\pi\Phi_{\mathrm S}}{\Phi_0} \cos \hat \varphi_{\mathrm {sl}}
- d\sin \frac{\pi\Phi_{\mathrm S}}{\Phi_0} \sin \hat \varphi_{\mathrm {sl}}\right ],\\
 &\hat H_{\mathrm {C, fl}} =  J_{\mathrm {sl}} \left(\hat n_{\mathrm {A}} - \hat n_{\mathrm {B}} \right) \hat n_{\mathrm {sl}},\\
 &\hat H_{\mathrm S}^{\mathrm{fl}} = - E_{\mathrm J_\Sigma} \cos\left(\frac{\pi\Phi_{\mathrm S}}{\Phi_0}\right)\cos \left(\frac{\hat{\varphi }_-}{2} - \hat \varphi_{\mathrm {sl}} \right)\\ 
 &\quad\quad\quad+ d E_{\mathrm J_\Sigma} \sin\left(\frac{\pi\Phi_{\mathrm S}}{\Phi_0}\right)\sin \left(\frac{\hat{\varphi }_-}{2} - \hat \varphi_{\mathrm {sl}} \right).
\end{split}
\label{eq:floating-bare-hamiltonians-and-couplings}
\end{equation}
In the expression above, $E_{\mathrm C_\mu}$ represents the charging energy for the mode $\mu$, $J_{\mathrm c}$ is the direct charge-charge coupling between the two fluxonium modes, and $J_{\mathrm {sl}}$ couples the sloshing mode charge operator $\hat n_\mathrm{sl}$ to the difference of the fluxonium charge operators $(\hat n_{\mathrm A} -\hat n_\mathrm{B})$. We express these quantities in terms of the circuit parameters below
\begin{equation}
    \begin{split}
        E_{\mathrm {C_A}}  = E_{\mathrm {C_B}} & = \frac{2\left(C_\Sigma^{2} -  C_{\mathrm c}^{2} - C_{\mathrm g}^{2}\right) -  C_{\mathrm c} C_{g}}{2C_{\mathrm Q}^3 }, \, J_{\mathrm c} = \frac{4 C_{\mathrm c} C_{\mathrm g}}{C_{\mathrm Q }^3}\\
         E_{\mathrm {C_{sl}}} & = \frac{C + C_\Sigma}{2\left(C_\Sigma^{2} -  C_{\mathrm c}^{2} - C_{\mathrm g}^{2}\right) }, \,J_{\mathrm {sl}} = \frac{4 C_{\mathrm c}}{C_\Sigma^{2} -  C_{\mathrm c}^{2} - C_{\mathrm g}^{2}},
    \end{split}
    \label{eq:floating-design-charging-energies-couplings}
\end{equation}
where $C_{\mathrm Q}^3 =  \left(2 C + C_{\mathrm g}\right)^{2} \left(C_{\mathrm c} + C_{\mathrm g}\right) + C_{\mathrm c} C_{\mathrm g} \left(2 C + C_{\mathrm g}\right)$, $C_{\mathrm {sl}}^2  = C_\Sigma^{2} -  C_{\mathrm c}^{2} - C_{\mathrm g}^{2}$ and $  C_\Sigma = C + C_{\mathrm g} + C_{\mathrm c}$.

For physical insight, we can rewrite $\hat H_{\mathrm S}^{\mathrm{fl}}$ as a sum of terms for each mode and couplings between them. To do so, we expand the trigonometric SQUID terms to all orders, then collect all mixed products of flux operators. The remaining terms contain the flux operator for a single mode only. For each mode, we re-sum the terms into cosine and sine functions~\cite{kounalakis2018tuneable}, which act as corrections to the bare qubit potential. 
 \begin{equation}
      \hat H_{\mathrm S}^{\mathrm{fl}} = \sum_\mu \hat H^{\mathrm{fl}}_{ \mathrm S,\mu} + \hat H_{ \mathrm {S,c}}^{\mathrm{fl, sym}} + \hat H_{\mathrm {S,c}}^{\mathrm{fl, asym}}.
 \end{equation}

and the corrections to individual fluxonium from the SQUID coupler are given by
\begin{equation}
\begin{split}
\hat H_{ \mathrm {S,A}}^{\mathrm{fl}} &= + E_{\mathrm J_\Sigma} \left[ \cos \left(\frac{\pi\Phi_{\mathrm S}}{\Phi_0}\right) \cos \frac{\hat \varphi_{\mathrm {A}}}{2} - d\sin \left(\frac{\pi\Phi_{\mathrm S}}{\Phi_0}\right) \sin \frac{\hat \varphi_{\mathrm {A}}}{2}\right ],\\
\hat H_{ \mathrm {S,B}}^{\mathrm{fl}} &= - E_{\mathrm J_\Sigma} \left[ \cos \left(\frac{\pi\Phi_{\mathrm S}}{\Phi_0}\right) \cos \frac{\hat \varphi_{\mathrm {B}}}{2} - d\sin \left(\frac{\pi\Phi_{\mathrm S}}{\Phi_0}\right) \sin \frac{\hat \varphi_{\mathrm {B}}}{2}\right ].
\end{split}
\label{eq:floating-coupler-correction-qubits}
\end{equation}
The inductive coupling Hamiltonians ${\hat H_{\mathrm {sym}}^{\mathrm {SQ}}}$ and ${\hat H_{\mathrm {asym}}^{\mathrm {SQ}}}$ contain only terms with mixed product of flux operators $\hat \varphi_{\mathrm {A}}, \hat \varphi_{\mathrm {B}}$, and $\hat \varphi_{\mathrm {sl}}$, labeled to distinguish terms arising due to a symmetric SQUID interaction from those due to junction asymmetry. These terms will be considered in detail below.  

\subsubsection{Even-order in flux operator inductive coupling terms}
The even-order inductive coupling terms arise from the expansion of the second cosine in the $- E_{\mathrm J_\Sigma} \cos\left(\pi\Phi_\mathrm{S}/\Phi_0\right)\cos \left(\hat{\varphi }_-/2 - \hat \varphi_{\mathrm {sl}} \right)$ term of Eq.~\eqref{eq:floating-bare-hamiltonians-and-couplings}.  For simplicity, we define the coupler-flux dependent coefficient as $g_{\mathrm {SQ}} = -E_{J_\Sigma} \cos\left(\pi\Phi_\mathrm{S}/\Phi_0\right)$, and rewrite $ \hat H_{ \mathrm {S,c}}^{\mathrm{fl, sym}}$ as 
\begin{align}
       &\hat H_{ \mathrm {S,c}}^{\mathrm{fl, sym}}\nonumber \\
       &=g_{\mathrm {SQ}} \left[\cos \left(\frac{\hat{\varphi }_-}{2}- \hat \varphi_{\mathrm {sl}} \right) - \cos \frac{\hat \varphi_{\mathrm {A}}}{2} - \cos \frac{\hat \varphi_{\mathrm {B}}}{2} - \cos \hat \varphi_{\mathrm {sl}}\right]\nonumber\\
        & =g_{\mathrm {SQ}} \left(\frac{\hat \varphi_{\mathrm {A}} \hat \varphi_{\mathrm {B}}}{4}  + \frac{\hat \varphi_{A}^{2} \hat \varphi_{\mathrm {B}}^{2}}{64}  - \frac{\hat \varphi_{\mathrm {A}}^{3} \hat \varphi_{\mathrm {B}}}{96} - \frac{\hat \varphi_{\mathrm {A}} \hat \varphi_{\mathrm {B}}^{3}}{96}\right.\nonumber\\
        & \left.~~~~ + \left(\frac{\hat{\varphi }_-}{2}\right) \hat \varphi_{\mathrm {sl}}  + \frac{1}{4}\left(\frac{\hat{\varphi }_-}{2}\right)^2 \hat \varphi_{\mathrm {sl}}^{2}- \frac{1}{6}\left(\frac{\hat{\varphi }_-}{2}\right)^3\hat \varphi_{\mathrm {sl}}\right.\nonumber\\
        &\left.~~~~ -\frac{1}{6}\left(\frac{\hat{\varphi }_-}{2}\right) \hat \varphi_{\mathrm {sl}}^{3}+ \mathcal {O}(\hat \varphi^5)\right).
    \label{eq:sym-term-expansion}
\end{align}
In the second step of Eq.~\eqref{eq:sym-term-expansion}, we expanded the cosine to include terms up to fourth order in flux operators. The terms in the second line correspond to direct inductive coupling between the two fluxonium modes. The first term is the linear inductive coupling, and the last three terms are fourth-order coupling responsible for the ZZ interaction. The last line shows how the sloshing mode couples to the difference of the fluxonium mode operators, containing both linear and higher-order terms. Note that at each order, the direct coupling terms have a smaller coefficient than terms coupling either mode to the sloshing mode by a factor of $1/2$. We include only up to second-order terms in Eq.~(\ref{eq:coupl_float}) of the main text.

\subsubsection{Odd-order in flux operator inductive coupling terms}
The odd order inductive coupling terms arise from the expansion of the sine term, $- dE_{\mathrm J_\Sigma} \sin\left(\pi\Phi_\mathrm{S}/\Phi_0\right)\sin \left[\hat{\varphi }_-/2 - \hat \varphi_{\mathrm {sl}} \right]$, in Eq.~\eqref{eq:floating-bare-hamiltonians-and-couplings}.  As before, we define $g^{\mathrm {asym}}_{\mathrm {SQ}} = -dE_{\mathrm J_\Sigma} \sin\left(\pi\Phi_\mathrm{S}/\Phi_0\right)$ and rewrite the term as 
\begin{align}
       \hat H_{ \mathrm {S,c}}^{\mathrm{fl}, asym} & = g^{\mathrm {asym}}_{\mathrm {SQ}} \Big[\sin \left(\frac{\hat{\varphi }_-}{2} - \hat \varphi_{\mathrm {sl}} \right) - \sin \frac{\hat \varphi_{\mathrm {A}}}{2} \nonumber\\
       &~~~~+ \sin \frac{\hat \varphi_{\mathrm {B}}}{2} + \sin \hat \varphi_{\mathrm {sl}}\Big]\nonumber \\
        & =g^{\mathrm {asym}}_{\mathrm {SQ}} \Big[ \frac{\hat \varphi_{\mathrm {A}} \hat \varphi_{\mathrm {B}}}{8}\left(\frac{\hat{\varphi }_-}{2}\right) + \frac{1}{2}\left(\frac{\hat{\varphi }_-}{2}\right)^{2} \hat \varphi_{\mathrm {sl}} \nonumber \\
        &~~~~- \frac{1}{2}\left(\frac{\hat{\varphi }_-}{2}\right) \hat \varphi_{\mathrm {sl}}^{2}+ \mathcal {O}(\hat \varphi^5)\Big]
    \label{eq:asym-term-expansion}
\end{align}
These coupling terms do not qualitatively change the nature of the coupling, but contribute to coherent errors. While beyond fourth-order terms are not analytically considered here, our numerical simulations capture the full nonlinearity of the coupling. 
\\

\bibliography{refs_sup}

\begin{thebibliography}{66}%
\makeatletter
\providecommand \@ifxundefined [1]{%
 \@ifx{#1\undefined}
}%
\providecommand \@ifnum [1]{%
 \ifnum #1\expandafter \@firstoftwo
 \else \expandafter \@secondoftwo
 \fi
}%
\providecommand \@ifx [1]{%
 \ifx #1\expandafter \@firstoftwo
 \else \expandafter \@secondoftwo
 \fi
}%
\providecommand \natexlab [1]{#1}%
\providecommand \enquote  [1]{``#1''}%
\providecommand \bibnamefont  [1]{#1}%
\providecommand \bibfnamefont [1]{#1}%
\providecommand \citenamefont [1]{#1}%
\providecommand \href@noop [0]{\@secondoftwo}%
\providecommand \href [0]{\begingroup \@sanitize@url \@href}%
\providecommand \@href[1]{\@@startlink{#1}\@@href}%
\providecommand \@@href[1]{\endgroup#1\@@endlink}%
\providecommand \@sanitize@url [0]{\catcode `\\12\catcode `\$12\catcode `\&12\catcode `\#12\catcode `\^12\catcode `\_12\catcode `\%12\relax}%
\providecommand \@@startlink[1]{}%
\providecommand \@@endlink[0]{}%
\providecommand \url  [0]{\begingroup\@sanitize@url \@url }%
\providecommand \@url [1]{\endgroup\@href {#1}{\urlprefix }}%
\providecommand \urlprefix  [0]{URL }%
\providecommand \Eprint [0]{\href }%
\providecommand \doibase [0]{https://doi.org/}%
\providecommand \selectlanguage [0]{\@gobble}%
\providecommand \bibinfo  [0]{\@secondoftwo}%
\providecommand \bibfield  [0]{\@secondoftwo}%
\providecommand \translation [1]{[#1]}%
\providecommand \BibitemOpen [0]{}%
\providecommand \bibitemStop [0]{}%
\providecommand \bibitemNoStop [0]{.\EOS\space}%
\providecommand \EOS [0]{\spacefactor3000\relax}%
\providecommand \BibitemShut  [1]{\csname bibitem#1\endcsname}%
\let\auto@bib@innerbib\@empty
\bibitem [{\citenamefont {Vool}\ and\ \citenamefont {Devoret}(2017)}]{vool2017introduction}%
  \BibitemOpen
  \bibfield  {author} {\bibinfo {author} {\bibfnamefont {U.}~\bibnamefont {Vool}}\ and\ \bibinfo {author} {\bibfnamefont {M.}~\bibnamefont {Devoret}},\ }\bibfield  {title} {\bibinfo {title} {Introduction to quantum electromagnetic circuits},\ }\href {https://onlinelibrary.wiley.com/doi/10.1002/cta.2359} {\bibfield  {journal} {\bibinfo  {journal} {International Journal of Circuit Theory and Applications}\ }\textbf {\bibinfo {volume} {45}},\ \bibinfo {pages} {897} (\bibinfo {year} {2017})}\BibitemShut {NoStop}%
\bibitem [{\citenamefont {Blais}\ \emph {et~al.}(2021)\citenamefont {Blais}, \citenamefont {Grimsmo}, \citenamefont {Girvin},\ and\ \citenamefont {Wallraff}}]{Blais2020circuit}%
  \BibitemOpen
  \bibfield  {author} {\bibinfo {author} {\bibfnamefont {A.}~\bibnamefont {Blais}}, \bibinfo {author} {\bibfnamefont {A.~L.}\ \bibnamefont {Grimsmo}}, \bibinfo {author} {\bibfnamefont {S.~M.}\ \bibnamefont {Girvin}},\ and\ \bibinfo {author} {\bibfnamefont {A.}~\bibnamefont {Wallraff}},\ }\bibfield  {title} {\bibinfo {title} {Circuit quantum electrodynamics},\ }\href {https://doi.org/10.1103/RevModPhys.93.025005} {\bibfield  {journal} {\bibinfo  {journal} {Rev. Mod. Phys.}\ }\textbf {\bibinfo {volume} {93}},\ \bibinfo {pages} {025005} (\bibinfo {year} {2021})}\BibitemShut {NoStop}%
\bibitem [{\citenamefont {Kjaergaard}\ \emph {et~al.}(2020)\citenamefont {Kjaergaard}, \citenamefont {Schwartz}, \citenamefont {Braum{\"u}ller}, \citenamefont {Krantz}, \citenamefont {Wang}, \citenamefont {Gustavsson},\ and\ \citenamefont {Oliver}}]{kjaergaard2020superconducting}%
  \BibitemOpen
  \bibfield  {author} {\bibinfo {author} {\bibfnamefont {M.}~\bibnamefont {Kjaergaard}}, \bibinfo {author} {\bibfnamefont {M.~E.}\ \bibnamefont {Schwartz}}, \bibinfo {author} {\bibfnamefont {J.}~\bibnamefont {Braum{\"u}ller}}, \bibinfo {author} {\bibfnamefont {P.}~\bibnamefont {Krantz}}, \bibinfo {author} {\bibfnamefont {J.~I.-J.}\ \bibnamefont {Wang}}, \bibinfo {author} {\bibfnamefont {S.}~\bibnamefont {Gustavsson}},\ and\ \bibinfo {author} {\bibfnamefont {W.~D.}\ \bibnamefont {Oliver}},\ }\bibfield  {title} {\bibinfo {title} {Superconducting qubits: Current state of play},\ }\href {https://www.annualreviews.org/content/journals/10.1146/annurev-conmatphys-031119-050605} {\bibfield  {journal} {\bibinfo  {journal} {Annual Review of Condensed Matter Physics}\ }\textbf {\bibinfo {volume} {11}},\ \bibinfo {pages} {369} (\bibinfo {year} {2020})}\BibitemShut {NoStop}%
\bibitem [{\citenamefont {Krantz}\ \emph {et~al.}(2019)\citenamefont {Krantz}, \citenamefont {Kjaergaard}, \citenamefont {Yan}, \citenamefont {Orlando}, \citenamefont {Gustavsson},\ and\ \citenamefont {Oliver}}]{krantz2019quantum}%
  \BibitemOpen
  \bibfield  {author} {\bibinfo {author} {\bibfnamefont {P.}~\bibnamefont {Krantz}}, \bibinfo {author} {\bibfnamefont {M.}~\bibnamefont {Kjaergaard}}, \bibinfo {author} {\bibfnamefont {F.}~\bibnamefont {Yan}}, \bibinfo {author} {\bibfnamefont {T.~P.}\ \bibnamefont {Orlando}}, \bibinfo {author} {\bibfnamefont {S.}~\bibnamefont {Gustavsson}},\ and\ \bibinfo {author} {\bibfnamefont {W.~D.}\ \bibnamefont {Oliver}},\ }\bibfield  {title} {\bibinfo {title} {A quantum engineer's guide to superconducting qubits},\ }\href {https://pubs.aip.org/aip/apr/article/6/2/021318/570326/A-quantum-engineer-s-guide-to-superconducting} {\bibfield  {journal} {\bibinfo  {journal} {Applied Physics Reviews}\ }\textbf {\bibinfo {volume} {6}} (\bibinfo {year} {2019})}\BibitemShut {NoStop}%
\bibitem [{\citenamefont {Rasmussen}\ \emph {et~al.}(2021)\citenamefont {Rasmussen}, \citenamefont {Christensen}, \citenamefont {Pedersen}, \citenamefont {Kristensen}, \citenamefont {B\ae{}kkegaard}, \citenamefont {Loft},\ and\ \citenamefont {Zinner}}]{rasmussen2021superconducting}%
  \BibitemOpen
  \bibfield  {author} {\bibinfo {author} {\bibfnamefont {S.}~\bibnamefont {Rasmussen}}, \bibinfo {author} {\bibfnamefont {K.}~\bibnamefont {Christensen}}, \bibinfo {author} {\bibfnamefont {S.}~\bibnamefont {Pedersen}}, \bibinfo {author} {\bibfnamefont {L.}~\bibnamefont {Kristensen}}, \bibinfo {author} {\bibfnamefont {T.}~\bibnamefont {B\ae{}kkegaard}}, \bibinfo {author} {\bibfnamefont {N.}~\bibnamefont {Loft}},\ and\ \bibinfo {author} {\bibfnamefont {N.}~\bibnamefont {Zinner}},\ }\bibfield  {title} {\bibinfo {title} {Superconducting circuit companion---an introduction with worked examples},\ }\href {https://doi.org/10.1103/PRXQuantum.2.040204} {\bibfield  {journal} {\bibinfo  {journal} {PRX Quantum}\ }\textbf {\bibinfo {volume} {2}},\ \bibinfo {pages} {040204} (\bibinfo {year} {2021})}\BibitemShut {NoStop}%
\bibitem [{\citenamefont {Grimm}\ \emph {et~al.}(2020)\citenamefont {Grimm}, \citenamefont {Frattini}, \citenamefont {Puri}, \citenamefont {Mundhada}, \citenamefont {Touzard}, \citenamefont {Mirrahimi}, \citenamefont {Girvin}, \citenamefont {Shankar},\ and\ \citenamefont {Devoret}}]{grimm2020stabilization}%
  \BibitemOpen
  \bibfield  {author} {\bibinfo {author} {\bibfnamefont {A.}~\bibnamefont {Grimm}}, \bibinfo {author} {\bibfnamefont {N.~E.}\ \bibnamefont {Frattini}}, \bibinfo {author} {\bibfnamefont {S.}~\bibnamefont {Puri}}, \bibinfo {author} {\bibfnamefont {S.~O.}\ \bibnamefont {Mundhada}}, \bibinfo {author} {\bibfnamefont {S.}~\bibnamefont {Touzard}}, \bibinfo {author} {\bibfnamefont {M.}~\bibnamefont {Mirrahimi}}, \bibinfo {author} {\bibfnamefont {S.~M.}\ \bibnamefont {Girvin}}, \bibinfo {author} {\bibfnamefont {S.}~\bibnamefont {Shankar}},\ and\ \bibinfo {author} {\bibfnamefont {M.~H.}\ \bibnamefont {Devoret}},\ }\bibfield  {title} {\bibinfo {title} {Stabilization and operation of a {K}err-cat qubit},\ }\href {https://www.nature.com/articles/s41586-020-2587-z} {\bibfield  {journal} {\bibinfo  {journal} {Nature}\ }\textbf {\bibinfo {volume} {584}},\ \bibinfo {pages} {205} (\bibinfo {year} {2020})}\BibitemShut {NoStop}%
\bibitem [{\citenamefont {R{\'e}glade}\ \emph {et~al.}(2024)\citenamefont {R{\'e}glade}, \citenamefont {Bocquet}, \citenamefont {Gautier}, \citenamefont {Cohen}, \citenamefont {Marquet}, \citenamefont {Albertinale}, \citenamefont {Pankratova}, \citenamefont {Hall{\'e}n}, \citenamefont {Rautschke}, \citenamefont {Sellem} \emph {et~al.}}]{reglade2024quantum}%
  \BibitemOpen
  \bibfield  {author} {\bibinfo {author} {\bibfnamefont {U.}~\bibnamefont {R{\'e}glade}}, \bibinfo {author} {\bibfnamefont {A.}~\bibnamefont {Bocquet}}, \bibinfo {author} {\bibfnamefont {R.}~\bibnamefont {Gautier}}, \bibinfo {author} {\bibfnamefont {J.}~\bibnamefont {Cohen}}, \bibinfo {author} {\bibfnamefont {A.}~\bibnamefont {Marquet}}, \bibinfo {author} {\bibfnamefont {E.}~\bibnamefont {Albertinale}}, \bibinfo {author} {\bibfnamefont {N.}~\bibnamefont {Pankratova}}, \bibinfo {author} {\bibfnamefont {M.}~\bibnamefont {Hall{\'e}n}}, \bibinfo {author} {\bibfnamefont {F.}~\bibnamefont {Rautschke}}, \bibinfo {author} {\bibfnamefont {L.-A.}\ \bibnamefont {Sellem}}, \emph {et~al.},\ }\bibfield  {title} {\bibinfo {title} {Quantum control of a cat qubit with bit-flip times exceeding ten seconds},\ }\href {https://www.nature.com/articles/s41586-024-07294-3} {\bibfield  {journal} {\bibinfo  {journal} {Nature}\ }\textbf {\bibinfo {volume} {629}},\ \bibinfo {pages} {778} (\bibinfo {year} {2024})}\BibitemShut {NoStop}%
\bibitem [{\citenamefont {Hajr}\ \emph {et~al.}(2024)\citenamefont {Hajr}, \citenamefont {Qing}, \citenamefont {Wang}, \citenamefont {Koolstra}, \citenamefont {Pedramrazi}, \citenamefont {Kang}, \citenamefont {Chen}, \citenamefont {Nguyen}, \citenamefont {J\"unger}, \citenamefont {Goss}, \citenamefont {Huang}, \citenamefont {Bhandari}, \citenamefont {Frattini}, \citenamefont {Puri}, \citenamefont {Dressel}, \citenamefont {Jordan}, \citenamefont {Santiago},\ and\ \citenamefont {Siddiqi}}]{hajr2024high}%
  \BibitemOpen
  \bibfield  {author} {\bibinfo {author} {\bibfnamefont {A.}~\bibnamefont {Hajr}}, \bibinfo {author} {\bibfnamefont {B.}~\bibnamefont {Qing}}, \bibinfo {author} {\bibfnamefont {K.}~\bibnamefont {Wang}}, \bibinfo {author} {\bibfnamefont {G.}~\bibnamefont {Koolstra}}, \bibinfo {author} {\bibfnamefont {Z.}~\bibnamefont {Pedramrazi}}, \bibinfo {author} {\bibfnamefont {Z.}~\bibnamefont {Kang}}, \bibinfo {author} {\bibfnamefont {L.}~\bibnamefont {Chen}}, \bibinfo {author} {\bibfnamefont {L.~B.}\ \bibnamefont {Nguyen}}, \bibinfo {author} {\bibfnamefont {C.}~\bibnamefont {J\"unger}}, \bibinfo {author} {\bibfnamefont {N.}~\bibnamefont {Goss}}, \bibinfo {author} {\bibfnamefont {I.}~\bibnamefont {Huang}}, \bibinfo {author} {\bibfnamefont {B.}~\bibnamefont {Bhandari}}, \bibinfo {author} {\bibfnamefont {N.~E.}\ \bibnamefont {Frattini}}, \bibinfo {author} {\bibfnamefont {S.}~\bibnamefont {Puri}}, \bibinfo {author} {\bibfnamefont {J.}~\bibnamefont {Dressel}}, \bibinfo {author} {\bibfnamefont {A.~N.}\ \bibnamefont {Jordan}},
  \bibinfo {author} {\bibfnamefont {D.~I.}\ \bibnamefont {Santiago}},\ and\ \bibinfo {author} {\bibfnamefont {I.}~\bibnamefont {Siddiqi}},\ }\bibfield  {title} {\bibinfo {title} {High-coherence {K}err-cat qubit in 2d architecture},\ }\href {https://doi.org/10.1103/PhysRevX.14.041049} {\bibfield  {journal} {\bibinfo  {journal} {Phys. Rev. X}\ }\textbf {\bibinfo {volume} {14}},\ \bibinfo {pages} {041049} (\bibinfo {year} {2024})}\BibitemShut {NoStop}%
\bibitem [{\citenamefont {Bhandari}\ \emph {et~al.}(2024)\citenamefont {Bhandari}, \citenamefont {Huang}, \citenamefont {Hajr}, \citenamefont {Yanik}, \citenamefont {Qing}, \citenamefont {Wang}, \citenamefont {Santiago}, \citenamefont {Dressel}, \citenamefont {Siddiqi},\ and\ \citenamefont {Jordan}}]{bhandari2024symmetrically}%
  \BibitemOpen
  \bibfield  {author} {\bibinfo {author} {\bibfnamefont {B.}~\bibnamefont {Bhandari}}, \bibinfo {author} {\bibfnamefont {I.}~\bibnamefont {Huang}}, \bibinfo {author} {\bibfnamefont {A.}~\bibnamefont {Hajr}}, \bibinfo {author} {\bibfnamefont {K.}~\bibnamefont {Yanik}}, \bibinfo {author} {\bibfnamefont {B.}~\bibnamefont {Qing}}, \bibinfo {author} {\bibfnamefont {K.}~\bibnamefont {Wang}}, \bibinfo {author} {\bibfnamefont {D.~I.}\ \bibnamefont {Santiago}}, \bibinfo {author} {\bibfnamefont {J.}~\bibnamefont {Dressel}}, \bibinfo {author} {\bibfnamefont {I.}~\bibnamefont {Siddiqi}},\ and\ \bibinfo {author} {\bibfnamefont {A.~N.}\ \bibnamefont {Jordan}},\ }\bibfield  {title} {\bibinfo {title} {Symmetrically threaded squids as next generation {K}err-cat qubits},\ }\href {https://arxiv.org/abs/2405.11375} {\bibfield  {journal} {\bibinfo  {journal} {arXiv:2405.11375}\ } (\bibinfo {year} {2024})}\BibitemShut {NoStop}%
\bibitem [{\citenamefont {Ding}\ \emph {et~al.}(2023)\citenamefont {Ding}, \citenamefont {Hays}, \citenamefont {Sung}, \citenamefont {Kannan}, \citenamefont {An}, \citenamefont {Di~Paolo}, \citenamefont {Karamlou}, \citenamefont {Hazard}, \citenamefont {Azar}, \citenamefont {Kim}, \citenamefont {Niedzielski}, \citenamefont {Melville}, \citenamefont {Schwartz}, \citenamefont {Yoder}, \citenamefont {Orlando}, \citenamefont {Gustavsson}, \citenamefont {Grover}, \citenamefont {Serniak},\ and\ \citenamefont {Oliver}}]{ding2023high}%
  \BibitemOpen
  \bibfield  {author} {\bibinfo {author} {\bibfnamefont {L.}~\bibnamefont {Ding}}, \bibinfo {author} {\bibfnamefont {M.}~\bibnamefont {Hays}}, \bibinfo {author} {\bibfnamefont {Y.}~\bibnamefont {Sung}}, \bibinfo {author} {\bibfnamefont {B.}~\bibnamefont {Kannan}}, \bibinfo {author} {\bibfnamefont {J.}~\bibnamefont {An}}, \bibinfo {author} {\bibfnamefont {A.}~\bibnamefont {Di~Paolo}}, \bibinfo {author} {\bibfnamefont {A.~H.}\ \bibnamefont {Karamlou}}, \bibinfo {author} {\bibfnamefont {T.~M.}\ \bibnamefont {Hazard}}, \bibinfo {author} {\bibfnamefont {K.}~\bibnamefont {Azar}}, \bibinfo {author} {\bibfnamefont {D.~K.}\ \bibnamefont {Kim}}, \bibinfo {author} {\bibfnamefont {B.~M.}\ \bibnamefont {Niedzielski}}, \bibinfo {author} {\bibfnamefont {A.}~\bibnamefont {Melville}}, \bibinfo {author} {\bibfnamefont {M.~E.}\ \bibnamefont {Schwartz}}, \bibinfo {author} {\bibfnamefont {J.~L.}\ \bibnamefont {Yoder}}, \bibinfo {author} {\bibfnamefont {T.~P.}\ \bibnamefont {Orlando}}, \bibinfo {author} {\bibfnamefont
  {S.}~\bibnamefont {Gustavsson}}, \bibinfo {author} {\bibfnamefont {J.~A.}\ \bibnamefont {Grover}}, \bibinfo {author} {\bibfnamefont {K.}~\bibnamefont {Serniak}},\ and\ \bibinfo {author} {\bibfnamefont {W.~D.}\ \bibnamefont {Oliver}},\ }\bibfield  {title} {\bibinfo {title} {High-fidelity, frequency-flexible two-qubit fluxonium gates with a transmon coupler},\ }\href {https://doi.org/10.1103/PhysRevX.13.031035} {\bibfield  {journal} {\bibinfo  {journal} {Phys. Rev. X}\ }\textbf {\bibinfo {volume} {13}},\ \bibinfo {pages} {031035} (\bibinfo {year} {2023})}\BibitemShut {NoStop}%
\bibitem [{\citenamefont {Acharya}\ \emph {et~al.}(2023)\citenamefont {Acharya} \emph {et~al.}}]{google2023suppressing}%
  \BibitemOpen
  \bibfield  {author} {\bibinfo {author} {\bibfnamefont {R.}~\bibnamefont {Acharya}} \emph {et~al.},\ }\bibfield  {title} {\bibinfo {title} {Suppressing quantum errors by scaling a surface code logical qubit},\ }\href {https://www.nature.com/articles/s41586-022-05434-1} {\bibfield  {journal} {\bibinfo  {journal} {Nature}\ }\textbf {\bibinfo {volume} {614}},\ \bibinfo {pages} {676} (\bibinfo {year} {2023})}\BibitemShut {NoStop}%
\bibitem [{\citenamefont {Livingston}\ \emph {et~al.}(2022)\citenamefont {Livingston}, \citenamefont {Blok}, \citenamefont {Flurin}, \citenamefont {Dressel}, \citenamefont {Jordan},\ and\ \citenamefont {Siddiqi}}]{livingston2022experimental}%
  \BibitemOpen
  \bibfield  {author} {\bibinfo {author} {\bibfnamefont {W.~P.}\ \bibnamefont {Livingston}}, \bibinfo {author} {\bibfnamefont {M.~S.}\ \bibnamefont {Blok}}, \bibinfo {author} {\bibfnamefont {E.}~\bibnamefont {Flurin}}, \bibinfo {author} {\bibfnamefont {J.}~\bibnamefont {Dressel}}, \bibinfo {author} {\bibfnamefont {A.~N.}\ \bibnamefont {Jordan}},\ and\ \bibinfo {author} {\bibfnamefont {I.}~\bibnamefont {Siddiqi}},\ }\bibfield  {title} {\bibinfo {title} {Experimental demonstration of continuous quantum error correction},\ }\href {https://www.nature.com/articles/s41467-022-29906-0} {\bibfield  {journal} {\bibinfo  {journal} {Nature communications}\ }\textbf {\bibinfo {volume} {13}},\ \bibinfo {pages} {2307} (\bibinfo {year} {2022})}\BibitemShut {NoStop}%
\bibitem [{\citenamefont {Nguyen}\ \emph {et~al.}(2024)\citenamefont {Nguyen}, \citenamefont {Kim}, \citenamefont {Hashim}, \citenamefont {Goss}, \citenamefont {Marinelli}, \citenamefont {Bhandari}, \citenamefont {Das}, \citenamefont {Naik}, \citenamefont {Kreikebaum}, \citenamefont {Jordan} \emph {et~al.}}]{nguyen2024programmable}%
  \BibitemOpen
  \bibfield  {author} {\bibinfo {author} {\bibfnamefont {L.~B.}\ \bibnamefont {Nguyen}}, \bibinfo {author} {\bibfnamefont {Y.}~\bibnamefont {Kim}}, \bibinfo {author} {\bibfnamefont {A.}~\bibnamefont {Hashim}}, \bibinfo {author} {\bibfnamefont {N.}~\bibnamefont {Goss}}, \bibinfo {author} {\bibfnamefont {B.}~\bibnamefont {Marinelli}}, \bibinfo {author} {\bibfnamefont {B.}~\bibnamefont {Bhandari}}, \bibinfo {author} {\bibfnamefont {D.}~\bibnamefont {Das}}, \bibinfo {author} {\bibfnamefont {R.~K.}\ \bibnamefont {Naik}}, \bibinfo {author} {\bibfnamefont {J.~M.}\ \bibnamefont {Kreikebaum}}, \bibinfo {author} {\bibfnamefont {A.~N.}\ \bibnamefont {Jordan}}, \emph {et~al.},\ }\bibfield  {title} {\bibinfo {title} {Programmable {H}eisenberg interactions between {F}loquet qubits},\ }\href {https://www.nature.com/articles/s41567-023-02326-7} {\bibfield  {journal} {\bibinfo  {journal} {Nature Physics}\ }\textbf {\bibinfo {volume} {20}},\ \bibinfo {pages} {240} (\bibinfo {year} {2024})}\BibitemShut {NoStop}%
\bibitem [{\citenamefont {Kounalakis}\ \emph {et~al.}(2018)\citenamefont {Kounalakis}, \citenamefont {Dickel}, \citenamefont {Bruno}, \citenamefont {Langford},\ and\ \citenamefont {Steele}}]{kounalakis2018tuneable}%
  \BibitemOpen
  \bibfield  {author} {\bibinfo {author} {\bibfnamefont {M.}~\bibnamefont {Kounalakis}}, \bibinfo {author} {\bibfnamefont {C.}~\bibnamefont {Dickel}}, \bibinfo {author} {\bibfnamefont {A.}~\bibnamefont {Bruno}}, \bibinfo {author} {\bibfnamefont {N.}~\bibnamefont {Langford}},\ and\ \bibinfo {author} {\bibfnamefont {G.}~\bibnamefont {Steele}},\ }\bibfield  {title} {\bibinfo {title} {Tuneable hopping and nonlinear cross-{K}err interactions in a high-coherence superconducting circuit},\ }\href {https://www.nature.com/articles/s41534-018-0088-9} {\bibfield  {journal} {\bibinfo  {journal} {npj Quantum Information}\ }\textbf {\bibinfo {volume} {4}},\ \bibinfo {pages} {38} (\bibinfo {year} {2018})}\BibitemShut {NoStop}%
\bibitem [{\citenamefont {Frey}\ and\ \citenamefont {Rachel}(2022)}]{frey2022realization}%
  \BibitemOpen
  \bibfield  {author} {\bibinfo {author} {\bibfnamefont {P.}~\bibnamefont {Frey}}\ and\ \bibinfo {author} {\bibfnamefont {S.}~\bibnamefont {Rachel}},\ }\bibfield  {title} {\bibinfo {title} {Realization of a discrete time crystal on 57 qubits of a quantum computer},\ }\href {https://www.science.org/doi/full/10.1126/sciadv.abm7652} {\bibfield  {journal} {\bibinfo  {journal} {Science advances}\ }\textbf {\bibinfo {volume} {8}},\ \bibinfo {pages} {7652} (\bibinfo {year} {2022})}\BibitemShut {NoStop}%
\bibitem [{\citenamefont {Bernien}\ \emph {et~al.}(2017)\citenamefont {Bernien}, \citenamefont {Schwartz}, \citenamefont {Keesling}, \citenamefont {Levine}, \citenamefont {Omran}, \citenamefont {Pichler}, \citenamefont {Choi}, \citenamefont {Zibrov}, \citenamefont {Endres}, \citenamefont {Greiner} \emph {et~al.}}]{bernien2017probing}%
  \BibitemOpen
  \bibfield  {author} {\bibinfo {author} {\bibfnamefont {H.}~\bibnamefont {Bernien}}, \bibinfo {author} {\bibfnamefont {S.}~\bibnamefont {Schwartz}}, \bibinfo {author} {\bibfnamefont {A.}~\bibnamefont {Keesling}}, \bibinfo {author} {\bibfnamefont {H.}~\bibnamefont {Levine}}, \bibinfo {author} {\bibfnamefont {A.}~\bibnamefont {Omran}}, \bibinfo {author} {\bibfnamefont {H.}~\bibnamefont {Pichler}}, \bibinfo {author} {\bibfnamefont {S.}~\bibnamefont {Choi}}, \bibinfo {author} {\bibfnamefont {A.~S.}\ \bibnamefont {Zibrov}}, \bibinfo {author} {\bibfnamefont {M.}~\bibnamefont {Endres}}, \bibinfo {author} {\bibfnamefont {M.}~\bibnamefont {Greiner}}, \emph {et~al.},\ }\bibfield  {title} {\bibinfo {title} {Probing many-body dynamics on a 51-atom quantum simulator},\ }\href {https://www.nature.com/articles/nature24622} {\bibfield  {journal} {\bibinfo  {journal} {Nature}\ }\textbf {\bibinfo {volume} {551}},\ \bibinfo {pages} {579} (\bibinfo {year} {2017})}\BibitemShut {NoStop}%
\bibitem [{\citenamefont {Manucharyan}\ \emph {et~al.}(2009)\citenamefont {Manucharyan}, \citenamefont {Koch}, \citenamefont {Glazman},\ and\ \citenamefont {Devoret}}]{manucharyan2009fluxonium}%
  \BibitemOpen
  \bibfield  {author} {\bibinfo {author} {\bibfnamefont {V.~E.}\ \bibnamefont {Manucharyan}}, \bibinfo {author} {\bibfnamefont {J.}~\bibnamefont {Koch}}, \bibinfo {author} {\bibfnamefont {L.~I.}\ \bibnamefont {Glazman}},\ and\ \bibinfo {author} {\bibfnamefont {M.~H.}\ \bibnamefont {Devoret}},\ }\bibfield  {title} {\bibinfo {title} {Fluxonium: Single {C}ooper-pair circuit free of charge offsets},\ }\href {https://www.science.org/doi/10.1126/science.1175552} {\bibfield  {journal} {\bibinfo  {journal} {Science}\ }\textbf {\bibinfo {volume} {326}},\ \bibinfo {pages} {113} (\bibinfo {year} {2009})}\BibitemShut {NoStop}%
\bibitem [{\citenamefont {Zhu}\ \emph {et~al.}(2013)\citenamefont {Zhu}, \citenamefont {Ferguson}, \citenamefont {Manucharyan},\ and\ \citenamefont {Koch}}]{zhu2013circuit}%
  \BibitemOpen
  \bibfield  {author} {\bibinfo {author} {\bibfnamefont {G.}~\bibnamefont {Zhu}}, \bibinfo {author} {\bibfnamefont {D.~G.}\ \bibnamefont {Ferguson}}, \bibinfo {author} {\bibfnamefont {V.~E.}\ \bibnamefont {Manucharyan}},\ and\ \bibinfo {author} {\bibfnamefont {J.}~\bibnamefont {Koch}},\ }\bibfield  {title} {\bibinfo {title} {Circuit qed with fluxonium qubits: Theory of the dispersive regime},\ }\href {https://doi.org/10.1103/PhysRevB.87.024510} {\bibfield  {journal} {\bibinfo  {journal} {Phys. Rev. B}\ }\textbf {\bibinfo {volume} {87}},\ \bibinfo {pages} {024510} (\bibinfo {year} {2013})}\BibitemShut {NoStop}%
\bibitem [{\citenamefont {Earnest}\ \emph {et~al.}(2018)\citenamefont {Earnest}, \citenamefont {Chakram}, \citenamefont {Lu}, \citenamefont {Irons}, \citenamefont {Naik}, \citenamefont {Leung}, \citenamefont {Ocola}, \citenamefont {Czaplewski}, \citenamefont {Baker}, \citenamefont {Lawrence}, \citenamefont {Koch},\ and\ \citenamefont {Schuster}}]{earnest2018realization}%
  \BibitemOpen
  \bibfield  {author} {\bibinfo {author} {\bibfnamefont {N.}~\bibnamefont {Earnest}}, \bibinfo {author} {\bibfnamefont {S.}~\bibnamefont {Chakram}}, \bibinfo {author} {\bibfnamefont {Y.}~\bibnamefont {Lu}}, \bibinfo {author} {\bibfnamefont {N.}~\bibnamefont {Irons}}, \bibinfo {author} {\bibfnamefont {R.~K.}\ \bibnamefont {Naik}}, \bibinfo {author} {\bibfnamefont {N.}~\bibnamefont {Leung}}, \bibinfo {author} {\bibfnamefont {L.}~\bibnamefont {Ocola}}, \bibinfo {author} {\bibfnamefont {D.~A.}\ \bibnamefont {Czaplewski}}, \bibinfo {author} {\bibfnamefont {B.}~\bibnamefont {Baker}}, \bibinfo {author} {\bibfnamefont {J.}~\bibnamefont {Lawrence}}, \bibinfo {author} {\bibfnamefont {J.}~\bibnamefont {Koch}},\ and\ \bibinfo {author} {\bibfnamefont {D.~I.}\ \bibnamefont {Schuster}},\ }\bibfield  {title} {\bibinfo {title} {Realization of a $\mathrm{\ensuremath{\Lambda}}$ system with metastable states of a capacitively shunted fluxonium},\ }\href {https://doi.org/10.1103/PhysRevLett.120.150504} {\bibfield  {journal}
  {\bibinfo  {journal} {Phys. Rev. Lett.}\ }\textbf {\bibinfo {volume} {120}},\ \bibinfo {pages} {150504} (\bibinfo {year} {2018})}\BibitemShut {NoStop}%
\bibitem [{\citenamefont {Nguyen}\ \emph {et~al.}(2019)\citenamefont {Nguyen}, \citenamefont {Lin}, \citenamefont {Somoroff}, \citenamefont {Mencia}, \citenamefont {Grabon},\ and\ \citenamefont {Manucharyan}}]{nguyen2019high}%
  \BibitemOpen
  \bibfield  {author} {\bibinfo {author} {\bibfnamefont {L.~B.}\ \bibnamefont {Nguyen}}, \bibinfo {author} {\bibfnamefont {Y.-H.}\ \bibnamefont {Lin}}, \bibinfo {author} {\bibfnamefont {A.}~\bibnamefont {Somoroff}}, \bibinfo {author} {\bibfnamefont {R.}~\bibnamefont {Mencia}}, \bibinfo {author} {\bibfnamefont {N.}~\bibnamefont {Grabon}},\ and\ \bibinfo {author} {\bibfnamefont {V.~E.}\ \bibnamefont {Manucharyan}},\ }\bibfield  {title} {\bibinfo {title} {High-coherence fluxonium qubit},\ }\href {https://doi.org/10.1103/PhysRevX.9.041041} {\bibfield  {journal} {\bibinfo  {journal} {Phys. Rev. X}\ }\textbf {\bibinfo {volume} {9}},\ \bibinfo {pages} {041041} (\bibinfo {year} {2019})}\BibitemShut {NoStop}%
\bibitem [{\citenamefont {Thibodeau}\ \emph {et~al.}(2024)\citenamefont {Thibodeau}, \citenamefont {Kou},\ and\ \citenamefont {Clark}}]{thibodeau2024the}%
  \BibitemOpen
  \bibfield  {author} {\bibinfo {author} {\bibfnamefont {M.}~\bibnamefont {Thibodeau}}, \bibinfo {author} {\bibfnamefont {A.}~\bibnamefont {Kou}},\ and\ \bibinfo {author} {\bibfnamefont {B.~K.}\ \bibnamefont {Clark}},\ }\bibfield  {title} {\bibinfo {title} {The {F}loquet fluxonium molecule: Driving down dephasing in coupled superconducting qubits},\ }\href {https://doi.org/10.1103/PRXQuantum.5.040314} {\bibfield  {journal} {\bibinfo  {journal} {PRX Quantum}\ }\textbf {\bibinfo {volume} {5}},\ \bibinfo {pages} {040314} (\bibinfo {year} {2024})}\BibitemShut {NoStop}%
\bibitem [{\citenamefont {Somoroff}\ \emph {et~al.}(2023)\citenamefont {Somoroff}, \citenamefont {Ficheux}, \citenamefont {Mencia}, \citenamefont {Xiong}, \citenamefont {Kuzmin},\ and\ \citenamefont {Manucharyan}}]{somoroff2023milli}%
  \BibitemOpen
  \bibfield  {author} {\bibinfo {author} {\bibfnamefont {A.}~\bibnamefont {Somoroff}}, \bibinfo {author} {\bibfnamefont {Q.}~\bibnamefont {Ficheux}}, \bibinfo {author} {\bibfnamefont {R.~A.}\ \bibnamefont {Mencia}}, \bibinfo {author} {\bibfnamefont {H.}~\bibnamefont {Xiong}}, \bibinfo {author} {\bibfnamefont {R.}~\bibnamefont {Kuzmin}},\ and\ \bibinfo {author} {\bibfnamefont {V.~E.}\ \bibnamefont {Manucharyan}},\ }\bibfield  {title} {\bibinfo {title} {Millisecond coherence in a superconducting qubit},\ }\href {https://doi.org/10.1103/PhysRevLett.130.267001} {\bibfield  {journal} {\bibinfo  {journal} {Phys. Rev. Lett.}\ }\textbf {\bibinfo {volume} {130}},\ \bibinfo {pages} {267001} (\bibinfo {year} {2023})}\BibitemShut {NoStop}%
\bibitem [{\citenamefont {Ding}\ \emph {et~al.}(2021)\citenamefont {Ding}, \citenamefont {Ku}, \citenamefont {Shi},\ and\ \citenamefont {Zhao}}]{ding2021free}%
  \BibitemOpen
  \bibfield  {author} {\bibinfo {author} {\bibfnamefont {D.}~\bibnamefont {Ding}}, \bibinfo {author} {\bibfnamefont {H.-S.}\ \bibnamefont {Ku}}, \bibinfo {author} {\bibfnamefont {Y.}~\bibnamefont {Shi}},\ and\ \bibinfo {author} {\bibfnamefont {H.-H.}\ \bibnamefont {Zhao}},\ }\bibfield  {title} {\bibinfo {title} {Free-mode removal and mode decoupling for simulating general superconducting quantum circuits},\ }\href {https://doi.org/10.1103/PhysRevB.103.174501} {\bibfield  {journal} {\bibinfo  {journal} {Phys. Rev. B}\ }\textbf {\bibinfo {volume} {103}},\ \bibinfo {pages} {174501} (\bibinfo {year} {2021})}\BibitemShut {NoStop}%
\bibitem [{\citenamefont {Nguyen}\ \emph {et~al.}(2022)\citenamefont {Nguyen}, \citenamefont {Koolstra}, \citenamefont {Kim}, \citenamefont {Morvan}, \citenamefont {Chistolini}, \citenamefont {Singh}, \citenamefont {Nesterov}, \citenamefont {J\"unger}, \citenamefont {Chen}, \citenamefont {Pedramrazi}, \citenamefont {Mitchell}, \citenamefont {Kreikebaum}, \citenamefont {Puri}, \citenamefont {Santiago},\ and\ \citenamefont {Siddiqi}}]{long2022blueprint}%
  \BibitemOpen
  \bibfield  {author} {\bibinfo {author} {\bibfnamefont {L.~B.}\ \bibnamefont {Nguyen}}, \bibinfo {author} {\bibfnamefont {G.}~\bibnamefont {Koolstra}}, \bibinfo {author} {\bibfnamefont {Y.}~\bibnamefont {Kim}}, \bibinfo {author} {\bibfnamefont {A.}~\bibnamefont {Morvan}}, \bibinfo {author} {\bibfnamefont {T.}~\bibnamefont {Chistolini}}, \bibinfo {author} {\bibfnamefont {S.}~\bibnamefont {Singh}}, \bibinfo {author} {\bibfnamefont {K.~N.}\ \bibnamefont {Nesterov}}, \bibinfo {author} {\bibfnamefont {C.}~\bibnamefont {J\"unger}}, \bibinfo {author} {\bibfnamefont {L.}~\bibnamefont {Chen}}, \bibinfo {author} {\bibfnamefont {Z.}~\bibnamefont {Pedramrazi}}, \bibinfo {author} {\bibfnamefont {B.~K.}\ \bibnamefont {Mitchell}}, \bibinfo {author} {\bibfnamefont {J.~M.}\ \bibnamefont {Kreikebaum}}, \bibinfo {author} {\bibfnamefont {S.}~\bibnamefont {Puri}}, \bibinfo {author} {\bibfnamefont {D.~I.}\ \bibnamefont {Santiago}},\ and\ \bibinfo {author} {\bibfnamefont {I.}~\bibnamefont {Siddiqi}},\ }\bibfield  {title} {\bibinfo
  {title} {Blueprint for a high-performance fluxonium quantum processor},\ }\href {https://doi.org/10.1103/PRXQuantum.3.037001} {\bibfield  {journal} {\bibinfo  {journal} {PRX Quantum}\ }\textbf {\bibinfo {volume} {3}},\ \bibinfo {pages} {037001} (\bibinfo {year} {2022})}\BibitemShut {NoStop}%
\bibitem [{\citenamefont {Koch}\ \emph {et~al.}(2007)\citenamefont {Koch}, \citenamefont {Yu}, \citenamefont {Gambetta}, \citenamefont {Houck}, \citenamefont {Schuster}, \citenamefont {Majer}, \citenamefont {Blais}, \citenamefont {Devoret}, \citenamefont {Girvin},\ and\ \citenamefont {Schoelkopf}}]{koch2007charge}%
  \BibitemOpen
  \bibfield  {author} {\bibinfo {author} {\bibfnamefont {J.}~\bibnamefont {Koch}}, \bibinfo {author} {\bibfnamefont {T.~M.}\ \bibnamefont {Yu}}, \bibinfo {author} {\bibfnamefont {J.}~\bibnamefont {Gambetta}}, \bibinfo {author} {\bibfnamefont {A.~A.}\ \bibnamefont {Houck}}, \bibinfo {author} {\bibfnamefont {D.~I.}\ \bibnamefont {Schuster}}, \bibinfo {author} {\bibfnamefont {J.}~\bibnamefont {Majer}}, \bibinfo {author} {\bibfnamefont {A.}~\bibnamefont {Blais}}, \bibinfo {author} {\bibfnamefont {M.~H.}\ \bibnamefont {Devoret}}, \bibinfo {author} {\bibfnamefont {S.~M.}\ \bibnamefont {Girvin}},\ and\ \bibinfo {author} {\bibfnamefont {R.~J.}\ \bibnamefont {Schoelkopf}},\ }\bibfield  {title} {\bibinfo {title} {Charge-insensitive qubit design derived from the {C}ooper pair box},\ }\href {https://doi.org/10.1103/PhysRevA.76.042319} {\bibfield  {journal} {\bibinfo  {journal} {Phys. Rev. A}\ }\textbf {\bibinfo {volume} {76}},\ \bibinfo {pages} {042319} (\bibinfo {year} {2007})}\BibitemShut {NoStop}%
\bibitem [{\citenamefont {Zhang}\ \emph {et~al.}(2021)\citenamefont {Zhang}, \citenamefont {Chakram}, \citenamefont {Roy}, \citenamefont {Earnest}, \citenamefont {Lu}, \citenamefont {Huang}, \citenamefont {Weiss}, \citenamefont {Koch},\ and\ \citenamefont {Schuster}}]{zhang2021universal}%
  \BibitemOpen
  \bibfield  {author} {\bibinfo {author} {\bibfnamefont {H.}~\bibnamefont {Zhang}}, \bibinfo {author} {\bibfnamefont {S.}~\bibnamefont {Chakram}}, \bibinfo {author} {\bibfnamefont {T.}~\bibnamefont {Roy}}, \bibinfo {author} {\bibfnamefont {N.}~\bibnamefont {Earnest}}, \bibinfo {author} {\bibfnamefont {Y.}~\bibnamefont {Lu}}, \bibinfo {author} {\bibfnamefont {Z.}~\bibnamefont {Huang}}, \bibinfo {author} {\bibfnamefont {D.~K.}\ \bibnamefont {Weiss}}, \bibinfo {author} {\bibfnamefont {J.}~\bibnamefont {Koch}},\ and\ \bibinfo {author} {\bibfnamefont {D.~I.}\ \bibnamefont {Schuster}},\ }\bibfield  {title} {\bibinfo {title} {Universal fast-flux control of a coherent, low-frequency qubit},\ }\href {https://doi.org/10.1103/PhysRevX.11.011010} {\bibfield  {journal} {\bibinfo  {journal} {Phys. Rev. X}\ }\textbf {\bibinfo {volume} {11}},\ \bibinfo {pages} {011010} (\bibinfo {year} {2021})}\BibitemShut {NoStop}%
\bibitem [{\citenamefont {Smith}\ \emph {et~al.}(2016)\citenamefont {Smith}, \citenamefont {Kou}, \citenamefont {Vool}, \citenamefont {Pop}, \citenamefont {Frunzio}, \citenamefont {Schoelkopf},\ and\ \citenamefont {Devoret}}]{smith2016quantization}%
  \BibitemOpen
  \bibfield  {author} {\bibinfo {author} {\bibfnamefont {W.~C.}\ \bibnamefont {Smith}}, \bibinfo {author} {\bibfnamefont {A.}~\bibnamefont {Kou}}, \bibinfo {author} {\bibfnamefont {U.}~\bibnamefont {Vool}}, \bibinfo {author} {\bibfnamefont {I.~M.}\ \bibnamefont {Pop}}, \bibinfo {author} {\bibfnamefont {L.}~\bibnamefont {Frunzio}}, \bibinfo {author} {\bibfnamefont {R.~J.}\ \bibnamefont {Schoelkopf}},\ and\ \bibinfo {author} {\bibfnamefont {M.~H.}\ \bibnamefont {Devoret}},\ }\bibfield  {title} {\bibinfo {title} {Quantization of inductively shunted superconducting circuits},\ }\href {https://doi.org/10.1103/PhysRevB.94.144507} {\bibfield  {journal} {\bibinfo  {journal} {Phys. Rev. B}\ }\textbf {\bibinfo {volume} {94}},\ \bibinfo {pages} {144507} (\bibinfo {year} {2016})}\BibitemShut {NoStop}%
\bibitem [{\citenamefont {Chen}\ \emph {et~al.}(2022)\citenamefont {Chen}, \citenamefont {Nesterov}, \citenamefont {Manucharyan},\ and\ \citenamefont {Vavilov}}]{chen2022fast}%
  \BibitemOpen
  \bibfield  {author} {\bibinfo {author} {\bibfnamefont {Y.}~\bibnamefont {Chen}}, \bibinfo {author} {\bibfnamefont {K.~N.}\ \bibnamefont {Nesterov}}, \bibinfo {author} {\bibfnamefont {V.~E.}\ \bibnamefont {Manucharyan}},\ and\ \bibinfo {author} {\bibfnamefont {M.~G.}\ \bibnamefont {Vavilov}},\ }\bibfield  {title} {\bibinfo {title} {Fast flux entangling gate for fluxonium circuits},\ }\href {https://doi.org/10.1103/PhysRevApplied.18.034027} {\bibfield  {journal} {\bibinfo  {journal} {Phys. Rev. Appl.}\ }\textbf {\bibinfo {volume} {18}},\ \bibinfo {pages} {034027} (\bibinfo {year} {2022})}\BibitemShut {NoStop}%
\bibitem [{\citenamefont {Nesterov}\ \emph {et~al.}(2018)\citenamefont {Nesterov}, \citenamefont {Pechenezhskiy}, \citenamefont {Wang}, \citenamefont {Manucharyan},\ and\ \citenamefont {Vavilov}}]{nesterov2018microwave}%
  \BibitemOpen
  \bibfield  {author} {\bibinfo {author} {\bibfnamefont {K.~N.}\ \bibnamefont {Nesterov}}, \bibinfo {author} {\bibfnamefont {I.~V.}\ \bibnamefont {Pechenezhskiy}}, \bibinfo {author} {\bibfnamefont {C.}~\bibnamefont {Wang}}, \bibinfo {author} {\bibfnamefont {V.~E.}\ \bibnamefont {Manucharyan}},\ and\ \bibinfo {author} {\bibfnamefont {M.~G.}\ \bibnamefont {Vavilov}},\ }\bibfield  {title} {\bibinfo {title} {Microwave-activated controlled-{Z} gate for fixed-frequency fluxonium qubits},\ }\href {https://doi.org/10.1103/PhysRevA.98.030301} {\bibfield  {journal} {\bibinfo  {journal} {Phys. Rev. A}\ }\textbf {\bibinfo {volume} {98}},\ \bibinfo {pages} {030301} (\bibinfo {year} {2018})}\BibitemShut {NoStop}%
\bibitem [{\citenamefont {Nesterov}\ \emph {et~al.}(2022)\citenamefont {Nesterov}, \citenamefont {Wang}, \citenamefont {Manucharyan},\ and\ \citenamefont {Vavilov}}]{nesterov2022cnot}%
  \BibitemOpen
  \bibfield  {author} {\bibinfo {author} {\bibfnamefont {K.~N.}\ \bibnamefont {Nesterov}}, \bibinfo {author} {\bibfnamefont {C.}~\bibnamefont {Wang}}, \bibinfo {author} {\bibfnamefont {V.~E.}\ \bibnamefont {Manucharyan}},\ and\ \bibinfo {author} {\bibfnamefont {M.~G.}\ \bibnamefont {Vavilov}},\ }\bibfield  {title} {\bibinfo {title} {{CNOT} gates for fluxonium qubits via selective darkening of transitions},\ }\href {https://doi.org/10.1103/PhysRevApplied.18.034063} {\bibfield  {journal} {\bibinfo  {journal} {Phys. Rev. Appl.}\ }\textbf {\bibinfo {volume} {18}},\ \bibinfo {pages} {034063} (\bibinfo {year} {2022})}\BibitemShut {NoStop}%
\bibitem [{\citenamefont {Nesterov}\ \emph {et~al.}(2021)\citenamefont {Nesterov}, \citenamefont {Ficheux}, \citenamefont {Manucharyan},\ and\ \citenamefont {Vavilov}}]{nesterov2021proposal}%
  \BibitemOpen
  \bibfield  {author} {\bibinfo {author} {\bibfnamefont {K.~N.}\ \bibnamefont {Nesterov}}, \bibinfo {author} {\bibfnamefont {Q.}~\bibnamefont {Ficheux}}, \bibinfo {author} {\bibfnamefont {V.~E.}\ \bibnamefont {Manucharyan}},\ and\ \bibinfo {author} {\bibfnamefont {M.~G.}\ \bibnamefont {Vavilov}},\ }\bibfield  {title} {\bibinfo {title} {Proposal for entangling gates on fluxonium qubits via a two-photon transition},\ }\href {https://doi.org/10.1103/PRXQuantum.2.020345} {\bibfield  {journal} {\bibinfo  {journal} {PRX Quantum}\ }\textbf {\bibinfo {volume} {2}},\ \bibinfo {pages} {020345} (\bibinfo {year} {2021})}\BibitemShut {NoStop}%
\bibitem [{\citenamefont {Dogan}\ \emph {et~al.}(2023)\citenamefont {Dogan}, \citenamefont {Rosenstock}, \citenamefont {Le~Guevel}, \citenamefont {Xiong}, \citenamefont {Mencia}, \citenamefont {Somoroff}, \citenamefont {Nesterov}, \citenamefont {Vavilov}, \citenamefont {Manucharyan},\ and\ \citenamefont {Wang}}]{dogan2023twofluxonium}%
  \BibitemOpen
  \bibfield  {author} {\bibinfo {author} {\bibfnamefont {E.}~\bibnamefont {Dogan}}, \bibinfo {author} {\bibfnamefont {D.}~\bibnamefont {Rosenstock}}, \bibinfo {author} {\bibfnamefont {L.}~\bibnamefont {Le~Guevel}}, \bibinfo {author} {\bibfnamefont {H.}~\bibnamefont {Xiong}}, \bibinfo {author} {\bibfnamefont {R.~A.}\ \bibnamefont {Mencia}}, \bibinfo {author} {\bibfnamefont {A.}~\bibnamefont {Somoroff}}, \bibinfo {author} {\bibfnamefont {K.~N.}\ \bibnamefont {Nesterov}}, \bibinfo {author} {\bibfnamefont {M.~G.}\ \bibnamefont {Vavilov}}, \bibinfo {author} {\bibfnamefont {V.~E.}\ \bibnamefont {Manucharyan}},\ and\ \bibinfo {author} {\bibfnamefont {C.}~\bibnamefont {Wang}},\ }\bibfield  {title} {\bibinfo {title} {Two-fluxonium cross-resonance gate},\ }\href {https://doi.org/10.1103/PhysRevApplied.20.024011} {\bibfield  {journal} {\bibinfo  {journal} {Phys. Rev. Appl.}\ }\textbf {\bibinfo {volume} {20}},\ \bibinfo {pages} {024011} (\bibinfo {year} {2023})}\BibitemShut {NoStop}%
\bibitem [{\citenamefont {Lin}\ \emph {et~al.}(2025{\natexlab{a}})\citenamefont {Lin}, \citenamefont {Cho}, \citenamefont {Chen}, \citenamefont {Vavilov}, \citenamefont {Wang},\ and\ \citenamefont {Manucharyan}}]{lin2025verifying}%
  \BibitemOpen
  \bibfield  {author} {\bibinfo {author} {\bibfnamefont {W.-J.}\ \bibnamefont {Lin}}, \bibinfo {author} {\bibfnamefont {H.}~\bibnamefont {Cho}}, \bibinfo {author} {\bibfnamefont {Y.}~\bibnamefont {Chen}}, \bibinfo {author} {\bibfnamefont {M.~G.}\ \bibnamefont {Vavilov}}, \bibinfo {author} {\bibfnamefont {C.}~\bibnamefont {Wang}},\ and\ \bibinfo {author} {\bibfnamefont {V.~E.}\ \bibnamefont {Manucharyan}},\ }\bibfield  {title} {\bibinfo {title} {Verifying the analogy between transversely coupled spin-1/2 systems and inductively-coupled fluxoniums},\ }\href {https://iopscience.iop.org/article/10.1088/1367-2630/adb77b/meta} {\bibfield  {journal} {\bibinfo  {journal} {New Journal of Physics}\ }\textbf {\bibinfo {volume} {27}},\ \bibinfo {pages} {033012} (\bibinfo {year} {2025}{\natexlab{a}})}\BibitemShut {NoStop}%
\bibitem [{\citenamefont {Lin}\ \emph {et~al.}(2025{\natexlab{b}})\citenamefont {Lin}, \citenamefont {Cho}, \citenamefont {Chen}, \citenamefont {Vavilov}, \citenamefont {Wang},\ and\ \citenamefont {Manucharyan}}]{lin202524days}%
  \BibitemOpen
  \bibfield  {author} {\bibinfo {author} {\bibfnamefont {W.-J.}\ \bibnamefont {Lin}}, \bibinfo {author} {\bibfnamefont {H.}~\bibnamefont {Cho}}, \bibinfo {author} {\bibfnamefont {Y.}~\bibnamefont {Chen}}, \bibinfo {author} {\bibfnamefont {M.~G.}\ \bibnamefont {Vavilov}}, \bibinfo {author} {\bibfnamefont {C.}~\bibnamefont {Wang}},\ and\ \bibinfo {author} {\bibfnamefont {V.~E.}\ \bibnamefont {Manucharyan}},\ }\bibfield  {title} {\bibinfo {title} {24 days-stable {CNOT} gate on fluxonium qubits with over 99.9$\%$ fidelity},\ }\href {https://doi.org/10.1103/PRXQuantum.6.010349} {\bibfield  {journal} {\bibinfo  {journal} {PRX Quantum}\ }\textbf {\bibinfo {volume} {6}},\ \bibinfo {pages} {010349} (\bibinfo {year} {2025}{\natexlab{b}})}\BibitemShut {NoStop}%
\bibitem [{\citenamefont {Sete}\ \emph {et~al.}(2021{\natexlab{a}})\citenamefont {Sete}, \citenamefont {Chen}, \citenamefont {Manenti}, \citenamefont {Kulshreshtha},\ and\ \citenamefont {Poletto}}]{sete2021floating}%
  \BibitemOpen
  \bibfield  {author} {\bibinfo {author} {\bibfnamefont {E.~A.}\ \bibnamefont {Sete}}, \bibinfo {author} {\bibfnamefont {A.~Q.}\ \bibnamefont {Chen}}, \bibinfo {author} {\bibfnamefont {R.}~\bibnamefont {Manenti}}, \bibinfo {author} {\bibfnamefont {S.}~\bibnamefont {Kulshreshtha}},\ and\ \bibinfo {author} {\bibfnamefont {S.}~\bibnamefont {Poletto}},\ }\bibfield  {title} {\bibinfo {title} {Floating tunable coupler for scalable quantum computing architectures},\ }\href {https://doi.org/10.1103/PhysRevApplied.15.064063} {\bibfield  {journal} {\bibinfo  {journal} {Phys. Rev. Appl.}\ }\textbf {\bibinfo {volume} {15}},\ \bibinfo {pages} {064063} (\bibinfo {year} {2021}{\natexlab{a}})}\BibitemShut {NoStop}%
\bibitem [{\citenamefont {Liang}\ \emph {et~al.}(2023)\citenamefont {Liang}, \citenamefont {Song}, \citenamefont {Deng}, \citenamefont {Gu}, \citenamefont {Yan}, \citenamefont {Mei}, \citenamefont {Zhao}, \citenamefont {Bu}, \citenamefont {Xiao}, \citenamefont {Yu}, \citenamefont {Wang}, \citenamefont {Liu}, \citenamefont {Shi}, \citenamefont {Zhang}, \citenamefont {Li}, \citenamefont {Li}, \citenamefont {Wang}, \citenamefont {Tian}, \citenamefont {Zhao}, \citenamefont {Xu}, \citenamefont {Fan}, \citenamefont {Xiang},\ and\ \citenamefont {Zheng}}]{liang2023tunable}%
  \BibitemOpen
  \bibfield  {author} {\bibinfo {author} {\bibfnamefont {G.-H.}\ \bibnamefont {Liang}}, \bibinfo {author} {\bibfnamefont {X.-H.}\ \bibnamefont {Song}}, \bibinfo {author} {\bibfnamefont {C.-L.}\ \bibnamefont {Deng}}, \bibinfo {author} {\bibfnamefont {X.-Y.}\ \bibnamefont {Gu}}, \bibinfo {author} {\bibfnamefont {Y.}~\bibnamefont {Yan}}, \bibinfo {author} {\bibfnamefont {Z.-Y.}\ \bibnamefont {Mei}}, \bibinfo {author} {\bibfnamefont {S.-L.}\ \bibnamefont {Zhao}}, \bibinfo {author} {\bibfnamefont {Y.-Z.}\ \bibnamefont {Bu}}, \bibinfo {author} {\bibfnamefont {Y.-X.}\ \bibnamefont {Xiao}}, \bibinfo {author} {\bibfnamefont {Y.-H.}\ \bibnamefont {Yu}}, \bibinfo {author} {\bibfnamefont {M.-C.}\ \bibnamefont {Wang}}, \bibinfo {author} {\bibfnamefont {T.}~\bibnamefont {Liu}}, \bibinfo {author} {\bibfnamefont {Y.-H.}\ \bibnamefont {Shi}}, \bibinfo {author} {\bibfnamefont {H.}~\bibnamefont {Zhang}}, \bibinfo {author} {\bibfnamefont {X.}~\bibnamefont {Li}}, \bibinfo {author} {\bibfnamefont {L.}~\bibnamefont {Li}}, \bibinfo
  {author} {\bibfnamefont {J.-Z.}\ \bibnamefont {Wang}}, \bibinfo {author} {\bibfnamefont {Y.}~\bibnamefont {Tian}}, \bibinfo {author} {\bibfnamefont {S.-P.}\ \bibnamefont {Zhao}}, \bibinfo {author} {\bibfnamefont {K.}~\bibnamefont {Xu}}, \bibinfo {author} {\bibfnamefont {H.}~\bibnamefont {Fan}}, \bibinfo {author} {\bibfnamefont {Z.-C.}\ \bibnamefont {Xiang}},\ and\ \bibinfo {author} {\bibfnamefont {D.-N.}\ \bibnamefont {Zheng}},\ }\bibfield  {title} {\bibinfo {title} {Tunable-coupling architectures with capacitively connecting pads for large-scale superconducting multiqubit processors},\ }\href {https://doi.org/10.1103/PhysRevApplied.20.044028} {\bibfield  {journal} {\bibinfo  {journal} {Phys. Rev. Appl.}\ }\textbf {\bibinfo {volume} {20}},\ \bibinfo {pages} {044028} (\bibinfo {year} {2023})}\BibitemShut {NoStop}%
\bibitem [{\citenamefont {Campbell}\ \emph {et~al.}(2023)\citenamefont {Campbell}, \citenamefont {Kamal}, \citenamefont {Ranzani}, \citenamefont {Senatore},\ and\ \citenamefont {LaHaye}}]{campbell2023modular}%
  \BibitemOpen
  \bibfield  {author} {\bibinfo {author} {\bibfnamefont {D.~L.}\ \bibnamefont {Campbell}}, \bibinfo {author} {\bibfnamefont {A.}~\bibnamefont {Kamal}}, \bibinfo {author} {\bibfnamefont {L.}~\bibnamefont {Ranzani}}, \bibinfo {author} {\bibfnamefont {M.}~\bibnamefont {Senatore}},\ and\ \bibinfo {author} {\bibfnamefont {M.~D.}\ \bibnamefont {LaHaye}},\ }\bibfield  {title} {\bibinfo {title} {Modular tunable coupler for superconducting circuits},\ }\href {https://doi.org/10.1103/PhysRevApplied.19.064043} {\bibfield  {journal} {\bibinfo  {journal} {Phys. Rev. Appl.}\ }\textbf {\bibinfo {volume} {19}},\ \bibinfo {pages} {064043} (\bibinfo {year} {2023})}\BibitemShut {NoStop}%
\bibitem [{\citenamefont {Yan}\ \emph {et~al.}(2018)\citenamefont {Yan}, \citenamefont {Krantz}, \citenamefont {Sung}, \citenamefont {Kjaergaard}, \citenamefont {Campbell}, \citenamefont {Orlando}, \citenamefont {Gustavsson},\ and\ \citenamefont {Oliver}}]{yan2018tunable}%
  \BibitemOpen
  \bibfield  {author} {\bibinfo {author} {\bibfnamefont {F.}~\bibnamefont {Yan}}, \bibinfo {author} {\bibfnamefont {P.}~\bibnamefont {Krantz}}, \bibinfo {author} {\bibfnamefont {Y.}~\bibnamefont {Sung}}, \bibinfo {author} {\bibfnamefont {M.}~\bibnamefont {Kjaergaard}}, \bibinfo {author} {\bibfnamefont {D.~L.}\ \bibnamefont {Campbell}}, \bibinfo {author} {\bibfnamefont {T.~P.}\ \bibnamefont {Orlando}}, \bibinfo {author} {\bibfnamefont {S.}~\bibnamefont {Gustavsson}},\ and\ \bibinfo {author} {\bibfnamefont {W.~D.}\ \bibnamefont {Oliver}},\ }\bibfield  {title} {\bibinfo {title} {Tunable coupling scheme for implementing high-fidelity two-qubit gates},\ }\href {https://doi.org/10.1103/PhysRevApplied.10.054062} {\bibfield  {journal} {\bibinfo  {journal} {Phys. Rev. Appl.}\ }\textbf {\bibinfo {volume} {10}},\ \bibinfo {pages} {054062} (\bibinfo {year} {2018})}\BibitemShut {NoStop}%
\bibitem [{\citenamefont {Moskalenko}\ \emph {et~al.}(2021)\citenamefont {Moskalenko}, \citenamefont {Besedin}, \citenamefont {Simakov},\ and\ \citenamefont {Ustinov}}]{moskalenko2021tunable}%
  \BibitemOpen
  \bibfield  {author} {\bibinfo {author} {\bibfnamefont {I.}~\bibnamefont {Moskalenko}}, \bibinfo {author} {\bibfnamefont {I.}~\bibnamefont {Besedin}}, \bibinfo {author} {\bibfnamefont {I.}~\bibnamefont {Simakov}},\ and\ \bibinfo {author} {\bibfnamefont {A.}~\bibnamefont {Ustinov}},\ }\bibfield  {title} {\bibinfo {title} {Tunable coupling scheme for implementing two-qubit gates on fluxonium qubits},\ }\href {https://pubs.aip.org/aip/apl/article-abstract/119/19/194001/1065130/Tunable-coupling-scheme-for-implementing-two-qubit?redirectedFrom=fulltext} {\bibfield  {journal} {\bibinfo  {journal} {Applied Physics Letters}\ }\textbf {\bibinfo {volume} {119}} (\bibinfo {year} {2021})}\BibitemShut {NoStop}%
\bibitem [{\citenamefont {Moskalenko}\ \emph {et~al.}(2022)\citenamefont {Moskalenko}, \citenamefont {Simakov}, \citenamefont {Abramov}, \citenamefont {Grigorev}, \citenamefont {Moskalev}, \citenamefont {Pishchimova}, \citenamefont {Smirnov}, \citenamefont {Zikiy}, \citenamefont {Rodionov},\ and\ \citenamefont {Besedin}}]{moskalenko2022high}%
  \BibitemOpen
  \bibfield  {author} {\bibinfo {author} {\bibfnamefont {I.~N.}\ \bibnamefont {Moskalenko}}, \bibinfo {author} {\bibfnamefont {I.~A.}\ \bibnamefont {Simakov}}, \bibinfo {author} {\bibfnamefont {N.~N.}\ \bibnamefont {Abramov}}, \bibinfo {author} {\bibfnamefont {A.~A.}\ \bibnamefont {Grigorev}}, \bibinfo {author} {\bibfnamefont {D.~O.}\ \bibnamefont {Moskalev}}, \bibinfo {author} {\bibfnamefont {A.~A.}\ \bibnamefont {Pishchimova}}, \bibinfo {author} {\bibfnamefont {N.~S.}\ \bibnamefont {Smirnov}}, \bibinfo {author} {\bibfnamefont {E.~V.}\ \bibnamefont {Zikiy}}, \bibinfo {author} {\bibfnamefont {I.~A.}\ \bibnamefont {Rodionov}},\ and\ \bibinfo {author} {\bibfnamefont {I.~S.}\ \bibnamefont {Besedin}},\ }\bibfield  {title} {\bibinfo {title} {High fidelity two-qubit gates on fluxoniums using a tunable coupler},\ }\href {https://www.nature.com/articles/s41534-022-00644-x} {\bibfield  {journal} {\bibinfo  {journal} {npj Quantum Information}\ }\textbf {\bibinfo {volume} {8}},\ \bibinfo {pages} {130} (\bibinfo {year}
  {2022})}\BibitemShut {NoStop}%
\bibitem [{\citenamefont {Kou}\ \emph {et~al.}(2017)\citenamefont {Kou}, \citenamefont {Smith}, \citenamefont {Vool}, \citenamefont {Brierley}, \citenamefont {Meier}, \citenamefont {Frunzio}, \citenamefont {Girvin}, \citenamefont {Glazman},\ and\ \citenamefont {Devoret}}]{kou2017fluxonium}%
  \BibitemOpen
  \bibfield  {author} {\bibinfo {author} {\bibfnamefont {A.}~\bibnamefont {Kou}}, \bibinfo {author} {\bibfnamefont {W.~C.}\ \bibnamefont {Smith}}, \bibinfo {author} {\bibfnamefont {U.}~\bibnamefont {Vool}}, \bibinfo {author} {\bibfnamefont {R.~T.}\ \bibnamefont {Brierley}}, \bibinfo {author} {\bibfnamefont {H.}~\bibnamefont {Meier}}, \bibinfo {author} {\bibfnamefont {L.}~\bibnamefont {Frunzio}}, \bibinfo {author} {\bibfnamefont {S.~M.}\ \bibnamefont {Girvin}}, \bibinfo {author} {\bibfnamefont {L.~I.}\ \bibnamefont {Glazman}},\ and\ \bibinfo {author} {\bibfnamefont {M.~H.}\ \bibnamefont {Devoret}},\ }\bibfield  {title} {\bibinfo {title} {Fluxonium-based artificial molecule with a tunable magnetic moment},\ }\href {https://doi.org/10.1103/PhysRevX.7.031037} {\bibfield  {journal} {\bibinfo  {journal} {Phys. Rev. X}\ }\textbf {\bibinfo {volume} {7}},\ \bibinfo {pages} {031037} (\bibinfo {year} {2017})}\BibitemShut {NoStop}%
\bibitem [{\citenamefont {Chen}\ \emph {et~al.}(2014)\citenamefont {Chen}, \citenamefont {Neill}, \citenamefont {Roushan}, \citenamefont {Leung}, \citenamefont {Fang}, \citenamefont {Barends}, \citenamefont {Kelly}, \citenamefont {Campbell}, \citenamefont {Chen}, \citenamefont {Chiaro}, \citenamefont {Dunsworth}, \citenamefont {Jeffrey}, \citenamefont {Megrant}, \citenamefont {Mutus}, \citenamefont {O'Malley}, \citenamefont {Quintana}, \citenamefont {Sank}, \citenamefont {Vainsencher}, \citenamefont {Wenner}, \citenamefont {White}, \citenamefont {Geller}, \citenamefont {Cleland},\ and\ \citenamefont {Martinis}}]{chen2014qubit}%
  \BibitemOpen
  \bibfield  {author} {\bibinfo {author} {\bibfnamefont {Y.}~\bibnamefont {Chen}}, \bibinfo {author} {\bibfnamefont {C.}~\bibnamefont {Neill}}, \bibinfo {author} {\bibfnamefont {P.}~\bibnamefont {Roushan}}, \bibinfo {author} {\bibfnamefont {N.}~\bibnamefont {Leung}}, \bibinfo {author} {\bibfnamefont {M.}~\bibnamefont {Fang}}, \bibinfo {author} {\bibfnamefont {R.}~\bibnamefont {Barends}}, \bibinfo {author} {\bibfnamefont {J.}~\bibnamefont {Kelly}}, \bibinfo {author} {\bibfnamefont {B.}~\bibnamefont {Campbell}}, \bibinfo {author} {\bibfnamefont {Z.}~\bibnamefont {Chen}}, \bibinfo {author} {\bibfnamefont {B.}~\bibnamefont {Chiaro}}, \bibinfo {author} {\bibfnamefont {A.}~\bibnamefont {Dunsworth}}, \bibinfo {author} {\bibfnamefont {E.}~\bibnamefont {Jeffrey}}, \bibinfo {author} {\bibfnamefont {A.}~\bibnamefont {Megrant}}, \bibinfo {author} {\bibfnamefont {J.~Y.}\ \bibnamefont {Mutus}}, \bibinfo {author} {\bibfnamefont {P.~J.~J.}\ \bibnamefont {O'Malley}}, \bibinfo {author} {\bibfnamefont {C.~M.}\ \bibnamefont
  {Quintana}}, \bibinfo {author} {\bibfnamefont {D.}~\bibnamefont {Sank}}, \bibinfo {author} {\bibfnamefont {A.}~\bibnamefont {Vainsencher}}, \bibinfo {author} {\bibfnamefont {J.}~\bibnamefont {Wenner}}, \bibinfo {author} {\bibfnamefont {T.~C.}\ \bibnamefont {White}}, \bibinfo {author} {\bibfnamefont {M.~R.}\ \bibnamefont {Geller}}, \bibinfo {author} {\bibfnamefont {A.~N.}\ \bibnamefont {Cleland}},\ and\ \bibinfo {author} {\bibfnamefont {J.~M.}\ \bibnamefont {Martinis}},\ }\bibfield  {title} {\bibinfo {title} {Qubit architecture with high coherence and fast tunable coupling},\ }\href {https://doi.org/10.1103/PhysRevLett.113.220502} {\bibfield  {journal} {\bibinfo  {journal} {Phys. Rev. Lett.}\ }\textbf {\bibinfo {volume} {113}},\ \bibinfo {pages} {220502} (\bibinfo {year} {2014})}\BibitemShut {NoStop}%
\bibitem [{\citenamefont {Weiss}\ \emph {et~al.}(2022)\citenamefont {Weiss}, \citenamefont {Zhang}, \citenamefont {Ding}, \citenamefont {Ma}, \citenamefont {Schuster},\ and\ \citenamefont {Koch}}]{weiss2022fast}%
  \BibitemOpen
  \bibfield  {author} {\bibinfo {author} {\bibfnamefont {D.}~\bibnamefont {Weiss}}, \bibinfo {author} {\bibfnamefont {H.}~\bibnamefont {Zhang}}, \bibinfo {author} {\bibfnamefont {C.}~\bibnamefont {Ding}}, \bibinfo {author} {\bibfnamefont {Y.}~\bibnamefont {Ma}}, \bibinfo {author} {\bibfnamefont {D.~I.}\ \bibnamefont {Schuster}},\ and\ \bibinfo {author} {\bibfnamefont {J.}~\bibnamefont {Koch}},\ }\bibfield  {title} {\bibinfo {title} {Fast high-fidelity gates for galvanically-coupled fluxonium qubits using strong flux modulation},\ }\href {https://doi.org/10.1103/PRXQuantum.3.040336} {\bibfield  {journal} {\bibinfo  {journal} {PRX Quantum}\ }\textbf {\bibinfo {volume} {3}},\ \bibinfo {pages} {040336} (\bibinfo {year} {2022})}\BibitemShut {NoStop}%
\bibitem [{\citenamefont {Zhang}\ \emph {et~al.}(2024)\citenamefont {Zhang}, \citenamefont {Ding}, \citenamefont {Weiss}, \citenamefont {Huang}, \citenamefont {Ma}, \citenamefont {Guinn}, \citenamefont {Sussman}, \citenamefont {Chitta}, \citenamefont {Chen}, \citenamefont {Houck}, \citenamefont {Koch},\ and\ \citenamefont {Schuster}}]{zhang2024tunable}%
  \BibitemOpen
  \bibfield  {author} {\bibinfo {author} {\bibfnamefont {H.}~\bibnamefont {Zhang}}, \bibinfo {author} {\bibfnamefont {C.}~\bibnamefont {Ding}}, \bibinfo {author} {\bibfnamefont {D.}~\bibnamefont {Weiss}}, \bibinfo {author} {\bibfnamefont {Z.}~\bibnamefont {Huang}}, \bibinfo {author} {\bibfnamefont {Y.}~\bibnamefont {Ma}}, \bibinfo {author} {\bibfnamefont {C.}~\bibnamefont {Guinn}}, \bibinfo {author} {\bibfnamefont {S.}~\bibnamefont {Sussman}}, \bibinfo {author} {\bibfnamefont {S.~P.}\ \bibnamefont {Chitta}}, \bibinfo {author} {\bibfnamefont {D.}~\bibnamefont {Chen}}, \bibinfo {author} {\bibfnamefont {A.~A.}\ \bibnamefont {Houck}}, \bibinfo {author} {\bibfnamefont {J.}~\bibnamefont {Koch}},\ and\ \bibinfo {author} {\bibfnamefont {D.~I.}\ \bibnamefont {Schuster}},\ }\bibfield  {title} {\bibinfo {title} {Tunable inductive coupler for high-fidelity gates between fluxonium qubits},\ }\href {https://doi.org/10.1103/PRXQuantum.5.020326} {\bibfield  {journal} {\bibinfo  {journal} {PRX Quantum}\ }\textbf {\bibinfo
  {volume} {5}},\ \bibinfo {pages} {020326} (\bibinfo {year} {2024})}\BibitemShut {NoStop}%
\bibitem [{\citenamefont {Jin}\ \emph {et~al.}(2023)\citenamefont {Jin}, \citenamefont {Cicak}, \citenamefont {Parrott}, \citenamefont {Kotler}, \citenamefont {Lecocq}, \citenamefont {Teufel}, \citenamefont {Aumentado}, \citenamefont {Kapit},\ and\ \citenamefont {Simmonds}}]{jin2023fast}%
  \BibitemOpen
  \bibfield  {author} {\bibinfo {author} {\bibfnamefont {X.}~\bibnamefont {Jin}}, \bibinfo {author} {\bibfnamefont {K.}~\bibnamefont {Cicak}}, \bibinfo {author} {\bibfnamefont {Z.}~\bibnamefont {Parrott}}, \bibinfo {author} {\bibfnamefont {S.}~\bibnamefont {Kotler}}, \bibinfo {author} {\bibfnamefont {F.}~\bibnamefont {Lecocq}}, \bibinfo {author} {\bibfnamefont {J.}~\bibnamefont {Teufel}}, \bibinfo {author} {\bibfnamefont {J.}~\bibnamefont {Aumentado}}, \bibinfo {author} {\bibfnamefont {E.}~\bibnamefont {Kapit}},\ and\ \bibinfo {author} {\bibfnamefont {R.}~\bibnamefont {Simmonds}},\ }\bibfield  {title} {\bibinfo {title} {Fast, tunable, high fidelity {CZ}-gates between superconducting qubits with parametric microwave control of {ZZ}-coupling},\ }\href {https://arxiv.org/abs/2305.02907} {\bibfield  {journal} {\bibinfo  {journal} {arXiv:2305.02907}\ } (\bibinfo {year} {2023})}\BibitemShut {NoStop}%
\bibitem [{\citenamefont {Noh}\ \emph {et~al.}(2023)\citenamefont {Noh}, \citenamefont {Xiao}, \citenamefont {Jin}, \citenamefont {Cicak}, \citenamefont {Doucet}, \citenamefont {Aumentado}, \citenamefont {Govia}, \citenamefont {Ranzani}, \citenamefont {Kamal},\ and\ \citenamefont {Simmonds}}]{noh2023strong}%
  \BibitemOpen
  \bibfield  {author} {\bibinfo {author} {\bibfnamefont {T.}~\bibnamefont {Noh}}, \bibinfo {author} {\bibfnamefont {Z.}~\bibnamefont {Xiao}}, \bibinfo {author} {\bibfnamefont {X.}~\bibnamefont {Jin}}, \bibinfo {author} {\bibfnamefont {K.}~\bibnamefont {Cicak}}, \bibinfo {author} {\bibfnamefont {E.}~\bibnamefont {Doucet}}, \bibinfo {author} {\bibfnamefont {J.}~\bibnamefont {Aumentado}}, \bibinfo {author} {\bibfnamefont {L.~C.}\ \bibnamefont {Govia}}, \bibinfo {author} {\bibfnamefont {L.}~\bibnamefont {Ranzani}}, \bibinfo {author} {\bibfnamefont {A.}~\bibnamefont {Kamal}},\ and\ \bibinfo {author} {\bibfnamefont {R.~W.}\ \bibnamefont {Simmonds}},\ }\bibfield  {title} {\bibinfo {title} {Strong parametric dispersive shifts in a statically decoupled two-qubit cavity {QED} system},\ }\href {https://www.nature.com/articles/s41567-023-02107-2} {\bibfield  {journal} {\bibinfo  {journal} {Nature Physics}\ }\textbf {\bibinfo {volume} {19}},\ \bibinfo {pages} {1445} (\bibinfo {year} {2023})}\BibitemShut {NoStop}%
\bibitem [{\citenamefont {Riwar}\ and\ \citenamefont {DiVincenzo}(2022)}]{riwar2022circuit}%
  \BibitemOpen
  \bibfield  {author} {\bibinfo {author} {\bibfnamefont {R.-P.}\ \bibnamefont {Riwar}}\ and\ \bibinfo {author} {\bibfnamefont {D.~P.}\ \bibnamefont {DiVincenzo}},\ }\bibfield  {title} {\bibinfo {title} {Circuit quantization with time-dependent magnetic fields for realistic geometries},\ }\href {https://www.nature.com/articles/s41534-022-00539-x} {\bibfield  {journal} {\bibinfo  {journal} {npj Quantum Information}\ }\textbf {\bibinfo {volume} {8}},\ \bibinfo {pages} {36} (\bibinfo {year} {2022})}\BibitemShut {NoStop}%
\bibitem [{\citenamefont {You}\ \emph {et~al.}(2019)\citenamefont {You}, \citenamefont {Sauls},\ and\ \citenamefont {Koch}}]{you2019circuit}%
  \BibitemOpen
  \bibfield  {author} {\bibinfo {author} {\bibfnamefont {X.}~\bibnamefont {You}}, \bibinfo {author} {\bibfnamefont {J.~A.}\ \bibnamefont {Sauls}},\ and\ \bibinfo {author} {\bibfnamefont {J.}~\bibnamefont {Koch}},\ }\bibfield  {title} {\bibinfo {title} {Circuit quantization in the presence of time-dependent external flux},\ }\href {https://doi.org/10.1103/PhysRevB.99.174512} {\bibfield  {journal} {\bibinfo  {journal} {Phys. Rev. B}\ }\textbf {\bibinfo {volume} {99}},\ \bibinfo {pages} {174512} (\bibinfo {year} {2019})}\BibitemShut {NoStop}%
\bibitem [{\citenamefont {Lin}\ \emph {et~al.}(2018)\citenamefont {Lin}, \citenamefont {Nguyen}, \citenamefont {Grabon}, \citenamefont {San~Miguel}, \citenamefont {Pankratova},\ and\ \citenamefont {Manucharyan}}]{lin2018demonstration}%
  \BibitemOpen
  \bibfield  {author} {\bibinfo {author} {\bibfnamefont {Y.-H.}\ \bibnamefont {Lin}}, \bibinfo {author} {\bibfnamefont {L.~B.}\ \bibnamefont {Nguyen}}, \bibinfo {author} {\bibfnamefont {N.}~\bibnamefont {Grabon}}, \bibinfo {author} {\bibfnamefont {J.}~\bibnamefont {San~Miguel}}, \bibinfo {author} {\bibfnamefont {N.}~\bibnamefont {Pankratova}},\ and\ \bibinfo {author} {\bibfnamefont {V.~E.}\ \bibnamefont {Manucharyan}},\ }\bibfield  {title} {\bibinfo {title} {Demonstration of protection of a superconducting qubit from energy decay},\ }\href {https://doi.org/10.1103/PhysRevLett.120.150503} {\bibfield  {journal} {\bibinfo  {journal} {Phys. Rev. Lett.}\ }\textbf {\bibinfo {volume} {120}},\ \bibinfo {pages} {150503} (\bibinfo {year} {2018})}\BibitemShut {NoStop}%
\bibitem [{\citenamefont {Nakamura}\ \emph {et~al.}(1999)\citenamefont {Nakamura}, \citenamefont {Pashkin},\ and\ \citenamefont {Tsai}}]{nakamura1999coherent}%
  \BibitemOpen
  \bibfield  {author} {\bibinfo {author} {\bibfnamefont {Y.}~\bibnamefont {Nakamura}}, \bibinfo {author} {\bibfnamefont {Y.~A.}\ \bibnamefont {Pashkin}},\ and\ \bibinfo {author} {\bibfnamefont {J.}~\bibnamefont {Tsai}},\ }\bibfield  {title} {\bibinfo {title} {Coherent control of macroscopic quantum states in a single-{C}ooper-pair box},\ }\href {https://www.nature.com/articles/19718} {\bibfield  {journal} {\bibinfo  {journal} {nature}\ }\textbf {\bibinfo {volume} {398}},\ \bibinfo {pages} {786} (\bibinfo {year} {1999})}\BibitemShut {NoStop}%
\bibitem [{\citenamefont {Bouchiat}\ \emph {et~al.}(1998)\citenamefont {Bouchiat}, \citenamefont {Vion}, \citenamefont {Joyez}, \citenamefont {Esteve},\ and\ \citenamefont {Devoret}}]{bouchiat1998quantum}%
  \BibitemOpen
  \bibfield  {author} {\bibinfo {author} {\bibfnamefont {V.}~\bibnamefont {Bouchiat}}, \bibinfo {author} {\bibfnamefont {D.}~\bibnamefont {Vion}}, \bibinfo {author} {\bibfnamefont {P.}~\bibnamefont {Joyez}}, \bibinfo {author} {\bibfnamefont {D.}~\bibnamefont {Esteve}},\ and\ \bibinfo {author} {\bibfnamefont {M.}~\bibnamefont {Devoret}},\ }\bibfield  {title} {\bibinfo {title} {Quantum coherence with a single {C}ooper pair},\ }\href {https://iopscience.iop.org/article/10.1238/Physica.Topical.076a00165} {\bibfield  {journal} {\bibinfo  {journal} {Physica Scripta}\ }\textbf {\bibinfo {volume} {1998}},\ \bibinfo {pages} {165} (\bibinfo {year} {1998})}\BibitemShut {NoStop}%
\bibitem [{\citenamefont {Sung}\ \emph {et~al.}(2021)\citenamefont {Sung}, \citenamefont {Ding}, \citenamefont {Braum\"uller}, \citenamefont {Veps\"al\"ainen}, \citenamefont {Kannan}, \citenamefont {Kjaergaard}, \citenamefont {Greene}, \citenamefont {Samach}, \citenamefont {McNally}, \citenamefont {Kim}, \citenamefont {Melville}, \citenamefont {Niedzielski}, \citenamefont {Schwartz}, \citenamefont {Yoder}, \citenamefont {Orlando}, \citenamefont {Gustavsson},\ and\ \citenamefont {Oliver}}]{sung2021realization}%
  \BibitemOpen
  \bibfield  {author} {\bibinfo {author} {\bibfnamefont {Y.}~\bibnamefont {Sung}}, \bibinfo {author} {\bibfnamefont {L.}~\bibnamefont {Ding}}, \bibinfo {author} {\bibfnamefont {J.}~\bibnamefont {Braum\"uller}}, \bibinfo {author} {\bibfnamefont {A.}~\bibnamefont {Veps\"al\"ainen}}, \bibinfo {author} {\bibfnamefont {B.}~\bibnamefont {Kannan}}, \bibinfo {author} {\bibfnamefont {M.}~\bibnamefont {Kjaergaard}}, \bibinfo {author} {\bibfnamefont {A.}~\bibnamefont {Greene}}, \bibinfo {author} {\bibfnamefont {G.~O.}\ \bibnamefont {Samach}}, \bibinfo {author} {\bibfnamefont {C.}~\bibnamefont {McNally}}, \bibinfo {author} {\bibfnamefont {D.}~\bibnamefont {Kim}}, \bibinfo {author} {\bibfnamefont {A.}~\bibnamefont {Melville}}, \bibinfo {author} {\bibfnamefont {B.~M.}\ \bibnamefont {Niedzielski}}, \bibinfo {author} {\bibfnamefont {M.~E.}\ \bibnamefont {Schwartz}}, \bibinfo {author} {\bibfnamefont {J.~L.}\ \bibnamefont {Yoder}}, \bibinfo {author} {\bibfnamefont {T.~P.}\ \bibnamefont {Orlando}}, \bibinfo {author}
  {\bibfnamefont {S.}~\bibnamefont {Gustavsson}},\ and\ \bibinfo {author} {\bibfnamefont {W.~D.}\ \bibnamefont {Oliver}},\ }\bibfield  {title} {\bibinfo {title} {Realization of high-fidelity {CZ} and {ZZ}-free i{SWAP} gates with a tunable coupler},\ }\href {https://doi.org/10.1103/PhysRevX.11.021058} {\bibfield  {journal} {\bibinfo  {journal} {Phys. Rev. X}\ }\textbf {\bibinfo {volume} {11}},\ \bibinfo {pages} {021058} (\bibinfo {year} {2021})}\BibitemShut {NoStop}%
\bibitem [{\citenamefont {Fors}\ \emph {et~al.}(2024)\citenamefont {Fors}, \citenamefont {Fern{\'a}ndez-Pend{\'a}s},\ and\ \citenamefont {Kockum}}]{fors2024comprehensive}%
  \BibitemOpen
  \bibfield  {author} {\bibinfo {author} {\bibfnamefont {S.~P.}\ \bibnamefont {Fors}}, \bibinfo {author} {\bibfnamefont {J.}~\bibnamefont {Fern{\'a}ndez-Pend{\'a}s}},\ and\ \bibinfo {author} {\bibfnamefont {A.~F.}\ \bibnamefont {Kockum}},\ }\bibfield  {title} {\bibinfo {title} {Comprehensive explanation of {ZZ} coupling in superconducting qubits},\ }\href {https://arxiv.org/abs/2408.15402} {\bibfield  {journal} {\bibinfo  {journal} {arXiv:2408.15402}\ } (\bibinfo {year} {2024})}\BibitemShut {NoStop}%
\bibitem [{\citenamefont {Kim}\ \emph {et~al.}(2025)\citenamefont {Kim}, \citenamefont {Hays}, \citenamefont {Rosen}, \citenamefont {An}, \citenamefont {Zhang}, \citenamefont {Goswami}, \citenamefont {Azar}, \citenamefont {Gertler}, \citenamefont {Niedzielski}, \citenamefont {Schwartz} \emph {et~al.}}]{kim2025emergent}%
  \BibitemOpen
  \bibfield  {author} {\bibinfo {author} {\bibfnamefont {J.}~\bibnamefont {Kim}}, \bibinfo {author} {\bibfnamefont {M.}~\bibnamefont {Hays}}, \bibinfo {author} {\bibfnamefont {I.~T.}\ \bibnamefont {Rosen}}, \bibinfo {author} {\bibfnamefont {J.}~\bibnamefont {An}}, \bibinfo {author} {\bibfnamefont {H.}~\bibnamefont {Zhang}}, \bibinfo {author} {\bibfnamefont {A.}~\bibnamefont {Goswami}}, \bibinfo {author} {\bibfnamefont {K.}~\bibnamefont {Azar}}, \bibinfo {author} {\bibfnamefont {J.~M.}\ \bibnamefont {Gertler}}, \bibinfo {author} {\bibfnamefont {B.~M.}\ \bibnamefont {Niedzielski}}, \bibinfo {author} {\bibfnamefont {M.~E.}\ \bibnamefont {Schwartz}}, \emph {et~al.},\ }\bibfield  {title} {\bibinfo {title} {Emergent harmonics in {J}osephson tunnel junctions due to series inductance},\ }\href {https://arxiv.org/abs/2507.08171} {\bibfield  {journal} {\bibinfo  {journal} {arXiv:2507.08171}\ } (\bibinfo {year} {2025})}\BibitemShut {NoStop}%
\bibitem [{\citenamefont {Foxen}\ \emph {et~al.}(2020)\citenamefont {Foxen}, \citenamefont {Neill}, \citenamefont {Dunsworth}, \citenamefont {Roushan}, \citenamefont {Chiaro}, \citenamefont {Megrant}, \citenamefont {Kelly}, \citenamefont {Chen}, \citenamefont {Satzinger}, \citenamefont {Barends}, \citenamefont {Arute}, \citenamefont {Arya}, \citenamefont {Babbush}, \citenamefont {Bacon}, \citenamefont {Bardin}, \citenamefont {Boixo}, \citenamefont {Buell}, \citenamefont {Burkett}, \citenamefont {Chen}, \citenamefont {Collins}, \citenamefont {Farhi}, \citenamefont {Fowler}, \citenamefont {Gidney}, \citenamefont {Giustina}, \citenamefont {Graff}, \citenamefont {Harrigan}, \citenamefont {Huang}, \citenamefont {Isakov}, \citenamefont {Jeffrey}, \citenamefont {Jiang}, \citenamefont {Kafri}, \citenamefont {Kechedzhi}, \citenamefont {Klimov}, \citenamefont {Korotkov}, \citenamefont {Kostritsa}, \citenamefont {Landhuis}, \citenamefont {Lucero}, \citenamefont {McClean}, \citenamefont {McEwen}, \citenamefont {Mi},
  \citenamefont {Mohseni}, \citenamefont {Mutus}, \citenamefont {Naaman}, \citenamefont {Neeley}, \citenamefont {Niu}, \citenamefont {Petukhov}, \citenamefont {Quintana}, \citenamefont {Rubin}, \citenamefont {Sank}, \citenamefont {Smelyanskiy}, \citenamefont {Vainsencher}, \citenamefont {White}, \citenamefont {Yao}, \citenamefont {Yeh}, \citenamefont {Zalcman}, \citenamefont {Neven},\ and\ \citenamefont {Martinis}}]{foxen2020demonstrating}%
  \BibitemOpen
  \bibfield  {author} {\bibinfo {author} {\bibfnamefont {B.}~\bibnamefont {Foxen}}, \bibinfo {author} {\bibfnamefont {C.}~\bibnamefont {Neill}}, \bibinfo {author} {\bibfnamefont {A.}~\bibnamefont {Dunsworth}}, \bibinfo {author} {\bibfnamefont {P.}~\bibnamefont {Roushan}}, \bibinfo {author} {\bibfnamefont {B.}~\bibnamefont {Chiaro}}, \bibinfo {author} {\bibfnamefont {A.}~\bibnamefont {Megrant}}, \bibinfo {author} {\bibfnamefont {J.}~\bibnamefont {Kelly}}, \bibinfo {author} {\bibfnamefont {Z.}~\bibnamefont {Chen}}, \bibinfo {author} {\bibfnamefont {K.}~\bibnamefont {Satzinger}}, \bibinfo {author} {\bibfnamefont {R.}~\bibnamefont {Barends}}, \bibinfo {author} {\bibfnamefont {F.}~\bibnamefont {Arute}}, \bibinfo {author} {\bibfnamefont {K.}~\bibnamefont {Arya}}, \bibinfo {author} {\bibfnamefont {R.}~\bibnamefont {Babbush}}, \bibinfo {author} {\bibfnamefont {D.}~\bibnamefont {Bacon}}, \bibinfo {author} {\bibfnamefont {J.~C.}\ \bibnamefont {Bardin}}, \bibinfo {author} {\bibfnamefont {S.}~\bibnamefont {Boixo}},
  \bibinfo {author} {\bibfnamefont {D.}~\bibnamefont {Buell}}, \bibinfo {author} {\bibfnamefont {B.}~\bibnamefont {Burkett}}, \bibinfo {author} {\bibfnamefont {Y.}~\bibnamefont {Chen}}, \bibinfo {author} {\bibfnamefont {R.}~\bibnamefont {Collins}}, \bibinfo {author} {\bibfnamefont {E.}~\bibnamefont {Farhi}}, \bibinfo {author} {\bibfnamefont {A.}~\bibnamefont {Fowler}}, \bibinfo {author} {\bibfnamefont {C.}~\bibnamefont {Gidney}}, \bibinfo {author} {\bibfnamefont {M.}~\bibnamefont {Giustina}}, \bibinfo {author} {\bibfnamefont {R.}~\bibnamefont {Graff}}, \bibinfo {author} {\bibfnamefont {M.}~\bibnamefont {Harrigan}}, \bibinfo {author} {\bibfnamefont {T.}~\bibnamefont {Huang}}, \bibinfo {author} {\bibfnamefont {S.~V.}\ \bibnamefont {Isakov}}, \bibinfo {author} {\bibfnamefont {E.}~\bibnamefont {Jeffrey}}, \bibinfo {author} {\bibfnamefont {Z.}~\bibnamefont {Jiang}}, \bibinfo {author} {\bibfnamefont {D.}~\bibnamefont {Kafri}}, \bibinfo {author} {\bibfnamefont {K.}~\bibnamefont {Kechedzhi}}, \bibinfo {author}
  {\bibfnamefont {P.}~\bibnamefont {Klimov}}, \bibinfo {author} {\bibfnamefont {A.}~\bibnamefont {Korotkov}}, \bibinfo {author} {\bibfnamefont {F.}~\bibnamefont {Kostritsa}}, \bibinfo {author} {\bibfnamefont {D.}~\bibnamefont {Landhuis}}, \bibinfo {author} {\bibfnamefont {E.}~\bibnamefont {Lucero}}, \bibinfo {author} {\bibfnamefont {J.}~\bibnamefont {McClean}}, \bibinfo {author} {\bibfnamefont {M.}~\bibnamefont {McEwen}}, \bibinfo {author} {\bibfnamefont {X.}~\bibnamefont {Mi}}, \bibinfo {author} {\bibfnamefont {M.}~\bibnamefont {Mohseni}}, \bibinfo {author} {\bibfnamefont {J.~Y.}\ \bibnamefont {Mutus}}, \bibinfo {author} {\bibfnamefont {O.}~\bibnamefont {Naaman}}, \bibinfo {author} {\bibfnamefont {M.}~\bibnamefont {Neeley}}, \bibinfo {author} {\bibfnamefont {M.}~\bibnamefont {Niu}}, \bibinfo {author} {\bibfnamefont {A.}~\bibnamefont {Petukhov}}, \bibinfo {author} {\bibfnamefont {C.}~\bibnamefont {Quintana}}, \bibinfo {author} {\bibfnamefont {N.}~\bibnamefont {Rubin}}, \bibinfo {author} {\bibfnamefont
  {D.}~\bibnamefont {Sank}}, \bibinfo {author} {\bibfnamefont {V.}~\bibnamefont {Smelyanskiy}}, \bibinfo {author} {\bibfnamefont {A.}~\bibnamefont {Vainsencher}}, \bibinfo {author} {\bibfnamefont {T.~C.}\ \bibnamefont {White}}, \bibinfo {author} {\bibfnamefont {Z.}~\bibnamefont {Yao}}, \bibinfo {author} {\bibfnamefont {P.}~\bibnamefont {Yeh}}, \bibinfo {author} {\bibfnamefont {A.}~\bibnamefont {Zalcman}}, \bibinfo {author} {\bibfnamefont {H.}~\bibnamefont {Neven}},\ and\ \bibinfo {author} {\bibfnamefont {J.~M.}\ \bibnamefont {Martinis}} (\bibinfo {collaboration} {Google AI Quantum}),\ }\bibfield  {title} {\bibinfo {title} {Demonstrating a continuous set of two-qubit gates for near-term quantum algorithms},\ }\href {https://doi.org/10.1103/PhysRevLett.125.120504} {\bibfield  {journal} {\bibinfo  {journal} {Phys. Rev. Lett.}\ }\textbf {\bibinfo {volume} {125}},\ \bibinfo {pages} {120504} (\bibinfo {year} {2020})}\BibitemShut {NoStop}%
\bibitem [{\citenamefont {Zanardi}\ \emph {et~al.}(2000)\citenamefont {Zanardi}, \citenamefont {Zalka},\ and\ \citenamefont {Faoro}}]{zanardi2000entangling}%
  \BibitemOpen
  \bibfield  {author} {\bibinfo {author} {\bibfnamefont {P.}~\bibnamefont {Zanardi}}, \bibinfo {author} {\bibfnamefont {C.}~\bibnamefont {Zalka}},\ and\ \bibinfo {author} {\bibfnamefont {L.}~\bibnamefont {Faoro}},\ }\bibfield  {title} {\bibinfo {title} {Entangling power of quantum evolutions},\ }\href {https://doi.org/10.1103/PhysRevA.62.030301} {\bibfield  {journal} {\bibinfo  {journal} {Phys. Rev. A}\ }\textbf {\bibinfo {volume} {62}},\ \bibinfo {pages} {030301} (\bibinfo {year} {2000})}\BibitemShut {NoStop}%
\bibitem [{\citenamefont {Ma}\ and\ \citenamefont {Wang}(2007)}]{ma2007matrix}%
  \BibitemOpen
  \bibfield  {author} {\bibinfo {author} {\bibfnamefont {Z.}~\bibnamefont {Ma}}\ and\ \bibinfo {author} {\bibfnamefont {X.}~\bibnamefont {Wang}},\ }\bibfield  {title} {\bibinfo {title} {Matrix realignment and partial-transpose approach to entangling power of quantum evolutions},\ }\href {https://doi.org/10.1103/PhysRevA.75.014304} {\bibfield  {journal} {\bibinfo  {journal} {Phys. Rev. A}\ }\textbf {\bibinfo {volume} {75}},\ \bibinfo {pages} {014304} (\bibinfo {year} {2007})}\BibitemShut {NoStop}%
\bibitem [{\citenamefont {Nielsen}(2002)}]{nielsen2002simple}%
  \BibitemOpen
  \bibfield  {author} {\bibinfo {author} {\bibfnamefont {M.~A.}\ \bibnamefont {Nielsen}},\ }\bibfield  {title} {\bibinfo {title} {A simple formula for the average gate fidelity of a quantum dynamical operation},\ }\href {https://www.sciencedirect.com/science/article/abs/pii/S0375960102012720} {\bibfield  {journal} {\bibinfo  {journal} {Physics Letters A}\ }\textbf {\bibinfo {volume} {303}},\ \bibinfo {pages} {249} (\bibinfo {year} {2002})}\BibitemShut {NoStop}%
\bibitem [{\citenamefont {Horodecki}\ \emph {et~al.}(1999)\citenamefont {Horodecki}, \citenamefont {Horodecki},\ and\ \citenamefont {Horodecki}}]{horodecki1999general}%
  \BibitemOpen
  \bibfield  {author} {\bibinfo {author} {\bibfnamefont {M.}~\bibnamefont {Horodecki}}, \bibinfo {author} {\bibfnamefont {P.}~\bibnamefont {Horodecki}},\ and\ \bibinfo {author} {\bibfnamefont {R.}~\bibnamefont {Horodecki}},\ }\bibfield  {title} {\bibinfo {title} {General teleportation channel, singlet fraction, and quasidistillation},\ }\href {https://doi.org/10.1103/PhysRevA.60.1888} {\bibfield  {journal} {\bibinfo  {journal} {Phys. Rev. A}\ }\textbf {\bibinfo {volume} {60}},\ \bibinfo {pages} {1888} (\bibinfo {year} {1999})}\BibitemShut {NoStop}%
\bibitem [{\citenamefont {Chow}\ \emph {et~al.}(2009)\citenamefont {Chow}, \citenamefont {Gambetta}, \citenamefont {Tornberg}, \citenamefont {Koch}, \citenamefont {Bishop}, \citenamefont {Houck}, \citenamefont {Johnson}, \citenamefont {Frunzio}, \citenamefont {Girvin},\ and\ \citenamefont {Schoelkopf}}]{chow2009randomized}%
  \BibitemOpen
  \bibfield  {author} {\bibinfo {author} {\bibfnamefont {J.~M.}\ \bibnamefont {Chow}}, \bibinfo {author} {\bibfnamefont {J.~M.}\ \bibnamefont {Gambetta}}, \bibinfo {author} {\bibfnamefont {L.}~\bibnamefont {Tornberg}}, \bibinfo {author} {\bibfnamefont {J.}~\bibnamefont {Koch}}, \bibinfo {author} {\bibfnamefont {L.~S.}\ \bibnamefont {Bishop}}, \bibinfo {author} {\bibfnamefont {A.~A.}\ \bibnamefont {Houck}}, \bibinfo {author} {\bibfnamefont {B.~R.}\ \bibnamefont {Johnson}}, \bibinfo {author} {\bibfnamefont {L.}~\bibnamefont {Frunzio}}, \bibinfo {author} {\bibfnamefont {S.~M.}\ \bibnamefont {Girvin}},\ and\ \bibinfo {author} {\bibfnamefont {R.~J.}\ \bibnamefont {Schoelkopf}},\ }\bibfield  {title} {\bibinfo {title} {Randomized benchmarking and process tomography for gate errors in a solid-state qubit},\ }\href {https://doi.org/10.1103/PhysRevLett.102.090502} {\bibfield  {journal} {\bibinfo  {journal} {Phys. Rev. Lett.}\ }\textbf {\bibinfo {volume} {102}},\ \bibinfo {pages} {090502} (\bibinfo {year}
  {2009})}\BibitemShut {NoStop}%
\bibitem [{\citenamefont {Pedersen}\ \emph {et~al.}(2007)\citenamefont {Pedersen}, \citenamefont {M{\o}ller},\ and\ \citenamefont {M{\o}lmer}}]{pedersen2007fidelity}%
  \BibitemOpen
  \bibfield  {author} {\bibinfo {author} {\bibfnamefont {L.~H.}\ \bibnamefont {Pedersen}}, \bibinfo {author} {\bibfnamefont {N.~M.}\ \bibnamefont {M{\o}ller}},\ and\ \bibinfo {author} {\bibfnamefont {K.}~\bibnamefont {M{\o}lmer}},\ }\bibfield  {title} {\bibinfo {title} {Fidelity of quantum operations},\ }\href {https://www.sciencedirect.com/science/article/abs/pii/S0375960107003271} {\bibfield  {journal} {\bibinfo  {journal} {Physics Letters A}\ }\textbf {\bibinfo {volume} {367}},\ \bibinfo {pages} {47} (\bibinfo {year} {2007})}\BibitemShut {NoStop}%
\bibitem [{\citenamefont {Campbell}\ \emph {et~al.}(2020)\citenamefont {Campbell}, \citenamefont {Shim}, \citenamefont {Kannan}, \citenamefont {Winik}, \citenamefont {Kim}, \citenamefont {Melville}, \citenamefont {Niedzielski}, \citenamefont {Yoder}, \citenamefont {Tahan}, \citenamefont {Gustavsson},\ and\ \citenamefont {Oliver}}]{campbell2020universal}%
  \BibitemOpen
  \bibfield  {author} {\bibinfo {author} {\bibfnamefont {D.~L.}\ \bibnamefont {Campbell}}, \bibinfo {author} {\bibfnamefont {Y.-P.}\ \bibnamefont {Shim}}, \bibinfo {author} {\bibfnamefont {B.}~\bibnamefont {Kannan}}, \bibinfo {author} {\bibfnamefont {R.}~\bibnamefont {Winik}}, \bibinfo {author} {\bibfnamefont {D.~K.}\ \bibnamefont {Kim}}, \bibinfo {author} {\bibfnamefont {A.}~\bibnamefont {Melville}}, \bibinfo {author} {\bibfnamefont {B.~M.}\ \bibnamefont {Niedzielski}}, \bibinfo {author} {\bibfnamefont {J.~L.}\ \bibnamefont {Yoder}}, \bibinfo {author} {\bibfnamefont {C.}~\bibnamefont {Tahan}}, \bibinfo {author} {\bibfnamefont {S.}~\bibnamefont {Gustavsson}},\ and\ \bibinfo {author} {\bibfnamefont {W.~D.}\ \bibnamefont {Oliver}},\ }\bibfield  {title} {\bibinfo {title} {Universal nonadiabatic control of small-gap superconducting qubits},\ }\href {https://doi.org/10.1103/PhysRevX.10.041051} {\bibfield  {journal} {\bibinfo  {journal} {Phys. Rev. X}\ }\textbf {\bibinfo {volume} {10}},\ \bibinfo {pages} {041051}
  (\bibinfo {year} {2020})}\BibitemShut {NoStop}%
\bibitem [{\citenamefont {Sete}\ \emph {et~al.}(2021{\natexlab{b}})\citenamefont {Sete}, \citenamefont {Didier}, \citenamefont {Chen}, \citenamefont {Kulshreshtha}, \citenamefont {Manenti},\ and\ \citenamefont {Poletto}}]{sete2021parametric}%
  \BibitemOpen
  \bibfield  {author} {\bibinfo {author} {\bibfnamefont {E.~A.}\ \bibnamefont {Sete}}, \bibinfo {author} {\bibfnamefont {N.}~\bibnamefont {Didier}}, \bibinfo {author} {\bibfnamefont {A.~Q.}\ \bibnamefont {Chen}}, \bibinfo {author} {\bibfnamefont {S.}~\bibnamefont {Kulshreshtha}}, \bibinfo {author} {\bibfnamefont {R.}~\bibnamefont {Manenti}},\ and\ \bibinfo {author} {\bibfnamefont {S.}~\bibnamefont {Poletto}},\ }\bibfield  {title} {\bibinfo {title} {Parametric-resonance entangling gates with a tunable coupler},\ }\href {https://doi.org/10.1103/PhysRevApplied.16.024050} {\bibfield  {journal} {\bibinfo  {journal} {Phys. Rev. Appl.}\ }\textbf {\bibinfo {volume} {16}},\ \bibinfo {pages} {024050} (\bibinfo {year} {2021}{\natexlab{b}})}\BibitemShut {NoStop}%
\bibitem [{\citenamefont {Ye}\ \emph {et~al.}(2021)\citenamefont {Ye}, \citenamefont {Peng}, \citenamefont {Naghiloo}, \citenamefont {Cunningham},\ and\ \citenamefont {O'Brien}}]{ye2021engineering}%
  \BibitemOpen
  \bibfield  {author} {\bibinfo {author} {\bibfnamefont {Y.}~\bibnamefont {Ye}}, \bibinfo {author} {\bibfnamefont {K.}~\bibnamefont {Peng}}, \bibinfo {author} {\bibfnamefont {M.}~\bibnamefont {Naghiloo}}, \bibinfo {author} {\bibfnamefont {G.}~\bibnamefont {Cunningham}},\ and\ \bibinfo {author} {\bibfnamefont {K.~P.}\ \bibnamefont {O'Brien}},\ }\bibfield  {title} {\bibinfo {title} {Engineering purely nonlinear coupling between superconducting qubits using a quarton},\ }\href {https://doi.org/10.1103/PhysRevLett.127.050502} {\bibfield  {journal} {\bibinfo  {journal} {Phys. Rev. Lett.}\ }\textbf {\bibinfo {volume} {127}},\ \bibinfo {pages} {050502} (\bibinfo {year} {2021})}\BibitemShut {NoStop}%
\bibitem [{\citenamefont {Ye}\ \emph {et~al.}(2025)\citenamefont {Ye}, \citenamefont {Kline}, \citenamefont {Yen}, \citenamefont {Cunningham}, \citenamefont {Tan}, \citenamefont {Zang}, \citenamefont {Gingras}, \citenamefont {Niedzielski}, \citenamefont {Stickler}, \citenamefont {Serniak} \emph {et~al.}}]{ye2025near}%
  \BibitemOpen
  \bibfield  {author} {\bibinfo {author} {\bibfnamefont {Y.}~\bibnamefont {Ye}}, \bibinfo {author} {\bibfnamefont {J.~B.}\ \bibnamefont {Kline}}, \bibinfo {author} {\bibfnamefont {A.}~\bibnamefont {Yen}}, \bibinfo {author} {\bibfnamefont {G.}~\bibnamefont {Cunningham}}, \bibinfo {author} {\bibfnamefont {M.}~\bibnamefont {Tan}}, \bibinfo {author} {\bibfnamefont {A.}~\bibnamefont {Zang}}, \bibinfo {author} {\bibfnamefont {M.}~\bibnamefont {Gingras}}, \bibinfo {author} {\bibfnamefont {B.~M.}\ \bibnamefont {Niedzielski}}, \bibinfo {author} {\bibfnamefont {H.}~\bibnamefont {Stickler}}, \bibinfo {author} {\bibfnamefont {K.}~\bibnamefont {Serniak}}, \emph {et~al.},\ }\bibfield  {title} {\bibinfo {title} {Near-ultrastrong nonlinear light-matter coupling in superconducting circuits},\ }\href {https://www.nature.com/articles/s41467-025-59152-z} {\bibfield  {journal} {\bibinfo  {journal} {Nature Communications}\ }\textbf {\bibinfo {volume} {16}},\ \bibinfo {pages} {3799} (\bibinfo {year} {2025})}\BibitemShut {NoStop}%
\bibitem [{\citenamefont {Ye}\ \emph {et~al.}(2024)\citenamefont {Ye}, \citenamefont {Kline}, \citenamefont {Chen}, \citenamefont {Yen},\ and\ \citenamefont {O’Brien}}]{ye2024ultrafast}%
  \BibitemOpen
  \bibfield  {author} {\bibinfo {author} {\bibfnamefont {Y.}~\bibnamefont {Ye}}, \bibinfo {author} {\bibfnamefont {J.~B.}\ \bibnamefont {Kline}}, \bibinfo {author} {\bibfnamefont {S.}~\bibnamefont {Chen}}, \bibinfo {author} {\bibfnamefont {A.}~\bibnamefont {Yen}},\ and\ \bibinfo {author} {\bibfnamefont {K.~P.}\ \bibnamefont {O’Brien}},\ }\bibfield  {title} {\bibinfo {title} {Ultrafast superconducting qubit readout with the quarton coupler},\ }\href {https://www.science.org/doi/full/10.1126/sciadv.ado9094} {\bibfield  {journal} {\bibinfo  {journal} {Science Advances}\ }\textbf {\bibinfo {volume} {10}},\ \bibinfo {pages} {9094} (\bibinfo {year} {2024})}\BibitemShut {NoStop}%
\end{thebibliography}%

\end{document}